\documentstyle[aas2pp4]{article}

\newcommand{\simless}{\stackrel{\scriptstyle <}{\scriptstyle \sim}}
\newcommand{\simgreat}{\stackrel{\scriptstyle >}{\scriptstyle \sim}}

\lefthead {Kurtz and Mink}
\righthead {RVSAO 2.0}

\begin {document}

\title{RVSAO 2.0: Digital Redshifts and Radial Velocities}

\author{Michael J. Kurtz and Douglas J. Mink}
\affil{Harvard-Smithsonian Center for Astrophysics, Cambridge, MA 02138}
\affil{email: kurtz@cfa.harvard.edu, dmink@cfa.harvard.edu}
\affil {{\rm Submitted for publication in} Publications of the
Astronomical Society of the Pacific, {\rm 13. March 1998}}

\begin {abstract}

RVSAO is a set of programs to obtain redshifts and radial velocities
from digital spectra.  RVSAO operates in the IRAF(Tody 1986, 1993)
environment.  The heart of the system is {\bf xcsao}, which implements
the cross-correlation method, and is a direct descendant of the system
built by Tonry and Davis (1979).  {\bf emsao} uses intelligent
heuristics to search for emission lines in spectra, then fits them to
obtain a redshift.  {\bf sumspec} shifts and sums spectra to build
templates for cross-correlation.  {\bf linespec} builds synthetic
spectra given a list of spectral lines.  {\bf bcvcorr} corrects
velocities for the motion of the earth.  We discuss in detail the
parameters necessary to run {\bf xcsao} and {\bf emsao} properly.

 We discuss the reliability and error associated with {\bf xcsao}
derived redshifts.  We develop an internal error estimator, and we
show how large, stable surveys can be used to develop more accurate
error estimators.

We develop a new methodology for building spectral templates for
galaxy redshifts, using the new templates for the FAST spectrograph
(Fabricant, et al, 1998) as an example.  We show how to obtain
correlation velocities using emission line templates.  Emission line
correlations are substantially more efficient than the previous
standard technique, automated emission line fitting.

Using this machinery the blunder rate for redshift measurements can be
kept near zero; the automation rate for FAST spectra is $\sim$95\%.

We use {\bf emsao} to measure the instrumental zero point offset and
instrumental stability of the Z-Machine and FAST spectrographs.  

We compare the use of RVSAO with new methods, which use Singular Value
Decomposition and $\chi^2$ fitting techniques, and conclude that the
methods we use are either equal or superior.  We show that a
two-dimensional spectral classification of galaxy spectra can be
developed using our emission and absorption line templates as
physically orthogonal basis vectors.

\end {abstract}

\section {\label{intro} Introduction}

Radial velocities are, along with position and brightness, among the
fundamental measured values of astronomy.  Recent technical advances are
substantially increasing our ability to aquire radial velocity data; in the
decade of the 1990's the rate at which radial velocity measurements are taken
will increase by two or three orders of magnitude.  Substantial effort is
required for these data to be reduced and analyzed in an accurate and timely
fashion; here we describe the current reduction methods which we have developed
for use by the Center for Astrophysics radial velocity and redshift programs,
as well as by others.

Doppler (1841) understood that radial velocities would affect the
color of stars (by analogy with the pitch of sound); Fizeau (1848,
1870) first recognized that this would mean a shift in the position of
the Fraunhofer lines.  Huggins (1868) made the first (visual) attempt
(in 1862) to observe the shifts.  Vogel (1892) made the first accurate
photographic measurements, and established most of the procedures
necessary to calibrate and reduce the measurements of line positions
to radial velocities.

Correlation methods for obtaining radial velocities were first
suggested by Fellgett (1953), who was influenced by radar studies
during World War II.  Griffin (1967) was the first to implement these
techniques.  Note that Griffin credits Evershed (1913) with inventing
the basic technique; Griffin further notes that Babcock (1955) had
already built a similar instrument.  Griffin's instrument performed
analog correlations by physically shifting a template spectrum in the
focal plane of the spectrograph, a technique which is still in heavy
use today (e.g. Baranne, et al, 1979).

Digital power spectrum techniques for the estimation of lag have long
been known (e.g. Blackman and Tukey, 1958, and references therein).
Their use first became practical with the advent of digital detectors,
fast digital computers, and the FFT algorithm (Cooley and Tukey,
1965).

Simkin (1974) first showed how Fourier techniques could be used to
obtain radial velocities and velocity dispersions from digital
spectra.  Several groups used power spectrum techniques to obtain
velocity dispersions (see Sargent, et al 1977, and references therein)
and obtained velocities as a byproduct of their analysis, but
apparently the first use of digital cross-correlation specifically to
obtain radial velocities was by Lacy (1977) who did not use Fourier
techniques, but used direct convolution with a digital mask, emulating
Griffin's (1967) analog technique.

Tonry and Davis (1979; hereafter TD79) studied the use of power
spectrum techniques to obtain redshifts from digital spectra, and
demonstrated the effectiveness of the method.  TD79 invented the $r$
statistic, which can be calibrated to give both the confidence and
error of a measurement.  The techniques and software described here
are directly descended from the TD79 system; in September 1990 radial
velocity reductions at the CfA were moved from the old Data General
Nova computer where the TD79 system resided onto a Unix workstation
and the IRAF (Tody, 1986, 1993) environment.

We began with the IRAF task XCOR by G. Kriss and routines from TD79,
translated into Fortran by J. Tonry (\cite{ton88}); these were
extensively modified, and resulted in {\bf xcsao} version 1.0 (Kurtz, et al,
1992; Paper 1). Additionally we used algorithms from the
REDUCE/INTERACT system (\cite{mak82}), as modified by J. Thorstensen.
The emission line finding programs had a somewhat different history.
When the radial velocity reductions were moved from the Novas onto
Unix the emission line programs were implemented as stand-alone C
programs translated from the FORTH of TD79 by W. Wyatt; work began in
1991 on a new IRAF task, resulting in {\bf emsao} (Mink and Wyatt, 1995).

The software described here has been used extensively; examples
include the redshift surveys of Huchra, et al (1995); da Costa, et al
(1994), Shectman, et al (LCRS:1996), Vettolani, et al (1997), and
Geller, et al (1997).  Stellar use revolves around the CfA Digital
Speedometry program (Latham, 1985) and its many projects and
collaborations.  Latham (1992) lists several of these projects.

Nordstr\"om, et al (1994) have described the techniques used by the
CfA stellar group in detail; we will concentrate on issues related to
galaxy redshifts, although some stellar data will be used in section
\ref{error} on error.  We will conform to the convention that {\it
user settable parameters} will be in {\it italics}, PARAMETERS in the
SPECTRUM HEADER will be in CAPITALS, and IRAF {\bf tasknames} will be
in {\bf lowercase bold}.  Extensive on-line documentation, help files,
and examples, as well as the source code and executables can be found
at http://tdc-www.harvard.edu/.

\section {\label{prac} Practical use of {\bf xcsao}}

{\bf xcsao} is the heart of the RVSAO system.  In this section we give
an overview of the correlation system.  We point out critical features
which investigators need to consider when setting up the reductions
for a new project, to maximize the efficiency of measurement,
and minimize systematic errors.  The RVSAO system consists of several
IRAF tasks, including {\bf xcsao}, the algorithmic details of each of
them are described in appendix \ref{appendix}.

\subsection {\label{prep} spectrum preparation}

Tokarz and Roll (1997) discuss the steps we take to obtain 1-D
wavelength calibrated spectra suitable for redshift measurements.
Once the 1-D spectra are in hand it is necessary to tell {\bf xcsao} the
wavelength range over which the data are good, using the parameters
{\it st\_lambda} and {\it end\_lambda}.  Depending on the details of the
instrumental set up substantial additional error can occur if these
parameters are not set, or are set incorrectly.  For example, if a
substantial portion of the spectrum is from a region of the detector
with little or no sensitivity, or is from a region of the spectrum
with poor sky subtraction and strong night sky lines, the final
redshift will be compromised.

\subsection {\label{sup} continuum and emission line suppression}

For all spectra the continuum must be removed; {\bf contpars} (see
section \ref{contpars}) performs these tasks here, and was only
slightly modified from the IRAF ONEDSPEC {\bf continuum} package, as
implemented in the RV package (\cite{fit93}).

The continuum can be subtracted or divided out.  Subtraction preserves
the correct relative amplitudes for the lines in the data, and thus
the correct signal to noise behavior in the cross-correlation.
Division preserves the correct equivalent widths of the lines, and
thus the correct parameterization of the spectrum.  For spectral
classification studies division is preferred (Kurtz, 1982); for radial
velocity measurements subtraction is superior.  Division by the
continuum results in amplified noise in the blue part (the low S/N
part) of the spectrum.  For moderate S/N spectra, typical of FAST,
the difference between the two techniques is small, but subtraction
shows smaller redshift residuals by about 25\%\ ; while for very high
S/N spectra on FAST, typical of our calibration spectra, the
division method gives slightly smaller residuals.

In normal use of {\bf xcsao} only continuum subtraction is allowed, but if
the parameter DIVCONT is set true (T) in the template spectrum header the
continua of both the object and template spectra will be divided out.

\begin{figure}[t]
     \plotone{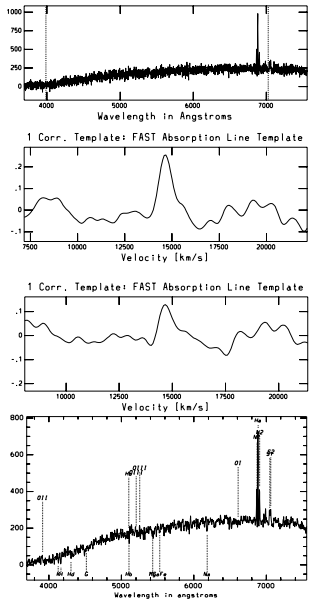}
     \caption{\label{emcut} 
     The effect of emission line suppression on absorption
     line correlation redshifts.  The upper panel shows the observed
     spectrum; the next panel down shows the correlation function with
     an absorption line template after emission line clipping; the
     next down shows the correlation function when the emission lines
     are not clipped; the bottom panel shows the observed spectrum
     after smoothing, and with the main lines marked.}

     \end{figure}

Many galaxy spectra show both emission and absorption lines; in
general the redshifts derived from the emission lines will be
different from the absorption line velocities.  To obtain a
correlation velocity from an absorption line template for a spectrum
with strong emission lines it is necessary to suppress the emission
lines; figure \ref{emcut} shows a typical spectrum and its
correlation function with one of our standard absorption templates
suppressing the emission lines  and without suppressing them,
the bottom panel shows the spectrum, smoothed and with the lines
marked.  The reduction with emission line suppression produces a
believable redshift, with an $r$ (TD79) value of 4.50, the reduction without
it, where the $r$ value is 1.88, does not.

Removing emission lines before correlating with an absorption line
template was a routine feature of the TD79 software; in {\bf xcsao} we
have extended and generalized the procedure.  Both emission and
absorption lines may now be removed, and the process may be controlled
by keywords in the template header.  Because we now obtain emission
line velocities using correlation methods (section \ref{emt}), the
same emission line suppression cannot be used for every template.

For routine redshift reductions at the CfA we use a subset of the {\bf
xcsao} capabilities, with emission lines replaced by the continuum
when we correlate against an absorption line template, and the
absorption lines replaced by the continuum when we correlate against
an emission line template.  The exact parameters used are stored in
each template, normally two sigma variations above(below) the
continuum are sufficient to remove emission(absorption) lines, with
the number of iterations and growing parameter as set in {\bf
contpars}(section \ref{contpars}). 

The emission(absorption) line suppression is controlled by the
parameters {\it s\_emchop} ({\it t\_emchop}) for the object (template)
spectra.  Other parameters are used as well, and when control is
given to the template or object spectrum header the interactions can
be complex.  They are more fully described in section \ref{xc}. 

In addition {\bf xcsao} permits the user to replace specific regions
of the spectrum with a simple linear approximation to the continuum.
This feature is typically used to prevent poor subtraction of the
bright night sky lines from compromising the results.

\subsection {\label{filt} apodization, zero padding, Fourier filtering}

Once the continuum has been removed the spectra are apodized, zero
padded, and bandpass filtered. Each of these operations has a
substantial effect on the final results.

Apodization is the simplest; essentially the goal is to remove any
ringing in the Fourier transform, by forcing the ends of the spectrum
smoothly to zero, while suppressing as little of the actual data as
possible.  Our apodization is performed by a cosine taper function,
which begins a symmetric set percentage of the spectrum from the ends.
For both FAST spectra and the earlier Z-Machine (Latham, 1982) spectra
0.05 is a reasonable value, i.e. the taper begins 5\%\ from the ends
(section \ref{xc}).

Zero padding the spectrum is intended to remove any artifacts caused
by computation of the correlations in Fourier space, thus using a
circular convolution.  The primary artifact which the zero padding
removes is the confusion of $H\alpha$ with $OII 3727$ due to the wrap
around of the convolution.  The zero padding has two side effects
which must be considered.  First, the relation of the TD79 $r$
statistic with error is changed, although the actual calculated error
remains correct (section \ref{error}).  Second, the envelope of noise
fluctuations of the correlation function, which is flat in the
non-zero padded case, is, in the zero padded case, a symmetric linear
function of the number of overlapping non-zero pixels, and is maximum
at the redshift of the template spectrum.  For the case of low S/N
spectra where the redshift is substantially different from the
template's redshift this structure in the noise can result in the
wrong correlation peak being chosen.  Zero padding can be controlled
via the parameter file, or the template header.  We only zero pad
spectra when correlating against the emission line template.

The design of the Fourier bandpass filter is critical to the optimal
measurement of redshifts.  As with the apodization we use a cosine
taper to suppress the ends of the (in this case) Fourier spectrum.
Several other filter techniques were tried (e.g. Oppenheim and
Schafer, 1975) but no difference was seen for any reasonable choice of
taper function (a sharp cutoff is not reasonable because of Gibbs
ringing).  In addition we tested a spectral weighting function shown
by Hassab and Boucher (1979) to produce the maximum likelihood
estimator for the lag (radial velocity) in the limit of infinitely
wide spectra; this weighting function had no positive effect on our
results, and we have not implemented it.

Removing high spatial frequency information, via a high-stop Fourier
filter, is intended to increase the S/N by removing information which
contains more noise than signal.  The design question is where to set
the high frequency turnoff; the method we use is to examine sets of
high S/N calibration spectra, e.g. our nightly exposures of NGC4486b.
We correlate each of these against the best match template (in this
case NGC7331), using a high pass Fourier filter which filters out ALL
the low frequency information, leaving only the high frequency noise.
If the turn-on frequency is set too high, the correlations all give an
incorrect redshift; if the turn-on frequency is set too low, all
correlations give the correct redshift.  We choose the turn-on
frequency where half the redshifts are correct and half incorrect, and
set this turn-on-frequency to the turn-off frequency for our high-stop
filter.  This procedure gives a turn-off frequency approximately equal
to that obtained by the ``optimal filter'' method of Brault and White
(1971) which chooses that point where the power of the signal is twice
that of the photon noise.  This frequency is less than half that which
corresponds to the projected slit width of the FAST.

The high-stop filter is only used for absorption line spectra.  For
emission line spectra the redshifts are seriously degraded if the high
spatial frequencies are removed from the data.  The key word FI-FLAG
in the template header controls the implementation of the high-stop
filter.  In addition, as it is possible to pre-filter the templates,
to save unnecessary computing, this flag also controls whether and how
to filter the templates.  Figure \ref{header} shows all the
possibilities.

Removing low spatial frequency information, via a low-stop Fourier
filter, is intended to remove any residual large-scale systematics
which remain following the continuum suppression.  Essentially this can
be viewed as a second continuum removal, equivalent to the continuum removal
technique of LaSala and Kurtz (1985), thus giving what Kurtz and
LaSala (1991) call a ``reflattened'' spectrum.  

The design question is where to put the low frequency turn-on point.
The difficulty with making this decision is that there is no
point where excluding all information with higher spatial
frequencies does not result in a redshift (i.e. even the lowest
spatial frequencies still contain accurate redshift information), and
there is no reasonable point where suppressing more low frequency
information does not result in lower residuals for high S/N sets of
spectra, such as our set of NGC4486b spectra.

We therefore set the low frequency turn-on point by a simple
heuristic.  We estimate the scale in wavelength of the broadest
spectral feature useful in estimating a redshift, in the case of FAST
galaxy spectra this is the change in the slope of the continuum around
the CaII H+K lines, and we suppress those spatial frequencies which
correspond to twice this scale, or greater.

The remaining design decision for the Fourier filter is the width of
the turn-on and turn-off ramps.  It may be expected that this is only
important at the low frequency turn-on point, as the power there is
typically two orders of magnitude above the high frequency turn-off
point.  The problem is that if the turn-on is too sharp, Gibbs ringing
will be introduced into the data.  A full turn-on width of 1.5\%\ of
the width of the power spectrum is sufficient to ameliorate this
effect.

The exact filter implemented, especially the exact implementation of
the low-stop filter, affects the resulting redshifts, their errors,
and the relation of the TD79 $r$ statistic with their errors.  For
example, for the set of NGC4486b spectra, the mean redshift obtained
using only the lowest spatial frequencies which we include in our
standard filter differs from the mean redshift obtained by only using
the highest included spatial frequencies by $61 km/s$, and different
reasonable choices for the Fourier filter can give redshifts which
differ in the mean by $10 km/s$.  These differences may be compared
with a typical variation about the mean of $15 km/s (1\sigma)$.  The
sign and amplitude of this effect changes with each object-template
pair, NGC4486b vs. NGC7331 is a typical result.  In addition the
relation of the TD79 $r$ statistic with error depends on the filter
(section \ref{error}).

\subsection {\label{bin} cross-correlation, rebinning, and 
redshift evaluation}

The cross correlation is the normal product of the Fourier transform
of the object spectrum with the conjugate of the transform of the
template spectrum, as described in TD79.

The object spectrum and the template spectra need to be pairwise
rebinned to have a common dispersion.  The number of bins {\it nbins}
is set by the user and must be a power of two; we recommend that {\it
nbins} always be larger than the number of observed pixels.  The
spectral region rebinned is set to obtain the maximum overlap between
the template spectrum and the portion of the object spectrum between
{\it st\_lambda} and {\it end\_lambda} in the rest frame.  On the
first pass the rest frame is determined by a user guess to the
redshift ({\it czguess}), or from a previous reduction (section
\ref{xc}).  On subsequent passes (if {\it nzpass}$>$0) the rest frame
is determined from the redshift obtained in the previous pass.  We
recommend that {\it czguess} be set to the approximate redshift
expected (normally a better guess than zero), and {\it nzpass}$=$2.

Next the correlation peak is determined and fit (section \ref{xc}).  The
type of fit ({\it pkmode}) has little effect on the result, but the
amount of the peak which is fit, {\it pkfrac}, is critical.  The fit
is performed from the top of the peak down to where the peak is {\it
pkfrac} of the maximum, thus more of the peak is fit if {\it
pkfrac}$=$0.5 than if {\it pkfrac}$=$0.7.  Because of side-lobes in the
correlation function due to the proximity of NII to H$\alpha$,
emission line correlation peaks cannot be fit as far down the peak as
absorption line correlations.  The template header parameter PEAKFRAC
overrides {\it pkfrac} on a template by template basis; we use this to
set the fit parameters for the emission line templates.

\section {\label{error}Error Analysis}

There are three main questions concerning the output of xcsao: are the
results reliable, can the process be automated?  what is the size and
nature (random or systematic) of the error? and, does what is measured
correspond to the physical property the investigator wants to measure?

\subsection {\label{rel}reliability}

To answer the first two of these questions we use a dataset
designed for this purpose; it contains 626 pairs of spectra
observed with the FAST spectrograph between 1994 and 1996.  Each pair
consists of two independent observations of the same object; about
half of these were observed to calibrate the velocity errors
for the 15R survey (Geller, et al. 1998), and about half are spectra
which were below the quality standard, and required a second
integration (these  would be summed in our normal reductions, but not
here).  All these spectra have been subjected to our normal processing
(Tokarz and Roll, 1996), and thus have had most of the cosmic rays
removed by a labor intensive process.

     \begin{figure}[t]
     \plotone{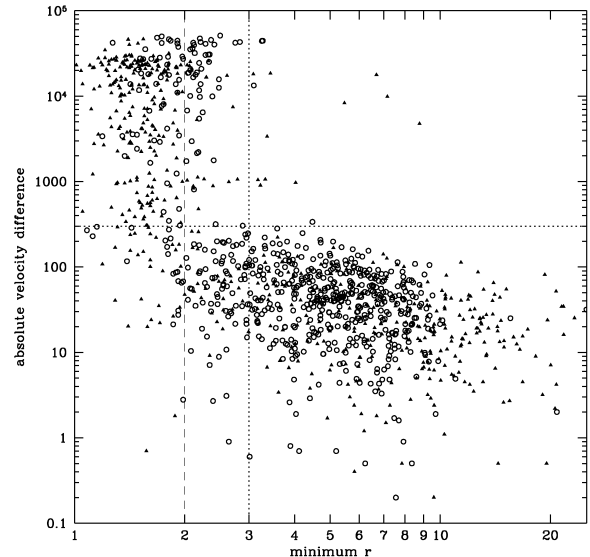}
     \caption{\label{blunders} The blunder diagram, see text for discussion
      The thick dotted lines are at $r=3$,
      and velocity differences of 300 $km/s$; the thin dashed line is at 
      $r=2$; both axes are on a log scale.}

     \end{figure} 

As discussed by TD79, the $r$ statistic can be calibrated as a
confidence measure.  We prefer to calibrate it empirically, rather
than use the prescription in TD79.  Figure \ref{blunders} shows the
results of correlating each of the 1252 spectra against each of two
templates.  Plotted are the absolute value of the velocity difference
between the two observations versus the minimum $r$ value for each
pair.  Each pair appears on the graph twice.  The open circles are
measurements using the NGC7331 template, and the filled triangles use
an emission line template, emtemp.

Figure \ref{blunders} shows two groupings, those with the absolute
velocity difference $\Delta v\simless 300 km/s$; which we will assume are
reliable observations (differences $\simless 300 km/s$ are consistent
with the expected random errors), and those where the velocity
difference is greater, which we will assume unreliable.  For the
moment we will adopt an $r$ value of 3.0 ($r_{min}$), above which we
expect the velocity determination to be reliable.  This is lower than
is typically used for FAST reductions.

We would then expect no points in the upper right quadrant of the
plot, where $r > r_{min}$, and the $\Delta v >$ 300 $km/s$; however
there are, 15 points in what we will call the ``blunder'' region.

We will examine each point to determine which rules would catch the
blunders in a fully automated reduction.  The two nearly co-incident
circles (circles are reductions with the NGC7331 template) at $r\sim
3.2$ and velocity difference about $50000 km/s$ are both objects with
strong emission lines.  It is known that the NGC7331 template
systematically returns velocities of $49000 km/s$ for some emission
line objects, so along with the high $r$ value emission line velocity
these measures could be discarded automatically.  The circle at $r\sim
3.1, \Delta v \sim 18000km/s$ is also an object with strong emission
lines.  A simple rule which requires that spectra with discordant, but
otherwise valid velocities be checked manually would catch this, but
the reduction could not be fully automatic.

The four triangles (triangles are the emission line template) between
$r\sim 3.2$ and $r\sim 4.2$ having a $\Delta v \sim 948km/s$ all have
good absorption line velocities, and would be caught by the rule of
discordance.  They would be caught by another rule, however, one which
does not require that the absorption line velocity be ``good;'' they
are all spectra where NII 6583\AA\ is stronger than $H\alpha$, and the
difference with the absorption line templates is about $948km/s$.  We
adopt the rule that all emission line velocities which differ from an
absorption line velocity (even if one with a low $r$ value) by about
$948km/s$ must be checked manually.

The circle near $r\sim 4.5$ and $\Delta v \sim 350km/s$ is probably
not a real blunder.  Examination of the POSS prints shows that the
object has two nucleii with $4''$ separation.  We assume that the
velocity difference is real, and that the two observations of this
object each correspond to a different nucleus.

All of the seven remaining objects in the ``blunder'' region of the
plot are emission line velocities.  One has the night sky line at
5577\AA\ mistaken for OII 5007\AA; we can eliminate this error by
either turning the badlines removal feature on to replace the region
around 5577\AA\ with the continuum, or by adopting the rule that all
emission line redshifts near $34152km/s$ be examined manually.

The remaining six are all spectra contaminated by cosmic rays.  Five
of these would be tagged by the rule of discordance, and the sixth has
an absorption line $r$ value of 2.77, so it would be tagged by only a
slightly more stringent rule of discordance; it cannot be assumed,
however, that the cosmic ray problem can be solved by looking for
absorption lines.  We therefore require that emission line redshifts
must all be checked using {\bf emsao} (section \ref{emsao}), and that
at least four lines must be found which correspond to the correlation
velocity, and at least two must be fit; ``blunders'' are made when
only three lines are found, or only one line is fit.  Using that
criterion all six spectra would be tagged for visual inspection as
well as 30 of the 178 emission line velocities ($r>3$) which do not
have a confirming absorption line velocity with $r>3$.

For the 610 objects (of 626 total objects observed) where at least one
of the two different template reductions gave a result with $r>3$,
xcsao yielded the correct result with no further problem for 595.  Of
the remaining 15 spectra 12 are easily discovered because two valid
redshift measures disagree; one is probably caused by source confusion
on the sky, and two are found by {\bf emsao}.

Looking at figure \ref {blunders}, it is clear that many spectra where
$2<r<3$ do indeed give the correct redshift.  Of the 16 objects which
have neither emission nor absorption reduction with $r>3$ four have
both emission and absorption redshifts equal (within normal errors)
and could be accepted (we do not currently do this).  Also 66 spectra
where the emission line $r$ value is $>3$ have confirming absorption
line velocities with $2<r<3$; if these are assumed correct (which we
also do not currently do), then the number of emission line spectra
which must be visually inspected after {\bf emsao} would drop from 30
to 19.  This would bring the total number which require visual
inspection to 31, or 5\%\ .

With the aid of {\bf emsao} for quality control, and partial manual
reduction of 31 spectra, {\bf xcsao} obtained the correct redshift for all
614 objects which yielded redshifts, save for the one object which was
probably an observational error.

A second experiment was made using 8606 emission line spectra from the
Z-Machine archive, which had $r$ values with the
emission line template above 3.  The results of the correlation with
the emission line template were compared with the stored redshift in
the archive, which was obtained by manually fitting the emission lines
with a precursor program to {\bf emsao}.  After sifting the results using
rules like those described above fifteen spectra (0.2\%\ ) had the
wrong redshift; essentially all these spectra were the victims of
very poor sky subtraction.  This may be compared with the twenty-four
spectra where the redshifts were incorrectly listed in the archive.
The Z-Machine spectra were substantially noisier than the FAST spectra;
nearly 15\%\ failed the sifting and would have had to be manually
reduced. 

Used carefully the RVSAO suite provides redshifts with a very low
blunder rate.  The automation rate obtainable with RVSAO is strongly
affected by the S/N of the observations.  Absorption line objects must
be observed long enough to have a fully reliable absorption
correlation velocity (we currently use $r \simgreat 4$, which is
conservative) or a confirming weak emission velocity.  Emission line
spectra must have a confirming weak absorption redshift, or be based
on at least four lines, and at least two of the four must be fit by
{\bf emsao} (section \ref{emt}).

\subsection {\label{est}error estimation}

Besides estimating the redshift of a spectrum {\bf xcsao} also
estimates the error in the redshift.  The error estimator can be
derived analytically following the discussion in section III.c.i of
TD79 with the additional assumption of sinusoidal noise, with the
halfwidth of the sinusoid equal to the halfwidth of the correlation
peak.  The derived error estimator is:

$$ error = {3 \over 8}{w \over (1 + r)} $$

where $error$ is the error in a single velocity measurement by {\bf xcsao},
$w$ is the FWHM of the correlation peak, and $r$ is as defined in TD79.

While the assumption of sinusoidal noise with halfwidth equal to the
correlation peak's halfwidth is reasonable, there is no compelling
argument for this assumption.  Therefore it is necessary to
demonstrate the effectiveness of the approximation by experiment.

We will use four datasets to examine the behavior of the error
estimator: the 610 duplicates described above; the 8606 Z-Machine
emission line spectra described above; 7810 synthetic spectra, each
identically Poisson sampled from a 45\AA\ section of a model
atmosphere for a 5500K dwarf star (Kurucz, 1992),  taken from the set
of synthetic stellar templates used by the CfA digital speedometry
program (Morse, et al 1991; Nordstr\"om, et al 1994); 50000 synthetic
spectra, using the same 5500K dwarf star template, each with a
different number of simulated photons (we confine ourselves to using
the 49880 spectra which, when correlated against the synthetic template,
achieved $r>3$).  

First we will look at the set of duplicate spectra.  We will limit
ourselves to cases where both reductions have $r$ values $>3.5$.  This
is $400 \pm 15$ spectra for the two absorption line combinations, and 297
for the emission line comparison.

     \begin{figure}[t]
     \plotone{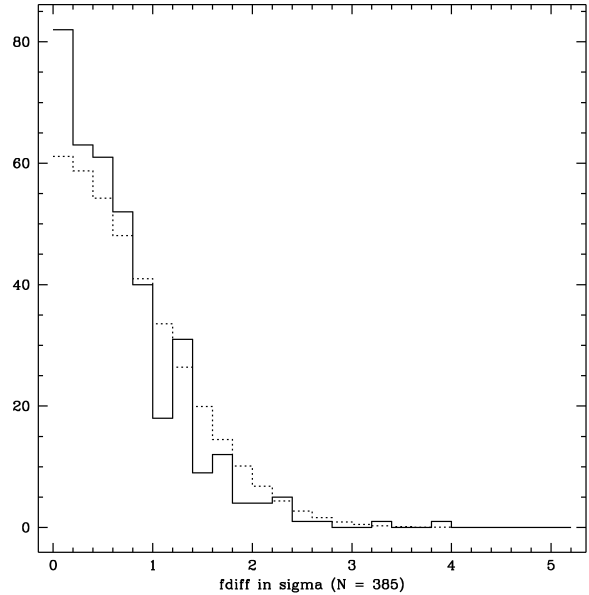}
     \caption{\label{fvelhisto} 
     The solid histogram is the
     distribution of velocity differences for duplicate observations
     of the same object, reduced with the fn7331temp template, divided
     by the {\bf xcsao} error estimate.  The dotted histogram is the
     expected Gaussian distribution.}

     \end{figure}

Figure \ref {fvelhisto} shows a typical result.  The solid line shows
a histogram of the absolute values of the differences between two
observations of the same object, both reduced in the same way using
the NGC7331 template, and divided by the sum in quadrature of the
errors calculated by {\bf xcsao} for the reductions.  The dotted
histogram is the expected Gaussian distribution; it is clearly broader
than the data.  {\bf xcsao} overestimated the error by $\sim 20\%\ $.

     \begin{figure}[htp]
     \plotone{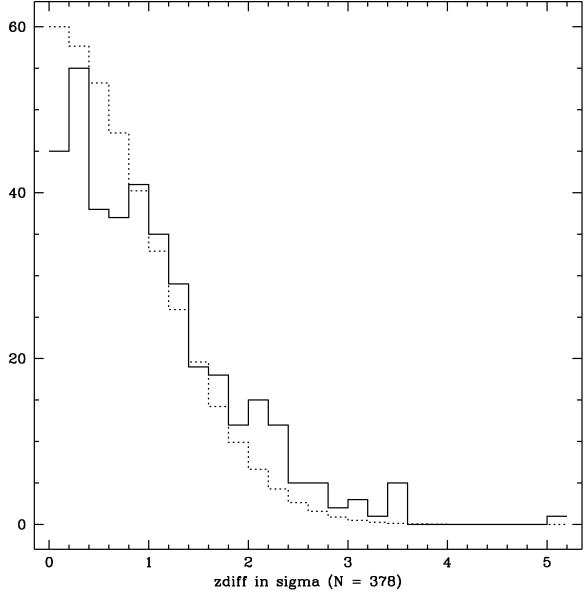}
     \caption{\label{zvelhisto} 
     The solid histogram is the
     distribution of velocity differences for duplicate observations
     of the same object, reduced with the ztemp template, divided
     by the {\bf xcsao} error estimate.  The dotted histogram is the
     expected Gaussian.}

     \end{figure}

Figure \ref {zvelhisto} is similar to Figure \ref {fvelhisto}.  Here
the ztemp template was used on the FAST data; ztemp is a combination
of bright galaxy spectra taken with the Z-Machine, and in use at the
CfA since the days of TD79.  ztemp has a restricted wavelength
coverage ($\lambda\lambda$ 4500--6200 \AA) compared with the FAST
spectra, has a different resolution, and has different residual
systematics.  In Figure \ref {zvelhisto} the (dotted) Gaussian is
narrower than the (solid) data.  {\bf xcsao} underestimated the error
by $\sim 20\%\ $.

     \begin{figure}[hbp]
     \plotone{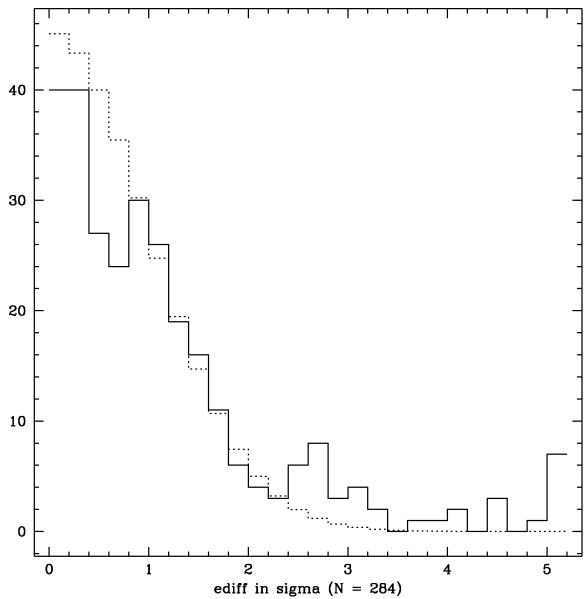}
     \caption{\label{evelhisto} 
     The solid histogram is the
     distribution of velocity differences for duplicate observations
     of the same object, reduced with the emtemp template, divided
     by the {\bf xcsao} error estimate.  The dotted histogram is the
     expected Gaussian.}

     \end{figure}

Figure \ref {evelhisto} shows a similar set of histograms for the
emission line template, emtemp.  emtemp is a synthetic spectrum made
before the creation of the {\bf linespec} task (section \ref{linespec}) to
match FAST emission line galaxy spectra.  Here the solid line which
represents the data cannot be transformed to match the dotted line
expected histogram by any multiplicative process (1.2 would be the
best multiplicative factor); it would still have more power in the
tail.  Adding 15 $km/s$ in quadrature helps remove power from the
tail.

     \begin{figure}[ht]
     \plotone{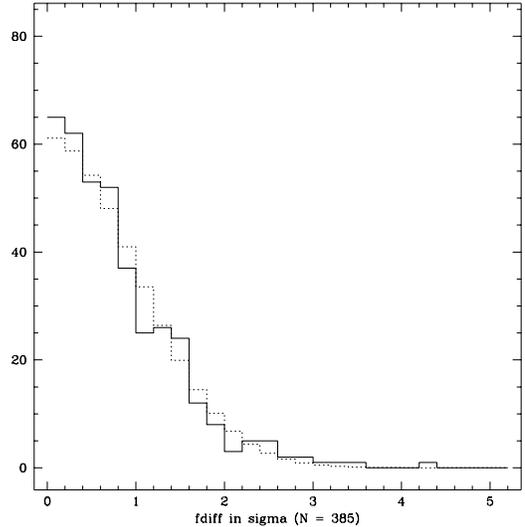}
     \caption{\label{fkvelhisto} 
     The solid histogram is the
     distribution of velocity differences for duplicate observations
     of the same object, reduced with the fn7331temp template, divided
     by the $k/(1+r)$ error estimate.  The dotted histogram is the
     expected Gaussian.}

     \end{figure}

     \begin{figure}[tp]
     \plotone{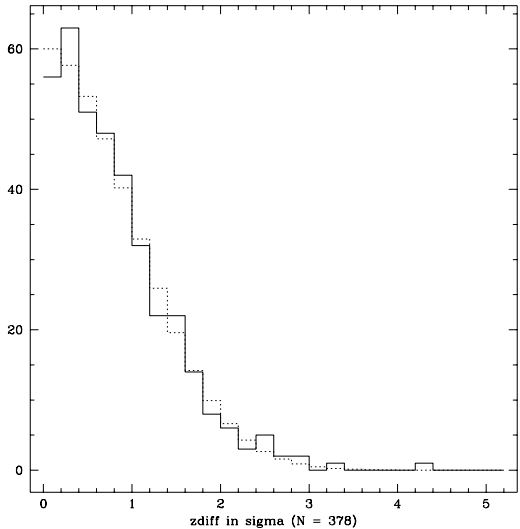}
     \caption{\label{zkvelhisto} 
     The solid histogram is the
     distribution of velocity differences for duplicate observations
     of the same object, reduced with the ztemp template, divided
     by the  $k/(1+r)$  error estimate.  The dotted histogram is the
     expected Gaussian.}

     \end{figure}

     \begin{figure}[bp]
     \plotone{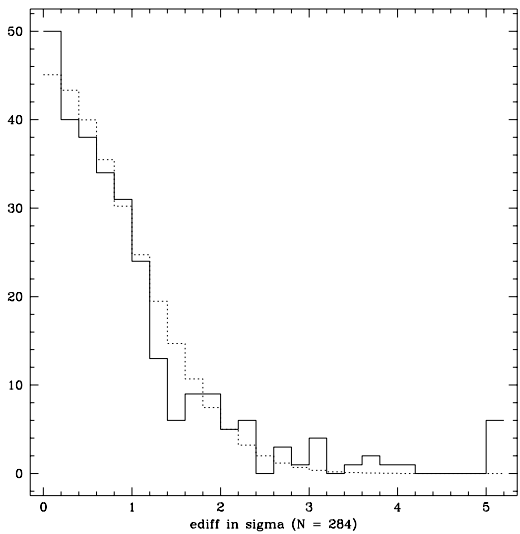}
     \caption{\label{ekvelhisto} 
     The solid histogram is the
     distribution of velocity differences for duplicate observations
     of the same object, reduced with the emtemp template, divided
     by the  $k/(1+r)$  error estimate.  The dotted histogram is the
     expected Gaussian.}

     \end{figure}

TD79, while giving a procedure to calculate the error, suggest that in
practice the error be calculated by calibrating $k$ in the equation
$error = k/(1+r)$ using external comparisons.  Paper 1 reiterates this
suggestion, noting that the measurement of $w$ has error, but when all
reduction parameters remain fixed, $w$ is essentially constant.  Figure
II of Paper 1 shows the effect of changing one of the reduction
parameters (the low frequency roll off of the Fourier filter), which
substantially changes the relation of $(1+r)$ to error, while $w$
scales correctly so that ${3w}\over {8(1+r)}$ still tracks the error.

Given a set of duplicate observations the constant $k$ can be
determined by {\it internal} comparisons.  The procedure is simply to
vary $k$ until the expected differences histogram matches the measured
one.  For the present case we obtain $k_{NGC7331}= 315km/s$,
$k_{ztemp}= 285km/s$, and $k_{emtemp}= 245km/s$.  Figures \ref
{fkvelhisto}, \ref {zkvelhisto}, and \ref {ekvelhisto} show the
distributions compared with the expected Gaussians; as expected, in all
cases the fit is better than with the unmodified {\bf xcsao} errors.
For large observing programs with stable reduction procedures, we
recommend using calibrated $k/(1+r)$ relations to estimate the error.

The duplicate spectra, as is typical of redshift survey data, do not
show a very large range of S/N, or $r$ value.  This is due to the fact
that one normally observes long enough to get a desired S/N, and no
longer.  If we continue to restrict the duplicate pairs to those where
both spectra achieved $r > 3.5$ (still lower than is normally required
for a FAST redshift) then it is not possible here to use the goodness
of fit to a Gaussian to prove that the error calculated by using $1+r$
is any better than a constant error, for each template.  If the error
were dominated by systematics one would expect the error to be
approximately constant.

     \begin{figure}[t]
     \plotone{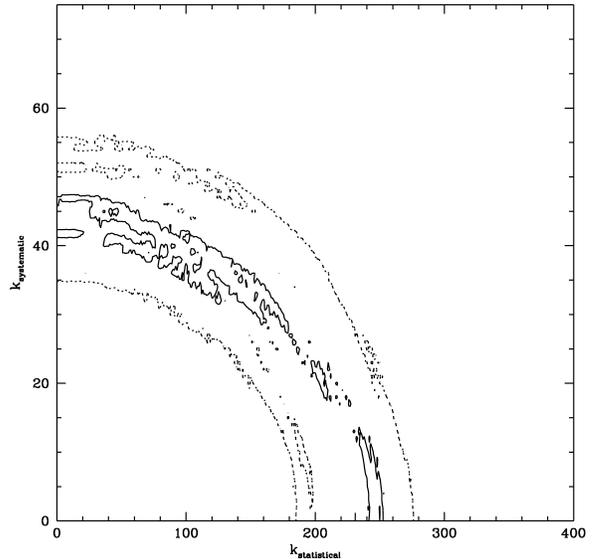}
     \caption{\label{emcontour} 
     Residuals in fit to Gaussian error distribution for pairs reduced
     with the emtemp template.  The contours represent lines of equal
     residual, position in the x,y space represents admixtures of the
     statistical and systematic error models; see text.}

     \end{figure}

We define a measure of error $$e = \sqrt{{k_{systematic}^2} +
{k_{statistical}^2\over (1 + r_1)^2} + { k_{statistical}^2\over (1 +
r_2)^2}}$$.  We then vary the values of $k_{systematic}$ and
$k_{statistical}$ and calculate the residuals of the fits to a
Gaussian. Note that $k_{systematic}$ is $\sqrt{2}$ times the error in
a single measurement.   Figure \ref {emcontour} shows the 
results for the duplicate pairs
reduced with emtemp.  The contours represent lines of equal residuals
in the fit to a Gaussian, in the $(k_{statistical}, k_{systematic})$
space.  The outer contour is 20\%\ larger than the inner contour.
Clearly either a fully systematic or a fully statistical error is
consistent with the data.

     \begin{figure}[t]
     \plotone{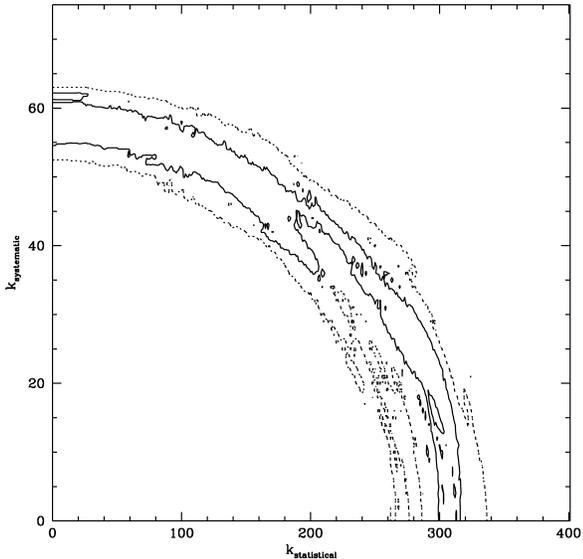}
     \caption{\label{n7contour} 
     Residuals in fit to Gaussian error distribution for pairs reduced
     with the fn7331temp template.  The contours represent lines of equal
     residual, position in the x,y space represents admixtures of the
     statistical and systematic error models; see text.}

     \end{figure}

Figure \ref {n7contour} shows the same diagram for the N7331 
template.  The inner
contour level here represents an absolute error half that of emtemp,
and the outer level is 33\%\ larger than the inner level.  Also here
one cannot rule out either a fully systematic or a fully statistical
error.   

In many cases it is not possible to reduce hundreds of duplicate
measurements using exactly the same reduction procedures to obtain an
improved error estimator; in these cases the {\bf xcsao} error estimator is
a reasonable choice.  The 20\%\ systematic deviations for the two
absorption line galaxy templates are the largest we have seen,
although Quintana, et al. (1996) suggest that for their data the error
is underestimated by $\sim 30\%$, by comparison with external
measurements.  While the {\bf xcsao} error estimator differs systematically 
from the true error for a particular combination of instrumental
set-up, reduction procedure, and template we have not seen any trend
for this to be a systematic over or under estimate.

While the 8606 Z-Machine emission line spectra cannot be used to
calibrate the error estimator, as to first order we are just comparing
the differences in two different methods of fitting H$\alpha$ in the
same spectrum, we can use them to look at any differences in zero
point, as a function of the fitting method.  The mean difference 
between the correlation velocity
and the velocity obtained by the semi-automated linefits is $0.56km/s
\pm .16$.

     \begin{figure}[t]
     \plotone{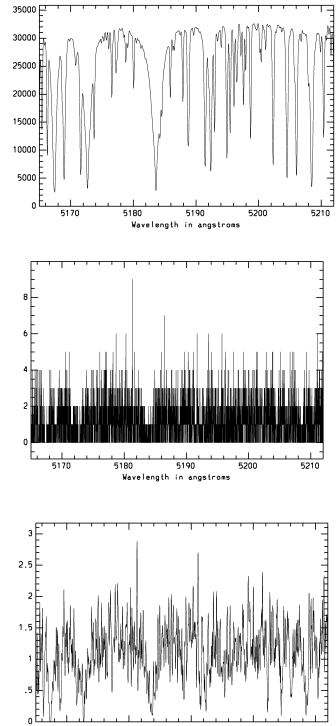}
     \caption{\label{g5} 
     The synthetic 5500K spectra.  Above is the template
     spectrum(Kurucz, 1992), in the middle is a typical poisson
     sampled spectrum, and below is the same spectrum smoothed.}
  
     \end{figure}

The 7810 synthetic spectra are each correlated against the template
from which they were identically randomly Poisson sampled; thus there is no
spectral type difference adding to the errors.  Figure \ref {g5} shows the
template, a typical sampled spectrum, and a smoothed version of the
typical spectrum.  By calculating the RMS
velocity about the expected velocity ($0 km/s$) we have a very
accurate measure of the error in a single measurement; the mean error
calculated by {\bf xcsao} is 3\%\ greater than this.  The distribution of
velocities is essentially Gaussian, and the deviation of the mean
velocity is within $1\sigma$ of zero.

     \begin{figure}[t]
     \plotone{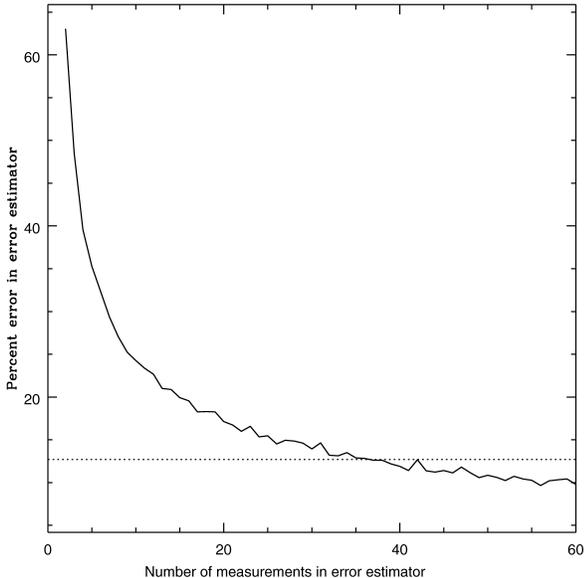}
     \caption{\label{error_est} 
     Variance in error estimation as a function of the number of
     measurements for the set of 7810 synthetic spectra.  The
     horizontal line is the variance (about the true value) of the
     xcsao error estimate.  See text for details.}

     \end{figure}

Using the 7810 spectra we can ask the question: ``how many independent
measurements of a spectrum are required to obtain a better measure of
the error in a single measurement than {\bf xcsao} provides?''
Nordstr\"om, et al (1994), on the basis of a study of echelle spectra
of rotating F stars give this answer as $\sim 7$.  Here we take as
many independent sets of N spectra as exist in 7810 spectra, for each
N we calculate the sample standard deviation about the sample mean,
and we compare it with the known error gotten by using all 7810
velocities.  The RMS of this difference is plotted as a function of N
in figure \ref {error_est}.  Also plotted (as a straight line) is the
RMS of the {\bf xcsao} error estimate about the true error.  In this
ideal case more than 30 independent measures are required before a
better error estimate is reached than the {\bf xcsao} error.

     \begin{figure}[t]
     \plotone{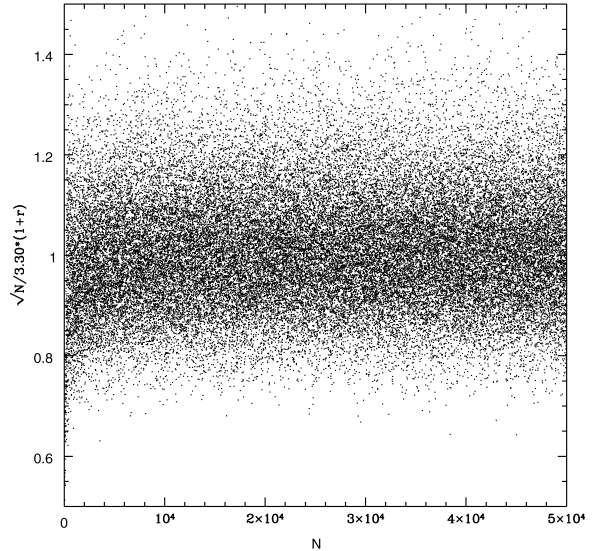}
     \caption{\label{rootN} 
     The relation between $r$ and $\sqrt{N}$  Each point represents
     the ratio of $\sqrt{N}$ with the best fit linear relation between
     $\sqrt{N}$ and $1+r$.}

     \end{figure}

The 50,000 synthetic spectra, with 1 to 50,000 counts, give similar
results.  The mean error is underestimated here by 8\%\ , and the mean
velocity is 1.1 $\sigma$  different from zero, using the 49880 spectra
where $r>3$.  The efficiency of $(1+r)$ in estimating signal to noise
is demonstrated in Figure \ref{rootN}; here we show the ratio of the square
root of the number of counts to the best fit linear relation with
$(1+r)$.  The $1\sigma$ scatter is $\sim 12\%\ $, independent of
N; this puts a limit on the inherent ability to estimate errors using
$(1+r)$.

\subsection {\label{sys}systematics}

There are many factors which can cause the redshifts and radial
velocitys measured by {\bf xcsao} to be other than those
desired.  Here we list several.

1.  Errors in the wavelength calibration.  For absorption line
spectra, where the signal is averaged over all the lines in a complex
spectrum, this effect should be tiny.  For emission line spectra,
where the signal comes from a couple of lines this effect could be as
large as the error in the pixel to wavelength calibration function.
For typical FAST galaxy spectra this error is $\sim 5 km/s$ averaged
over the entire wavelength range, it could be larger in small regions
(section \ref{zero}).

2.  Offsets in the calibration lamp illumination.  The light from the
calibration lamps does not follow the same exact optical path as the
light from the sky.  This can cause systematic errors in the
wavelength scale (section \ref{sky}).  

3.  Variations with the Fourier filter.  As noted above (section
\ref{filt}) different reasonable choices of the Fourier filter can
change the measured velocity of a typical FAST galaxy spectrum by
$10km/s$, comparable to the error in our highest S/N observations.
Unreasonable choices for the filter parameters can make more of a
difference.

4.  Spectral type mismatch.  It has long been known that there are
systematic velocity effects when the template spectrum does not match
the observed spectrum.  Nordstr\"om, et al. (1994), for example, show
the effect of rotational velocity mismatch for their echelle spectra.
For typical FAST galaxy spectra this effect is $\sim 20 km/s $, and
is discussed in section \ref{zero}.

5.  HII regions in spiral galaxies.  The rotation velocities of disks
and the finite number of HII regions on the slit can combine to yield
an $H{\alpha}$ velocity which is different from the mean velocity of
the stars in the bulge; this is especially true for galaxies which are
distorted.  Thus, while the measurement error in an emission line
velocity for a particular galaxy may be substantially smaller than for
the absorption line velocity, the systematic deviation from the
desired quantity, the cosmological redshift, may be substantially
larger.

6.  Two-lined systems.  {\bf xcsao} assumes that the template spectrum is a
reasonable spectral match to the observed spectrum.  For the case of
two (or more) lined spectroscopic binaries this condition is clearly
violated.  As demonstrated by Latham, et al. (1996) substantially
improved results may be obtained by using methods which explicitly
model the two lined case, such as TODCOR (Zucker and Mazeh, 1994).
{\bf xcsao} may be used to obtain the input data for TODCOR.

\section {\label{temp}Templates}

Accurate redshifts require the existence of very high signal-to-noise
templates, which have well determined velocities, and are good matches
to the scientific program objects being measured.  Here we describe
methods of  creating and maintaining systems of templates.  

Templates must be very high signal to noise spectra; there are two
basic ways to create them: 1) sum a number of observations; 2) build a
computer model of a spectrum.  The CfA Digital Speedometry group has,
over the past decade, switched entirely from using observed spectra to
models (see Latham, et al, 1996 and references therein for details).
For galaxy redshift studies we use both techniques.

The vast majority of nearby galaxies can be well matched by a typical
absorption line spectrum (like NGC7331), by a typical emission line
spectrum, or both.  Unusual spectra, as one obtains for QSOs,
H$\delta$ strong galaxies, galaxies with extremely high or low
internal velocity dispersions, etc. require special templates;
although in most cases special templates only lower the error in the
redshift; the correct redshift is normally obtained using standard
templates.

Creating templates is an iterative process: good templates are
required as a prerequisite for making better templates.  New templates
were made for FAST when it saw first light, and again when it
received a new thinned CCD in September 1994.  In 1997, sufficient new
observations having been made, we were able to make a set of
substantially improved templates.

\subsection {\label{emt}emission line template}

Until 1995, redshifts for emission line galaxies were obtained with
{\bf emsao} and its precursor programs.  Then a template was made by
placing Gaussians with approximately correct line widths and line
ratios at the emission line rest wavelengths.  This template, emtemp,
was tested against existing FAST observations, and against the
Z-Machine archive.  Typical results are for the Z-Machine comparison
with 8606 emission line spectra reduced by hand: the $1 \sigma$
difference is $13 km s^{-1}$, about half of the calculated error for a
typical Z-Machine emission line velocity.  As noted above the zero
point offset is $0.56km/s \pm .16$, about a two hundredth of a pixel.
The median difference of 3929 FAST spectra with $r_{emtemp}>3$ and the
$H_{\alpha}$ velocity obtained by {\bf emsao} is $0.24km/s$.

Essentially for all spectra where {\bf emsao} can obtain a redshift
{\bf xcsao} obtains a redshift using emtemp.  For low S/N spectra {\bf
xcsao} plus emtemp is much more sensitive than {\bf emsao}.  For a
sample of 2088 emission line galaxy spectra taken with FAST only 42\%\
of spectra where {\bf xcsao} plus emtemp obtained a redshift with
$3<r_{emtemp}<4$ could be reduced automatically using {\bf emsao},
64\%\ for $4<r_{emtemp}<5$, and 93\%\ for $r_{emtemp}>5$.  As noted
above the systematic effect of cosmic rays places severe constraints
on the unsupervised use of {\bf xcsao} with an emission line template.

     \begin{figure}
     \plotone{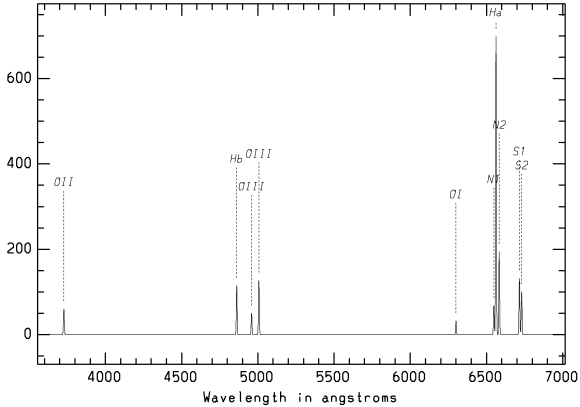}
     \caption{\label{femtemp97} 
     The emission line template femtemp97.}

     \end{figure}

To make a better emission line template we have taken 6498 FAST
spectra where $r_{emtemp}>5$, put them through {\bf emsao}, and
obtained 434 spectra where {\bf emsao} found 9 or more lines.  For
these we measured the ratio of line heights with the SII 6731\AA\
line, and the line widths for each line.  Then for each line we took
the median of these quantities, and, along with the laboratory rest
wavelengths for the lines put the data into the program {\bf linespec}
to produce a synthetic template which we call femtemp97 (figure \ref
{femtemp97}).

     \begin{figure}
     \plotone{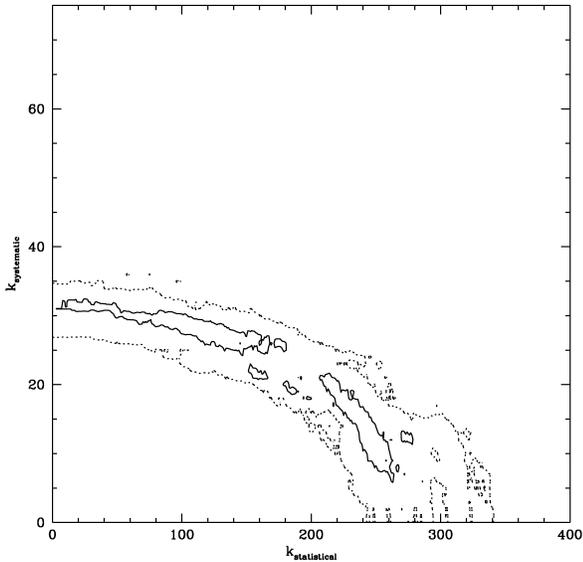}
     \caption{\label{femcontour} 
     Residuals in fit to Gaussian error distribution for pairs reduced
     with the femtemp97 template.  The contours represent lines of equal
     residual, position in the x,y space represents admixtures of the
     statistical and systematic error models; see text.}

     \end{figure} 

Femtemp97 is indeed a better template than emtemp.  Using the set of
626 duplicate spectra described in section \ref{rel}, we can compare
the velocity differences between pairs directly; the median difference
using femtemp97 is $ \sim 9\%\ $ smaller than in the reduction using
emtemp. The errors for femtemp97 are more Gaussian than for emtemp;
Figure \ref {femcontour} shows the fit to the two parameter error
model (section \ref{est}), and may be compared with figure \ref
{emcontour}.  The interior contour level here is half that in figure
\ref {emcontour}.  Also note that the y intercept is about $ 32km/s$,
substantially less than the $45km/s$ in Figure \ref{emcontour} ,
implying a greater reduction in error than the direct comparison of
velocity differences would indicate.  The median ratio $(1 +
r_{femtemp97})/ (1 + r_{emtemp}) \sim 1.33$; femtemp97 yields good
velocities for a substantial number of spectra where emtemp fails.  A
comparison of $k_{femtemp97}\over (1+r_{femtemp97})$ with the ${3
\over 8}{w \over (1 + r) } $ error estimator shows that ${3 \over 8}{w
\over (1 + r)} $ underestimates the error by 9\%\ .

\subsection {\label{abt}absorption line template}

The first absorption line templates used on the FAST data were the
TD79 vintage ztemp, and the NGC4486b template from the MMT
spectrograph, as well as some secondary MMT templates.  Over the next
year a program of template observations provided several high S/N
observations of candidate templates.  These were summed on an object
by object basis, to form extremely high S/N templates.  The best of
these is the NGC7331 template (fn7331temp) (section \ref{rel}).

We are now in a position to create a better template.  Using
fn7331temp and femtemp97, we selected galaxy spectra where the $r$
value for the reduction with fn7331temp was greater than eight, and
the $r$ value for the reduction with femtemp97 was less than 3.  These
are 1959 moderate to high S/N absorption line spectra.  This set of
spectra still contains spectra of objects with substantial emission.
NGC7331 shows clear emission in the NII 6583\AA\ line, for example.
We examined the differences between the velocities derived using
fn7331temp and femtemp97, where (1) the difference was near zero,
meaning that there was enough emission present to get a correct
redshift (about 100 spectra) and where (2) the difference was near
$948 km/s$, meaning that NII 6583\AA\ was confused with $H{\alpha} $
(about 300 spectra); we removed those spectra from the sample.

\begin{figure}
     \plotone{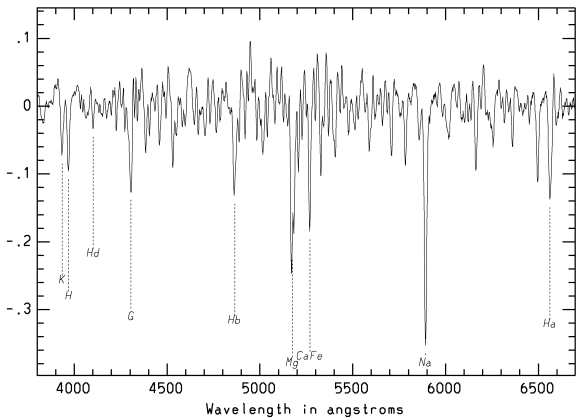}
     \caption{\label{fabtemp97} 
     The absorption line template fabtemp97.}

     \end{figure}

The remaining 1489 spectra were shifted to a common rest velocity,
using the fn7331temp velocity; they were normalized to the same number
of counts; their continua were subtracted, using a moderate order
spline; and finally they were summed.  The {\bf sumspec} task (section
\ref{sumspec}) performed these tasks.  The resulting spectrum had its
residual continuum subtracted using a high order spline, and was
normalized to represent the average spectrum.  This is the final new
absorption line FAST template, fabtemp97, shown in figure \ref
{fabtemp97}.

Fabtemp97 is a better template than fn7331temp.  The median difference
between pairs of spectra from the 628 duplicates of section \ref{rel}
is actually $\sim 2\%\ $ larger (insignificant) using fabtemp97,
because fn7331temp is better able to match emission line objects (it
has NII visibly in emission, and other lines buried in the noise); if
one restricts the comparison to pairs of spectra where $r_{femtemp97}
< 3$ for each spectrum, fabtemp97 shows a median difference $\sim
12\%\ $ smaller than fn7331temp.  As with the comparison of femtemp97
and emtemp, the differences between the new and the old templates are
not large.

     \begin{figure}[t]
     \plotone{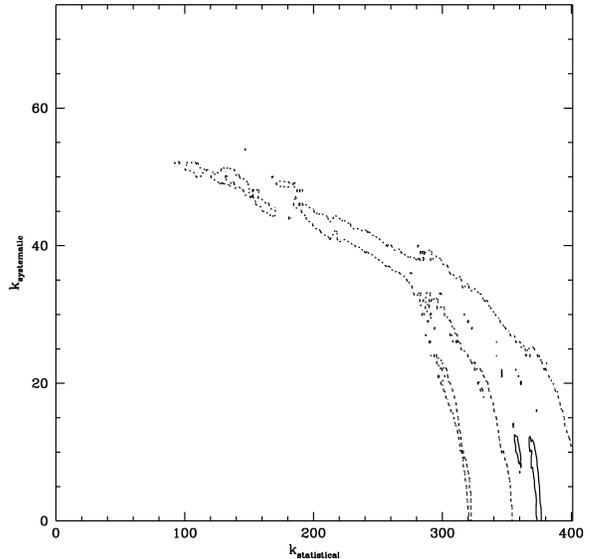}
     \caption{\label{fabcontour} 
     Residuals in fit to Gaussian error distribution for pairs reduced
     with the fabtemp97 template.  The contours represent lines of equal
     residual, position in the x,y space represents admixtures of the
     statistical and systematic error models; see text.}

     \end{figure}

Figure \ref {fabcontour} shows the fit to the two parameter error
model in section \ref{est} (compare with figure \ref {n7contour} for
fn7331temp).  The contours in figure \ref {fabcontour} are the same as
in figure \ref {n7contour}; note that even the outer contour does not
reach the y axis.  This suggests that the error for these spectra
cannot be modeled by a single number, $k_{systematic}$, independent of
$r$; this is not true for fn7331temp, or for the emission line
templates.  A comparison of $k_{fabtemp97}\over (1+r_{fabtemp97})$
with the ${3 \over 8}{w \over (1 + r)} $ error estimator shows that
${3 \over 8}{w \over (1 + r)} $ overestimates the error by 4\%\ .

The analyses of the four templates, shown in figures \ref {emcontour},
\ref{n7contour}, \ref {femcontour}, and \ref {fabcontour} all are
consistent with a model where $\sim 20 km/s$ constant systematic error
is combined with a statistical error determined by the value of the
$r$ statistic.

We also attempted to make a template for narrow lined absorption line
objects.  We built a template by summing, after shifting to a common
rest frame, several high S/N spectra from a set of M31 globular
clusters; we call this template fglotemp.  Next we extracted from the
FAST database all spectra where $r_{fglotemp} > r_{fn7331temp}$ and
$r_{fglotemp} > 6$ and $r_{femtemp97} < 3$.  More than 90\%\ of these
spectra were calibration stars; we excluded them, and the globular
clusters themselves, and were left with $\sim 300$ galaxy spectra.
These we shifted and summed with {\bf sumspec} in the same manner as
for fabtemp97, giving us a new narrowlined template.  This template
and fabtemp97 were correlated against the 300 spectra.  There was no
significant difference in the results.  From this we conclude that for
typical redshift survey spectra, from a spectrograph with resolution
$R \sim 1500$ there is no need to have a narrow lined template.

\subsection {\label{zero}velocity zero point}

We have developed a new methodology for defining the velocity zero
point.  Previous methods have depended on external calibrators, such
as 21 cm measurements; our new methods are fully internal, and should
minimize velocity offsets due to spectral type differences.

     \begin{figure}[t]
     \plotone{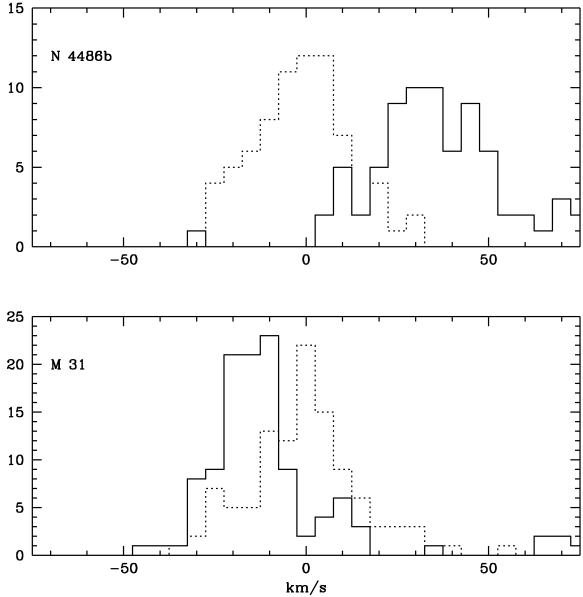}
     \caption{\label{spdiff} 
     Spectral type differences in determining velocity zero points.
     The top panel shows velocities for 75 observations of NGC4486b,
     the solid histogram using the fn7331temp template, and the dotted
     histogram using the ztemp template.  The x axis has been shifted
     so that the median redshift obtained using ztemp is zero.  The
     bottom panel shows the same information for 116 different spectra
     of M31.  No shifting of zero points can make both sets of
     histograms agree.}

     \end{figure}

Figure \ref {spdiff} illustrates the problem with spectral type
difference.  With the FAST spectrograph we observed M31 on 116
occasions, and NGC 4486b on 75 occasions.  Each of these spectra was
reduced using ztemp and fn7331temp.  The resulting velocities were
shifted so that, for each galaxy, the median redshift obtained by
using the ztemp template was zero.  The bottom panel of figure \ref
{spdiff} shows the results for M31; the solid histogram shows the
result of correlating the 116 spectra with fn7331temp, the dotted
histogram shows the result of correlating these spectra with ztemp.
Similarly, the top panel of figure \ref {spdiff} shows the results for
NGC 4486b, with the solid histogram representing 75 fn7331temp
correlations, and the dotted histogram 75 ztemp correlations.

No change in the zero point of either template can make both sets of
histograms agree; any setting of the zero point by matching velocities
for one object would make the systematic difference between different
template reductions for the other object worse.  

We choose to define the zero point of our velocity system to minimize
the systematic difference between the two main types of galaxy
spectra: emission line spectra and absorption line spectra.  To insure
that our internal procedure matches the ``true'' system we need only
make the reasonable assumption that we accurately know the rest
velocities of the main spectral lines in galaxies, such as $H\alpha$
and [O III].  We must also be certain that the internal wavelength
system of the spectrograph matches the external wavelength system of
the sky (section \ref{sky}).

Our basic procedure is to force the median difference between the
emission line velocity and the absorption line velocity, for those
galaxies which strongly show both sets of features, to zero.  Because
the absorption line velocity comes from the K giant stars in the
bulge, primarily, while the emission line velocity comes mainly from
HII regions in the disk, which are moving at $\sim \pm 200 km/s$ with
respect to the central bulge, there is no reason to expect that the
absorption line velocity and the emission line velocity should be the
same for any particular object.  These differences should be randomly
distributed about zero, however.

We began by setting the velocity for the fn7331temp template so that
the median velocity difference between the fn7331temp velocity and the
femtemp97 velocity, for spectra where $r_{femtemp97} > 5$ and
$r_{fn7331temp} > 5$ was identically zero.  This yields a velocity of
$797 km/s$ for NGC 7331, which may be compared with $820\pm3 km/s$
from the 21 cm observations (\cite{bot90}).

As described in section \ref{abt}, fabtemp97 was created by shifting 1489
spectra to the rest frame defined by their individual correlations
with fn7331temp, then summing them.  A comparison of fabtemp97
velocities with emission line velocities should give an indication of
the stability of the zero point technique.

     \begin{figure}
     \plotone{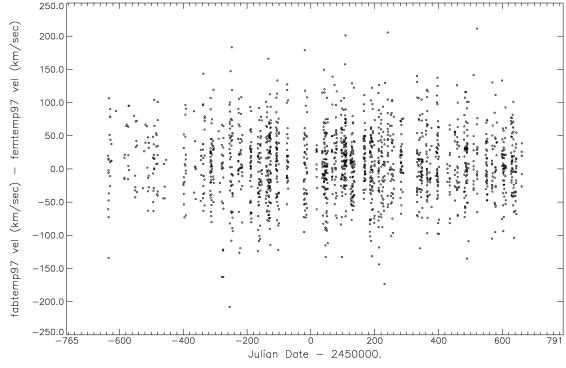}
     \caption{\label{fabvfem} 
     The difference between velocities obtained with the femtemp97
     template and the fabtemp97 template, for 1787 FAST spectra where
     both reductions had $r > 5$, as a function of observation date.}

     \end{figure}

     \begin{figure}
     \plotone{fab5ha.epsi}
     \caption{\label{fabha} 
     The difference between velocities obtained by fitting H$\alpha$
     and correlating with the fabtemp97 template, for 1514 spectra
     where $r > 5$ and H$\alpha$ was fit, as a function of observation
     date.}

     \end{figure}

     \begin{figure}
     \plotone{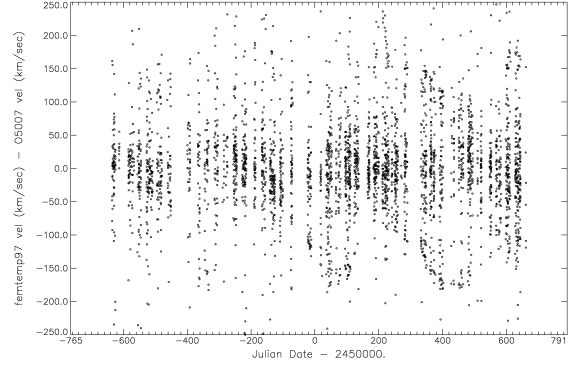}
     \caption{\label{femo3} 
     The difference between velocities obtained by fitting O$[III]
     5007\AA$ 
     and correlating with the femtemp97 template, for 3527 spectra
     where $r > 5$ and O$[III]$ was fit, as a function of observation
     date.}

     \end{figure} 

     \begin{figure}
     \plotone{fem5ha.epsi}
     \caption{\label{femha} 
     The difference between velocities obtained by fitting H$\alpha$
     and correlating with the femtemp97 template, for 4833 spectra
     where $r > 5$ and H$\alpha$ was fit, as a function of observation
     date.}

     \end{figure} 

     \begin{figure}[t]
     \plotone{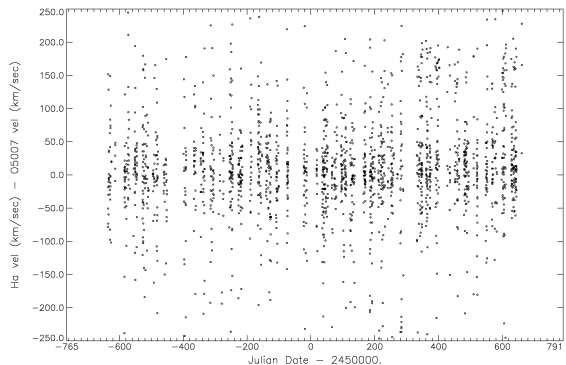}
     \caption{\label{fhao3} 
     The difference between velocities obtained by fitting O$[III]
     5007\AA$ and fitting H$\alpha$, for 2008 spectra
     where H$\alpha$ and O$[III]$ were both fit, as a function of observation
     date.}

     \end{figure}

Figure \ref {fabvfem} shows the difference in velocities between
fabtemp97 and femtemp97 as a function of observation date, for 1787
FAST spectra where $r_{femtemp97} > 5$ and $r_{fabtemp97} > 5$.  The
median difference is $7.1 km/s$, with an interquartile range of $62
km/s$.  Selecting subsets of the spectra with higher S/N ratios has no
significant effect on this result.  This difference is $\sim 0.1$
pixel, and is probably due to systematic, non-linear errors in our
wavelength scale.  Figure \ref{fabha} shows the difference between the
fabtemp97 velocity and the velocity found by fitting $H\alpha$ in
emsao for a representative subset of the data.  The median difference
is $-2.1 km/s$, consistent with zero.  Figure \ref {femha} shows the
difference between the femtemp97 velocity and the $H\alpha$ velocity,
$-9.51 km/s$, and figure \ref {femo3} shows the difference between the
femtemp97 velocity and the [OIII] 5007\AA velocity, $-0.16 km/s$.
These differences are very similar to the median differences between
individual lines measured in the same spectrum with emsao, Figure
\ref{fhao3} shows the difference between fits to H$\alpha$ and fits to
O[III], the median difference is $6.4 km/s$.

We conclude that our construction of fabtemp97 gives a 
velocity zero point equal to the velocity zero point of the emission
line system; this equivalence is as accurate as our ability
to define the wavelength system using standard lamps and polynomial
fits.

\section{\label{sky}Testing spectrograph zero point and stability using 
{\bf emsao} and night sky spectra}

Before 1995, we measured redshifts for emission line spectra using
{\bf emsao}.  Now we cross-correlate against an emission line template
with {\bf xcsao} (section \ref{emt}). {\bf emsao} is used to check
for errors in emission-line correlations, to interactively reduce
emission-line spectra, to automatically measure equivalent widths,
line heights, and line widths for sets of emission line spectra, and
to do various custom projects using alternate line lists (e.g. to
study QSOs).

We used {\bf emsao} to calibrate the Z-Machine and FAST spectrographs
for stability and zero point by measuring the apparent velocity of the
night sky lines.  These are from forbidden oxygen airglow lines from
the upper atmosphere and mercury and sodium emission from artificial
sources such as street lights. Sodium is a blended doublet and we do not
know the effective wavelength {\it a priori}; so we cannot use it to
establish a zero point.  Using the zero point defined by O[I] 
5577\AA\ we measure an effective wavelength of 5891.2\AA\ using 
FAST, with the Z-Machine data in agreement.  We also obtain a
systematic difference between the oxygen airglow lines and the mercury
streetlamp lines of about $20km/s$ with both spectrographs.  We assume
that this effect is due to differences between the effective wavelengths of
Hg in streetlamps and calibration lamps.  We therefore choose to
calibrate our zero points with the oxygen airglow lines.

\begin{figure}[p]
     \plotone{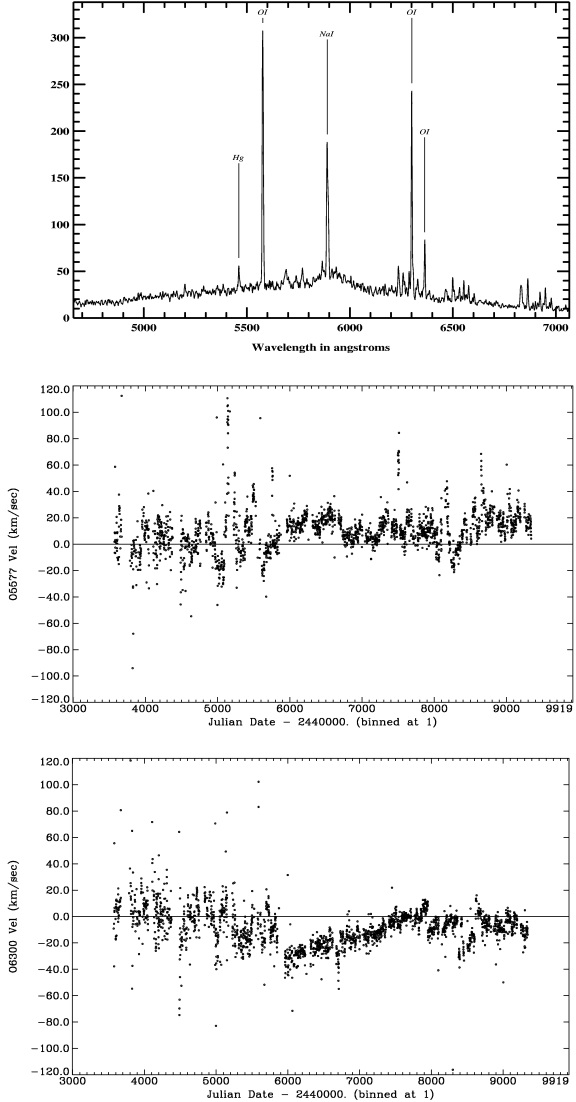}
     \caption{\label{zsky}
     {\bf emsao} reduction of night sky lines from the Z-Machine.  The 
     top panel shows a typical night sky spectrum, with the lines marked.
     The middle panel shows the apparent velocity of O[I] 5577\AA\ as
     a function of observation date.  Each point is the nightly median,
     where nights with fewer than 10 observations are excluded.  The
     bottom panel is similar to the middle, but shows the apparent 
     velocity of the O[I] 6300\AA\ line.}

     \end{figure}

\begin{figure}[p]
     \plotone{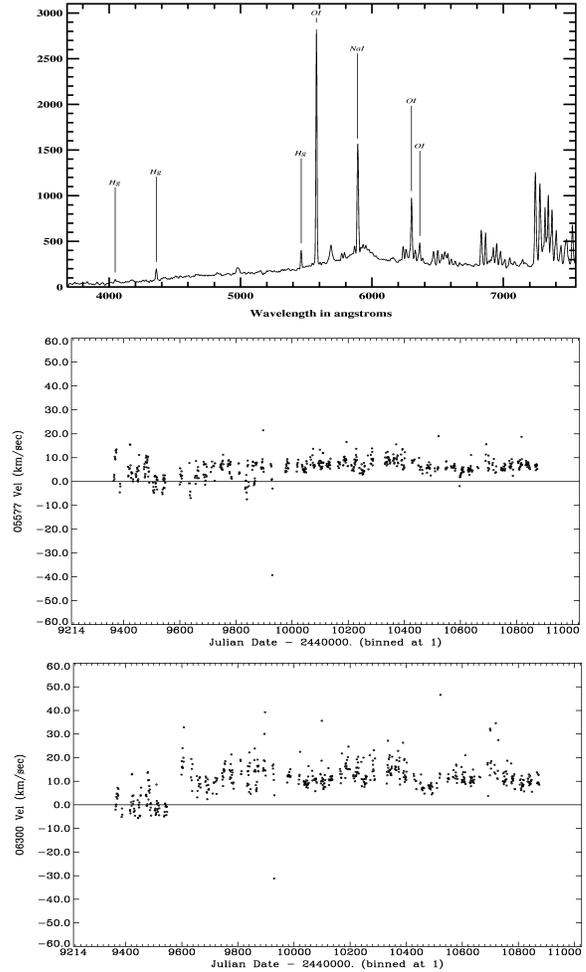}
     \caption{\label{fsky}
     {\bf emsao} reduction of night sky lines from FAST.  The top
     panel shows a typical night sky spectrum, with the lines marked.
     The middle panel shows the apparent velocity of O[I] 5577\AA\ as
     a function of observation date.  Each point is the nightly median,
     where nights with fewer than 10 observations are excluded.  The
     bottom panel is similar to the middle, but shows the apparent 
     velocity of the O[I] 6300\AA\ line.  Note that the scale of the 
     ordinate is half that in figure \ref{zsky}}

     \end{figure}

Figure \ref{zsky} shows the results for the Z-Machine.  The top panel is a
typical sky spectrum, the middle panel shows the apparent velocity of
the O[I] 5577\AA\ line, and the bottom panel shows the apparent
velocity of O[I] 6300\AA\ .  The velocities are shown as a
function of observation date, over 15 years.
Each point is the median of a single nights observations; nights with
fewer than 10 observations were excluded. 

The night sky emission lines were routinely used to supplement the
HeNeAr calibration lamp lines in the Z-Machine reductions, and a
seventh order polynomial was used to fit the lines.  The O[I] 5577\AA\
line does not show the same general pattern as the O[I] 6300\AA\ line.
All the other lines (Hg 5461, NaD, O[I] 6363) do show the same
behavior as the O[I] 6300 line.  O[I] 5577\AA\ is in a portion of the
spectrum with no strong HeNeAr calibration lines, which, along with
the high order polynomial, locally matches the calibration to the
position of O[I], instead of the system defined by the calibration
lamp.

While the O[I] line at 6300\AA\ was also used in the wavelength
calibration, it is in a region with many strong Ne lines, and should
have negligible effect on that calibration.  The apparent velocity of
this line should be a good measure of the difference between the
instrumental zero point and the true sky.  The scatter and long-term
changes in this line measure the stability of the instrument and data
reduction procedures.  Many of the large jumps in apparent velocity
can be attributed to known hardware changes; the largest feature in
figure \ref{zsky}, the 5.5 year slow increase in velocity of O[I]
6300\AA\ from $\sim -30 km/s$ to $\sim +10 km/s$ is of unknown origin.

Figure \ref{fsky} shows the results for FAST.  Again the top is a
typical sky spectrum, the middle shows the apparent velocity of the
O[I] 5577\AA\ line, and the bottom the apparent velocity of the O[I]
6300\AA\ line.  The abscissa covers four years, and the scale of the
ordinate is half that of figure \ref{zsky}.

The night sky lines are not used in the wavelength calibration of
FAST, and the HeNeArFe calibration lamp lines are fit with a third
order polynomial.  Both O[I] 5577\AA\ and O[I] 6300\AA\ should be good
measures of the instrumental zero point and the stability of the
instrument and reduction techniques.

Both lines show essentially the same behavior, especially since the
CCD was changed in September 1994.  The scatter in these line
positions is less than 20\%\ of the scatter for the same lines in the
Z-Machine.  Over the four years of FAST operation an additive offset
of $7.5\pm 2 km/s$ brings the instrumental system into agreement with
the true system of the sky.

FAST is a remarkably stable instrument.  The scatter in the O[I]
5577\AA\ line can provide a good measure of the errors due to a
combination of instrumental instability, wavelength calibration, and
line fitting.  We measure the error in fitting a single line by
dividing the scatter in the apparent velocity difference between O[I]
6300\AA\ and O[I] 6363\AA\ by $\sqrt{2}$.  Subtracting it in
quadrature from the scatter in O[I] 5577\AA\ , yields the error due to
the interaction of instrumental instability with the wavelength
calibration. For the entire four year period this is $2km/s$, where
most of the error comes from systematic changes associated with
changing the CCD, changing the linelist for the HeNeArFe calibration,
and changing the dewer.  For the most recent 18 months, since the
dewer change, the error from instrumental unstability is unmeasurably
small, is consistent with zero, and has a $2\sigma$ upper limit, by
bootstrap resampling, of $0.7km/s$.

\section {\label{other}Other Methods and Comments}

Press (1995) has suggested a new methodology for determining
redshifts.  One first reduces a set of galaxy spectra (shifted to a
common rest frame) to a set of orthogonal basis vectors, using
singular value decomposition, SVD.  Next, using the fast numerical
methods of Rybicki and Press (1995) the most significant of these
vectors are repeatedly fit to a spectrum with unknown redshift, with
each fit being at a different redshift.  For each fit $\chi^2$ is
calculated; the redshift corresponding to a minimum $\chi^2$ 
is the redshift of the galaxy.

Recently Glazebrook, et al (1998) have developed a nearly identical
scheme; the difference is that rather than
calculate $\chi^2$ exactly, they use the simplifying assumption that
the correlation function may substitute for $\chi^2$.  Thus they can
treat the (SVD derived) eigenvectors as templates in a cross
correlation program (such as {\bf xcsao}), and obtain a final redshift by
summing the correlation functions in quadrature, weighted by the
eigenvalues of each eigenvector-template.

Both Press (1995) and Glazebrook, et al (1998) have suggested that the
coefficients of the fits can be used to classify the spectra, and that
the position of a spectrum in coefficient space may be used to develop
confidence measures in the derived redshift.  Recently Bromley, et al
(1997) used this technique to classify spectra from the LCRS
(Shectman, et al, 1996).

We have not adopted the SVD method for creating templates, nor the
$\chi^2$ minimization technique for determining redshifts.  We have,
following Press's (1995) suggestion, developed methods to use best fit
parameters for rough classification and blunder discovery.

We do not implement the Glazebrook, et al (1998) method for a number
of reasons.  Their approximation assumes the variance in a
spectrum is not a function of wavelength.  This assumption is clearly
incorrect, and is one of the reasons why continuum subtraction is
superior to continuum division in the low S/N regime (section
\ref{sup}).  The use of the eigenvalue as a weight in the combination of
correlation functions overweights emission line correlations when
reducing absorption line spectra, thus making them even more
susceptible to shot noise.  But finally, as the exact methods are
available (Press, 1995), we see no advantage in the approximation.

Press's (1995) $\chi^2$ minimization technique, enabled by the Rybicki
and Press (1995) algorithm, shows substantial promise.  Especially for
the case where the rest frame of the template is much different from
that of the unknown galaxy we expect that a rigorous accounting of
observational errors as a function of wavelength will be important;
the new, deep redshift surveys of the next decade, e.g. with Hectospec
(Fabricant, et al, 1994), will test whether the $\chi^2$ techniques
will obtain better results for high redshift objects.  We do not now
implement the $\chi^2$ minimization because we believe that current
redshift survey data, with $z\simless 0.2$ would not benefit; some
confirmation of this view comes from the preliminary results of Press
(1995, 1997), who finds no improved ability to obtain redshifts from
low S/N spectra in the LCRS (Shectman, et al 1996) compared with the
original reduction, done by H. Lin using RVSAO 1.0.

We believe our template creation techniques are superior to a SVD
decomposition of a group of spectra for underlying physical reasons.
Emission line spectra and absorption line spectra arise from
independent physical causes in different physical locations within a
galaxy; basis vectors which are admixtures of absorption and emission
line spectra make no physical sense.  Additionally emission line
velocities and absorption line velocities are not identical for any
particular galaxy, as a perusal of optical rotation curves
(e.g. Barton, et al, 1998) clearly shows.

The simultaneous use of more than two templates or basis vectors to
obtain redshifts is unnecessary.  We performed a SVD decomposition
on the 1489 pure absorption line spectra used to create fabtemp97; the
first eigenvector was essentially identical to fabtemp97, the next
three represented small differences in the continuum subtractions, the
fifth eigenvector shows the H+K lines, which are systematically weaker
in the early data, before a blue sensitive chip was installed, the
next couple of eigenvectors also represent continuum differences.  The
systematic residuals from the reduction dominate the higher dimensions
of the SVD decomposition.

As a practical matter the limiting factor in obtaining redshifts for
faint objects is the ability to get redshifts for weak lined
absorption line systems.  Essentially, if one observes long enough with
a fiber instrument to get redshifts for the absorption line spectra,
the spectra with emission lines will all yield redshifts with almost
any technique.

     \begin{figure}[t]
     \plotone{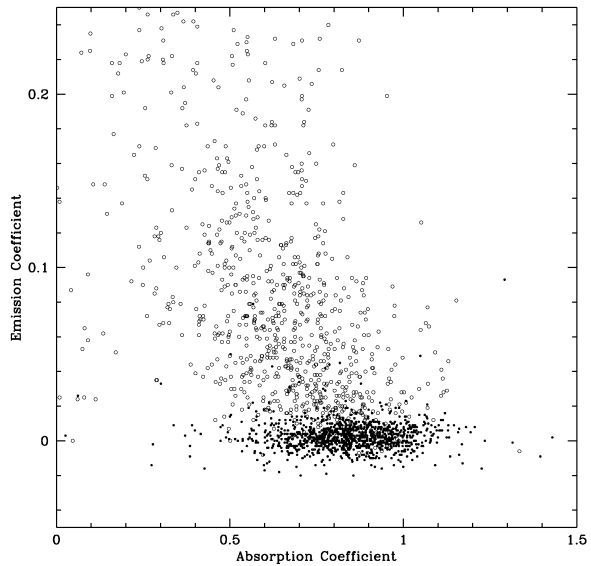}
     \caption{\label{fabfem} 
     The coefficient of the fit to fabtemp97 vs the coefficient of the
     fit to femtemp97, for 2000 spectra from the 15R survey (Geller,
     et al (1998)).  About a dozen objects have coefficients not in
     this range, and would be looked at manually.  The solid dots are
     spectra where $r_{femtemp}<3$, the open circles are for spectra
     where $r_{femtemp} > 3$.  The diagram can be
     viewed as a crude absorption line strength vs emission line
     strength classification diagram.}

     \end{figure}

We therefore will use fabtemp97 and femtemp97 as our fitting
functions, we fit each spectrum as a linear combination of these two
templates.  We take as a representative sample the last 2000 spectra
observed as part of the 15R survey (Geller, et al, 1998).  Figure \ref
{fabfem} shows the absorption line vs. emission line fit coefficients.
The solid dots are for $r_{fem} < 3$ and the open circles are for
$r_{fem} > 3$.  There is a clear locus, outliers could be easily
discovered and removed from any fully automatic data reduction
process.  Combined with more traditional measures, e.g. equivalent
widths and line ratios, figure \ref {fabfem} can form the basis for a
spectroscopic classification.  The x and y axes roughly measure
absorption line strength (metallicity perhaps) and emission line
strength.

     \begin{figure}
     \plotone{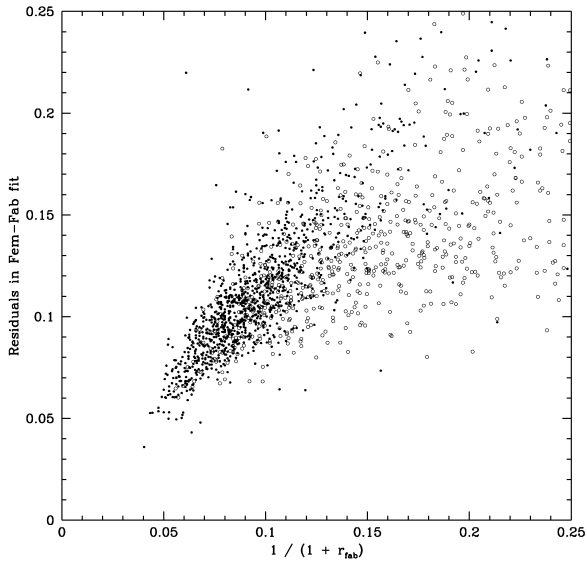}
     \caption{\label{absred} 
     The relation of the residuals to the fit to fabtemp97 to the
     inverse of $1+r_{fabtemp97}$  The symbols have the same meaning
     as in figure \ref {fabfem}.}

     \end{figure}

Figure \ref {absred} shows the relation between the fit residuals and
the inverse of the $1 + r_{fab}$ statistic for the fabtemp97
reduction. The symbols have the same meaning as in figure
\ref{fabfem}.  There is a very good correlation of fit residuals with
$1/(1+r)$ for spectra where $r_{fem} < 3$; for objects with emission
lines the correlation is much weaker.

     \begin{figure}
     \plotone{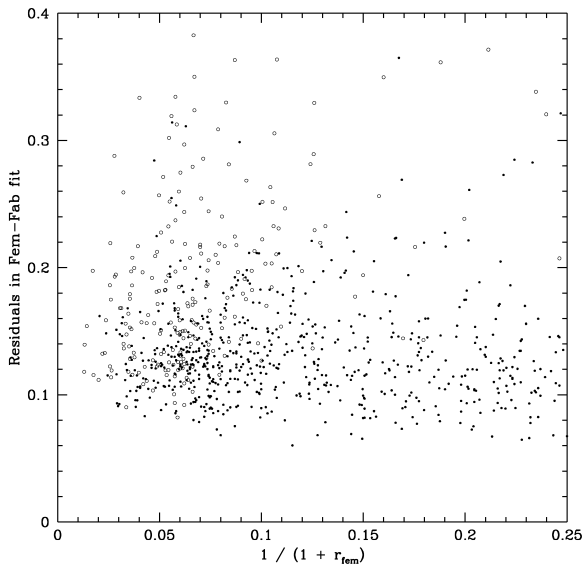}
     \caption{\label{emred} 
     The relation of the residuals to the fit to femtemp97 to the
     inverse of $1+r_{femtemp97}$.  The solid dots are for spectra
     where $r_{fabtemp97} > 3$ and the open circles are for spectra
     where $r_{fabtemp97} < 3$.}

     \end{figure}

Figure \ref {emred} shows the relation between the fit residuals and the
inverse of the $1 + r_{fem}$ statistic for the femtemp97 reduction, 
the symbols here are dots if $r_{fab} > 3$ and open circles if
$r_{fab} < 3$.  The correlation of fit residuals with $1/(1+r)$ for 
spectra where $r_{fab} < 3$ is weak, and for objects with absorption
line velocities nonexistent.

These two plots, along with the figures \ref {femcontour} and \ref
{fabcontour} in section \ref{temp}, which show the statistical and
systematic error models, demonstrate that for typical redshift survey
spectra, residuals for absorption line objects are mainly
determined by signal to noise; for emission line objects, the
residuals are mainly systematic.

\section{Conclusion}

We have demonstrated the techniques which we have developed in the
RVSAO suite to create a system for the accurate, automated reduction
of spectra for galaxy redshifts and stellar radial velocities.  More
than half of all published redshifts have been measured using these
techniques, as well as a large number of stellar radial velocities.

The correlation method for obtaining redshifts can be successfully
extended from absorption line spectra to emission line spectra, with a
substantial improvement in effectiveness over the previous method for
obtaining emission line redshifts, automated line fitting.  The
reduction of emission-line spectra requires different reduction steps
than absorption line correlations.  Emission line correlation
redshifts are susceptible to blunders due to the presence of cosmic
rays. However, using automated line fitting ({\bf emsao})
and absorption line correlation velocities the blunder rate can be
kept near zero, with the degree of automation kept high.

We have developed new techniques for calibrating and characterizing
the blunder rate and the individual errors in redshift measurements.
The blunder rate for RVSAO reductions can be kept near zero by the use
of some simple heuristics to identify possible mistakes.  For typical
redshift survey data from the FAST spectrograph the automation rate is
95\%.  Our self calibrating internal error estimator is accurate to
$\sim$ 20\%.  Large, stable surveys enable development of more
accurate and stable error estimators.

We have developed new methods for creating, calibrating, and using
galaxy redshift templates.  We have created an emission line template,
femtemp97, having the median properties of a large set of strong
emission line spectra.  We have created an absorption line template,
fabtemp97, having the mean properties of a large set of absorption
line spectra showing no sign of emission.  These spectra arise from
physically distinct processes, and can be used to form a pair of basis
vectors to perform a 2-D spectral classification.  We have developed a
new method for establishing the zero point for redshift observations,
a method which minimizes the systematic differences between emission
and absorption line redshifts.  This zero point is determined as
accurately as we can establish the wavelength calibration using
standard HeNeArFe lamps.  We have shown a technique for measuring and
eliminating differences between the instrumental zero point and the
true zero point.

We have shown improved techniques for a number of the substeps
necessary to obtain accurate redshifts, including: removal of
emission(absorption) lines when correlating against an
absorption(emission) line template; suppression of the night sky
lines; supression of the continuum; design of the Fourier filter; and
zero padding of spectra.

The rapid development of large aperture telescopes with multi-object
spectrographs presents substantial challenges for redshift and radial
velocity reductions.  Reducing one or two orders of magnitude more
spectra of objects one or two orders of magnitude fainter while
maintaining high quality control standards and minimal personnel costs
is clearly a difficult problem. RVSAO provides a solid methodological
and software basis to meet these new challenges.

\section{Acknowledgments}

A large number of astronomers have sent us suggestions, complaints,
bug reports, and kudos over the years.  We should like to thank all of
them, and ask that this continue.  We thank W. Press and B. Bromley
for permitting us to quote from their work in advance of publication.
We have benefited greatly from a number of detailed scientific and
technical discussions with D. Fabricant, J. Huchra, D. Latham, and
G. Torres.  E. Falco carefully read the manuscript, and made several 
suggestions to improve the clarity of the text.

Susan Tokarz has reduced tens of thousands of spectra using RVSAO; her
experience, patience, and friendly collaboration have been crucial to
this project.  Margaret Geller's support has enabled RVSAO to achieve
its high degree of robust effectiveness.

\appendix 
\section{\label{appendix} The Elements of RVSAO}

The RVSAO package consists of six IRAF tasks {\bf xcsao}, {\bf emsao},
{\bf linespec}, {\bf sumspec}, {\bf contpars}, and {\bf bcvcorr}.
Each of these tasks is controlled by a set of user settable
parameters.  In this appendix we list and describe all the relevant
parameters for these tasks, and demonstrate their use in a series of
examples.

\subsection {\label{xc} Cross-Correlating a Spectrum in {\bf xcsao}}

\begin{figure}
     \plotone{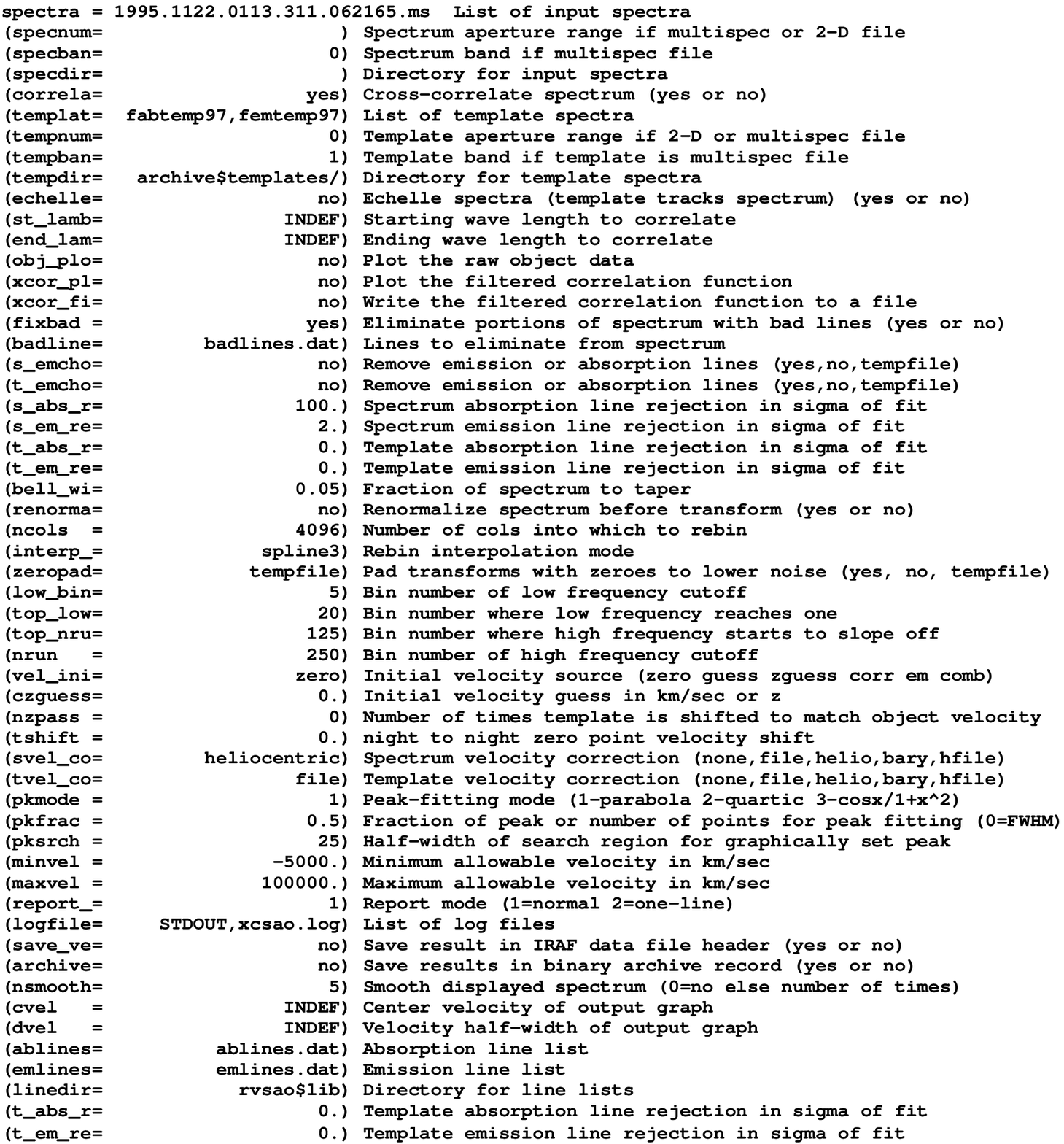}
     \caption{\label{xcsao0a} 
     Parameter list for {\bf xcsao}.}

     \end{figure}

\begin{figure}
     \plotone{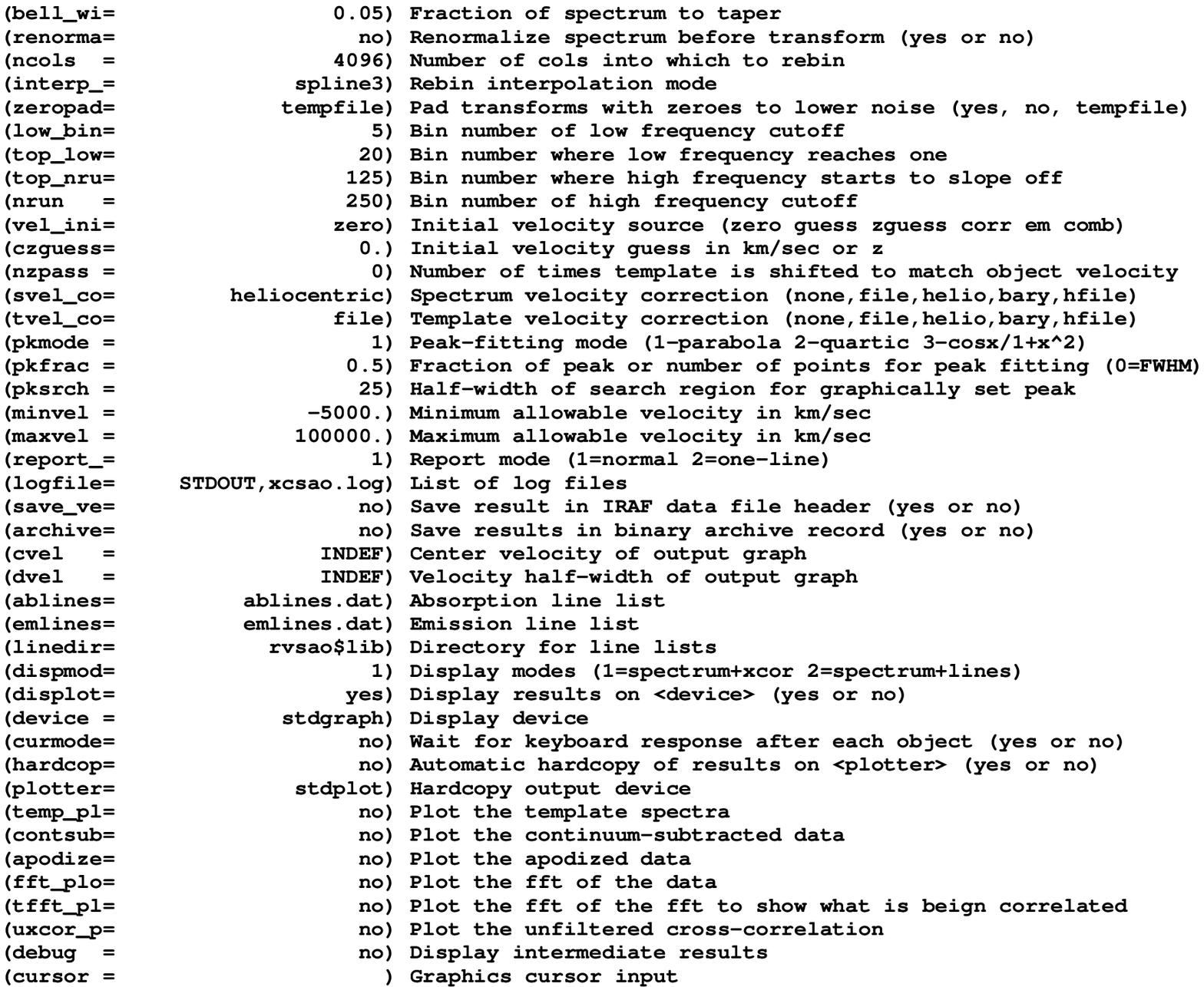}
     \caption{\label{xcsao0b} 
     Parameter list for {\bf xcsao} continued.}

     \end{figure}

\begin{figure}
     \plotone{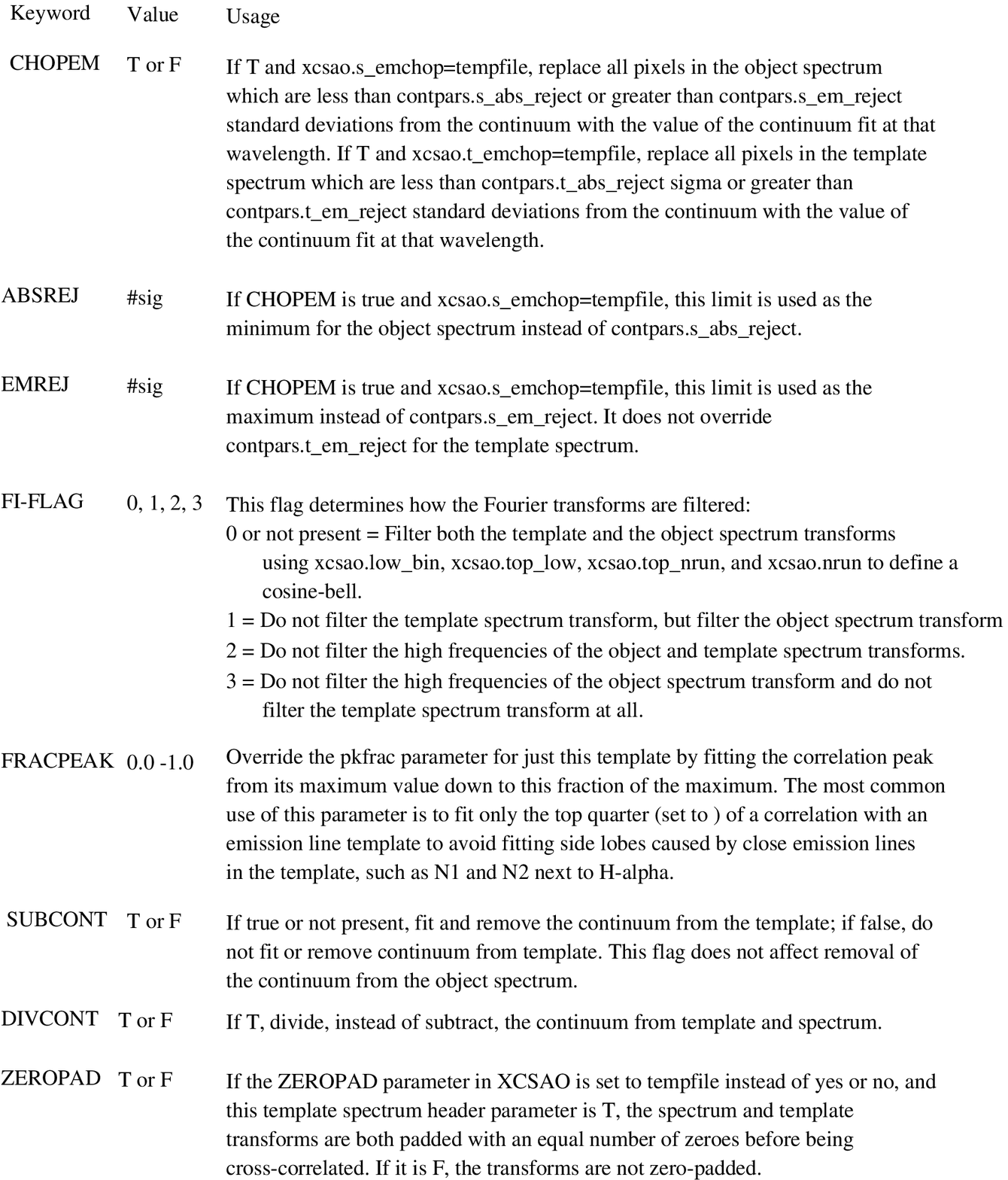}
     \caption{\label{header} 
      These template spectrum header parameters can be used to control
      the cross-correlation process to correlate object spectra
      against both absorption line templates and emission line templates
      in one {\bf xcsao} run.}

     \end{figure}

Digital cross-correlations in the RVSAO system are performed by {\bf
xcsao}, basically following the prescription of TD79, with a large
number of refinements and additions.  {\bf xcsao} is capable of
reducing a wide variety of input spectra, but must have its many
parameters properly set.  Parameters for {\bf xcsao} are set in its
parameter list (Figures \ref{xcsao0a} and \ref{xcsao0b}), in the {\bf
contpars} program's parameter list (Figure \ref{contpars1}), and by
special instructions to the program in the headers of the template
spectra (Figure \ref{header}).  In this section we examine in detail
the elements of {\bf xcsao}, and we describe how the parameters are
used in the code, along with the
algorithmic details of their use.  We follow the cross-correlation of
a single spectrum against two very different templates, one showing
primarily absorption features and one showing only emission features,
to demonstrate the ways in which processing varies in response to the
template spectra being used.

\begin{figure}
     \plotone{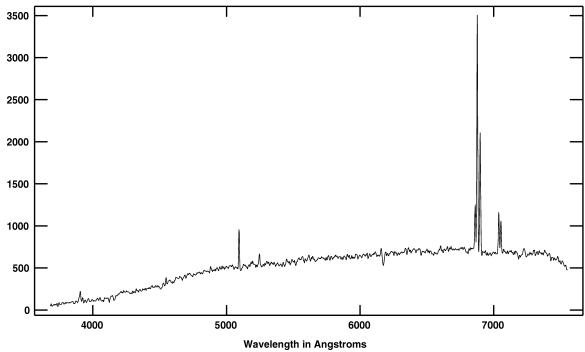} 
     \caption{\label{xcsao1} Object spectrum showing both emission
     and absorption lines as plotted by {\bf xcsao} if {\it obj\_plot} is yes.}

     \end{figure}

For each object spectrum file from the input list {\it spectra},
and/or each aperture specified in the aperture list {\it specnum}, the
fitting subroutine is called. Spectrum files are all read from
the directory specified by {\it specdir}, but full or relative
pathnames may be used in {\it spectra}, and {\it specdir} may be null.
Spectra which are in flux units instead of counts should be renormalized
by setting {\it renormalize} to yes.  If {\it obj\_plot} is yes, the
object spectrum is plotted as in Figure \ref{xcsao1}, and the plot is
kept on the screen available for zooming and editing until a ``q" is
typed.  If {\it fixbad} is ``yes,'' regions specified in the file named
by {\it badlines} are replaced by straight lines connecting the adjacent
pixels, and the image is plotted again if {\it obj\_plot} is ``yes.''

Band {\it tempband} of each template spectrum in the list {\it
templates} and list of multispec apertures, {\it tempnum}, is
loaded. Template files are all read from the directory specified by
{\it tempdir}, but full or relative pathnames may be used in {\it
templates}, and {\it tempdir} may be null. If {\it echelle} is yes,
t{\it empnum} is ignored, and the multispec lines used for templates
track those used for object spectra.  If {\it temp\_plot} is yes, the
template spectrum is plotted.  

A zero-point redshift is computed by adding the solar system
barycentric velocity correction, from a source specified by
{\it svel\_corr}, the redshift of the template from the VELOCITY parameter
in the template header, an optional template-dependent velocity shift
from the TSHIFT parameter in the header, and an optional constant
velocity shift from the {\it tshift} parameter.  The template spectrum's
barycentric velocity correction, from a source specified by {\it tvel\_corr},
is subtracted because the template spectrum's observed velocity, not
its corrected one, gives that spectrum the redshift which we are comparing.
An initial redshift source may be specified by {\it vel\_init}; if
``guess", this is from {\it czguess}; if ``zero'' the initial velocity
is 0; otherwise it can be read from an object spectrum header
parameter VELOCITY (``combination"), CZXC (``correlation"), or CZEM
(``emission"). If such an initial redshift has been called for, or if
this is the second pass or greater ({\it nzpass} $>$ 1), the template
log-wavelength limits are shifted by that initial redshift (on the
first pass) or the current correlation redshift. The wavelength region
over which the template and object spectra overlap is computed. If the
wavelength in Angstroms specified by {\it st\_lambda}, is greater than
the blue limit of the overlap region, it becomes the new limit. If the
wavelength in Angstroms specified by {\it end\_lambda}, is less than
the red limit of the overlap region, it becomes the new limit. The two
spectra are rebinned into log-wavelength with a number of pixels, set
by {\it ncols} using an interpolation mode specified by {\it
interp\_mode}.

\begin{figure}
     \plotone{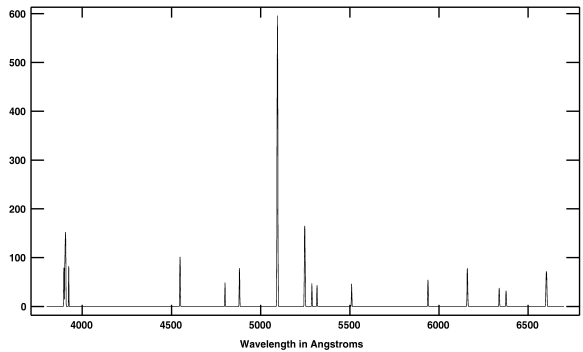}
     \caption{\label{xcsao3} 
     Emission lines removed from the object spectrum before cross-correlation
     against an absorption line template spectrum in {\bf xcsao}.}

     \end{figure}

\begin{figure}
     \plotone{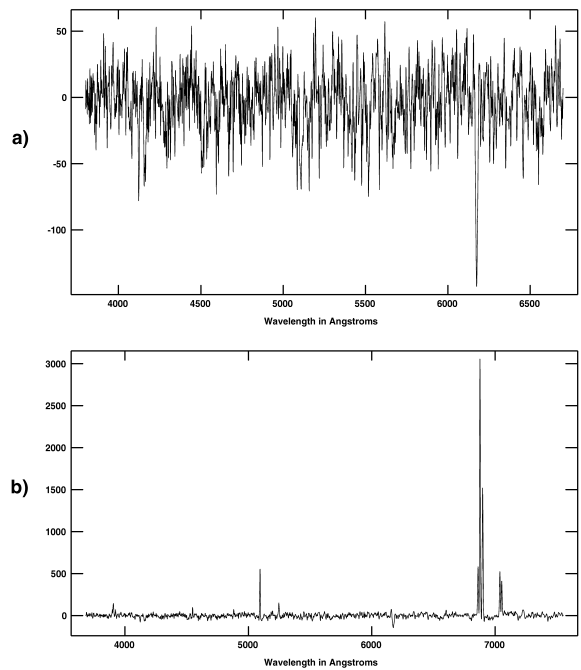}
     \caption{\label{xcsao4} 
     The object spectrum with a) its continuum subtracted and emission lines
     removed for cross-correlation against an absorption line template and
     b) its continuum subtracted, but its emission lines kept for
     cross-correlation against an emission line template spectrum
     in {\bf xcsao}.}
 
     \end{figure}

First, the continuum, and, optionally, emission and/or emission lines,
are removed from the rebinned object and template spectra.  Parameters
for fitting the continuum are in the IRAF pset task named {\bf contpars}
(see section \ref{contpars}).  Emission and/or absorption lines may be
removed from either or both of each object-template pair. {\it
s\_emchop} controls whether lines will be removed from the object
spectrum. If the template spectrum header parameter SUBCONT is
present, its value overrides that of {\it s\_emchop}. {\it s\_absrej}
and {\it s\_emrej} set the lower and upper acceptable limits for
object spectrum pixels to be used in the continuum fit.  If lines are
rejected, as they are when the object spectrum we are following is to
be correlated against an absorption line template, the rejected data
points are plotted as in figure \ref{xcsao3} if {\it contsub\_plot} is
yes.  A graph of the continuum-removed data, as shown in Figure \ref{xcsao4},
is displayed if the {\it contsub\_plot} parameter is yes.

Template pixels with values outside of the lower {\it t\_absrej} and
upper {\it t\_emrej} acceptable limits, in standard deviations from the
continuum fit, are replaced by continuum values in the template spectrum
if {\it t\_emchop} is set to yes.  The continuum is then subtracted from
the template spectra.

\begin{figure}[tp]
     \plotone{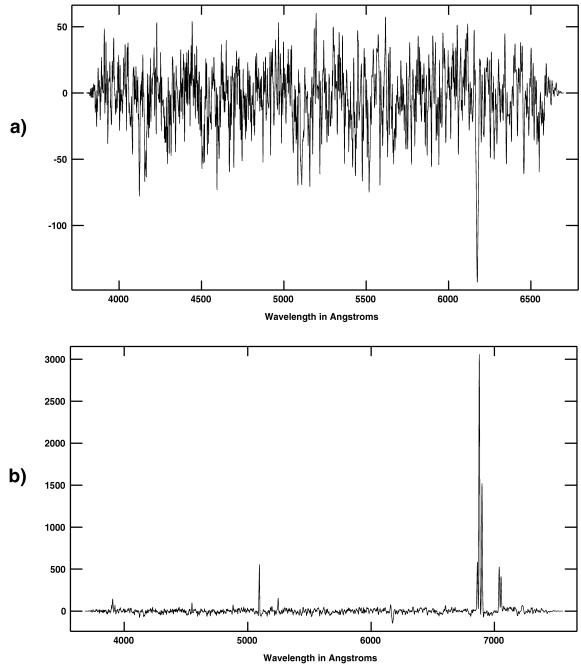}
     \caption{\label{xcsao5} 
     The object spectrum apodized at each end, ready to be Fourier transformed
     for cross-correlation against a) absorption and b) emission templates
     in {\bf xcsao}.}

     \end{figure}

Template and object spectra are then apodized, tapered at each end for
the fraction {\it bell\_window} of the entire spectrum. If {\it zeropad}
is yes, both spectra are padded with an equal length of null (zero)
spectrum.  A graph of the continuum-removed, apodized object spectrum is
displayed, as shown in Figure \ref{xcsao5}, if the {\it apodize\_plot}
parameter is yes.

\begin{figure}[t]
     \plotone{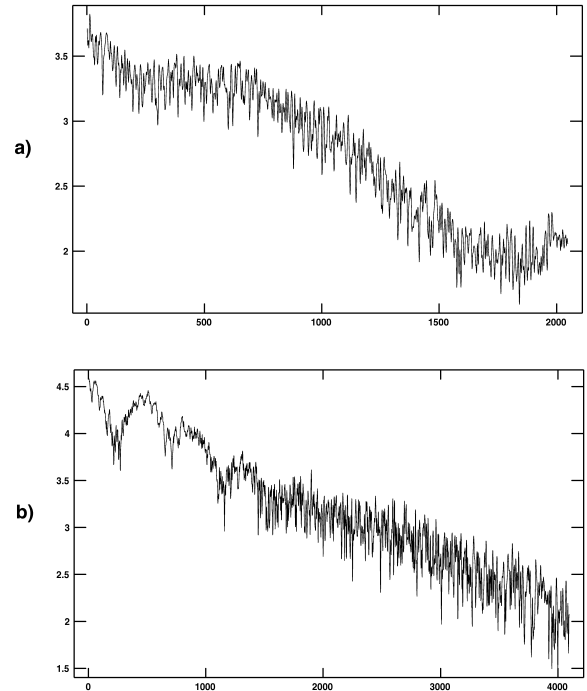}
     \caption{\label{xcsao6} 
     The object spectrum Fourier-transformed over the wavelength ranges of
     the a) absorption and b) emission template spectra in {\bf xcsao}.}

     \end{figure}

\begin{figure}[tp]
     \plotone{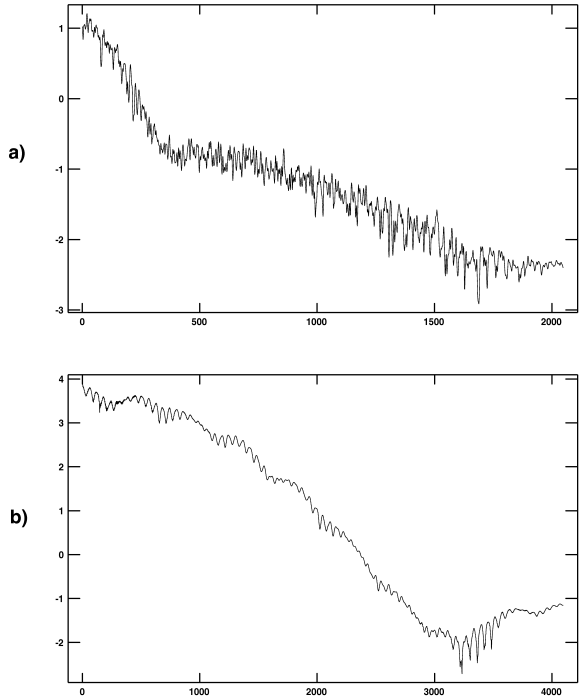}
     \caption{\label{xcsao7}
     Fourier transforms of a) absorption and b) emission template spectra.
     in {\bf xcsao}.}

     \end{figure}

The spectra are then Fourier transformed.
The Fourier power spectra of the object and template spectra are displayed
as in Figures \ref{xcsao6} and \ref{xcsao7} if {\it fft\_plot} is set to yes.

\begin{figure}[t]
     \plotone{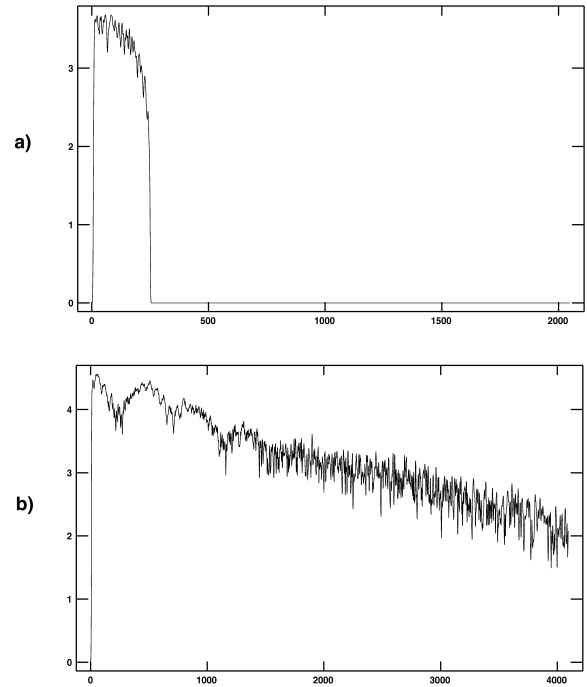}
     \caption{\label{xcsao8}
     The object spectrum transform filtered for correlation against
     a) absorption and b) emission templates in {\bf xcsao}.  Removing
     high frequencies adversely affects the shape of narrow emission lines,
     so no high frequency filtering is done for emission line correlations}
 
     \end{figure}

\begin{figure}[tp]
     \plotone{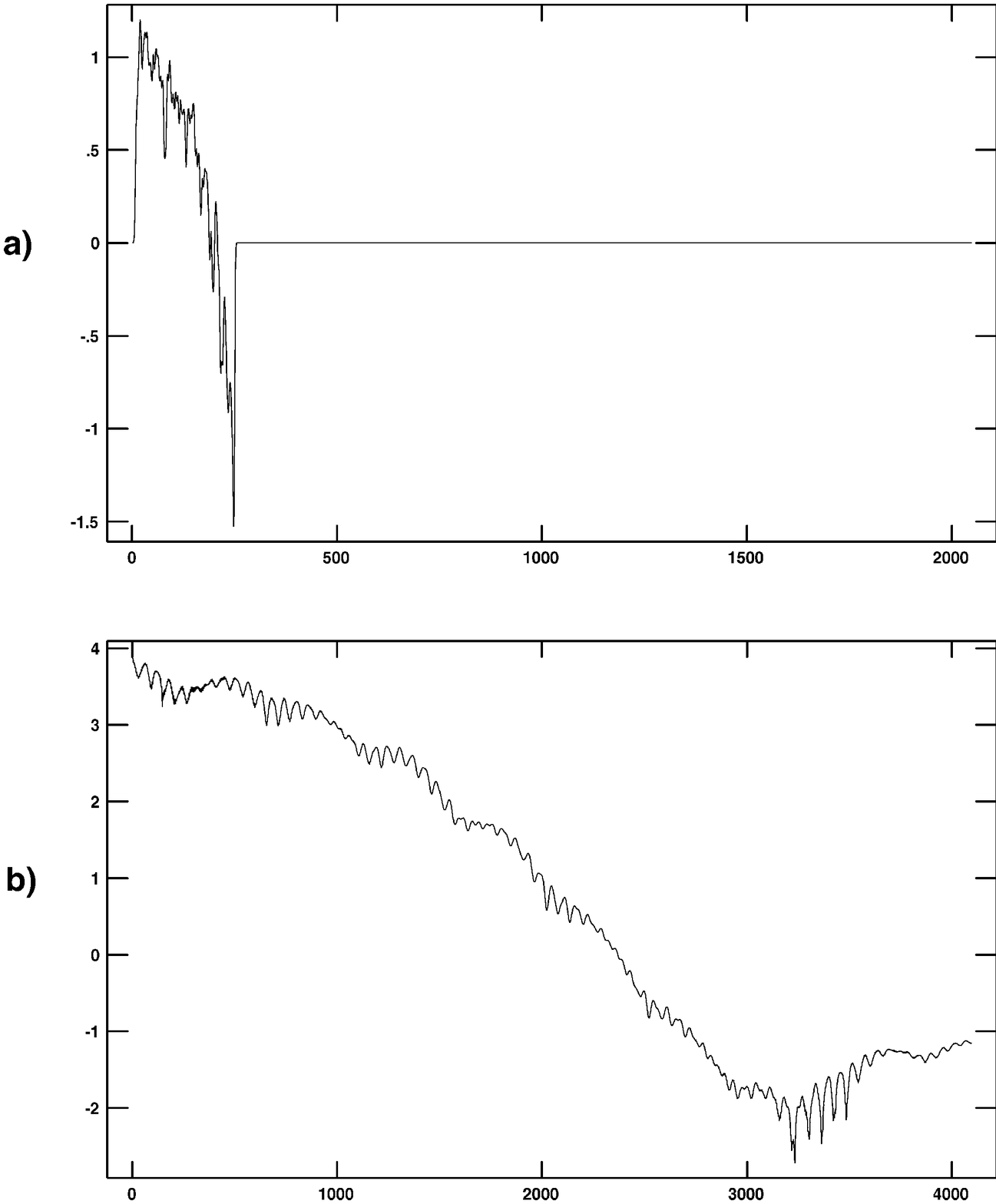}
     \caption{\label{xcsao9}
     Filtered Fourier transforms of a) absorption and b) emission templates
     in {\bf xcsao}.  No high frequencies are filtered out of the emission
     template transform.}

     \end{figure}

The transformed spectra are then filtered with a cosine-bell filter.
The low frequencies are filtered from {\it low\_bin} to {\it top\_low}
and the high frequencies are filtered from {\it top\_nrun} to {\it
nrun}. The template header parameter FI-FLAG controls whether the
template transform is filtered and whether the high-frequency filter
is turned off for both template and object transforms to leave in
emission lines.  If {\it tfft\_plot} is set to yes, the filtered
Fourier power spectra of the object and template spectra are displayed
as in Figures \ref{xcsao8} and \ref{xcsao9}.

\begin{figure}
     \plotone{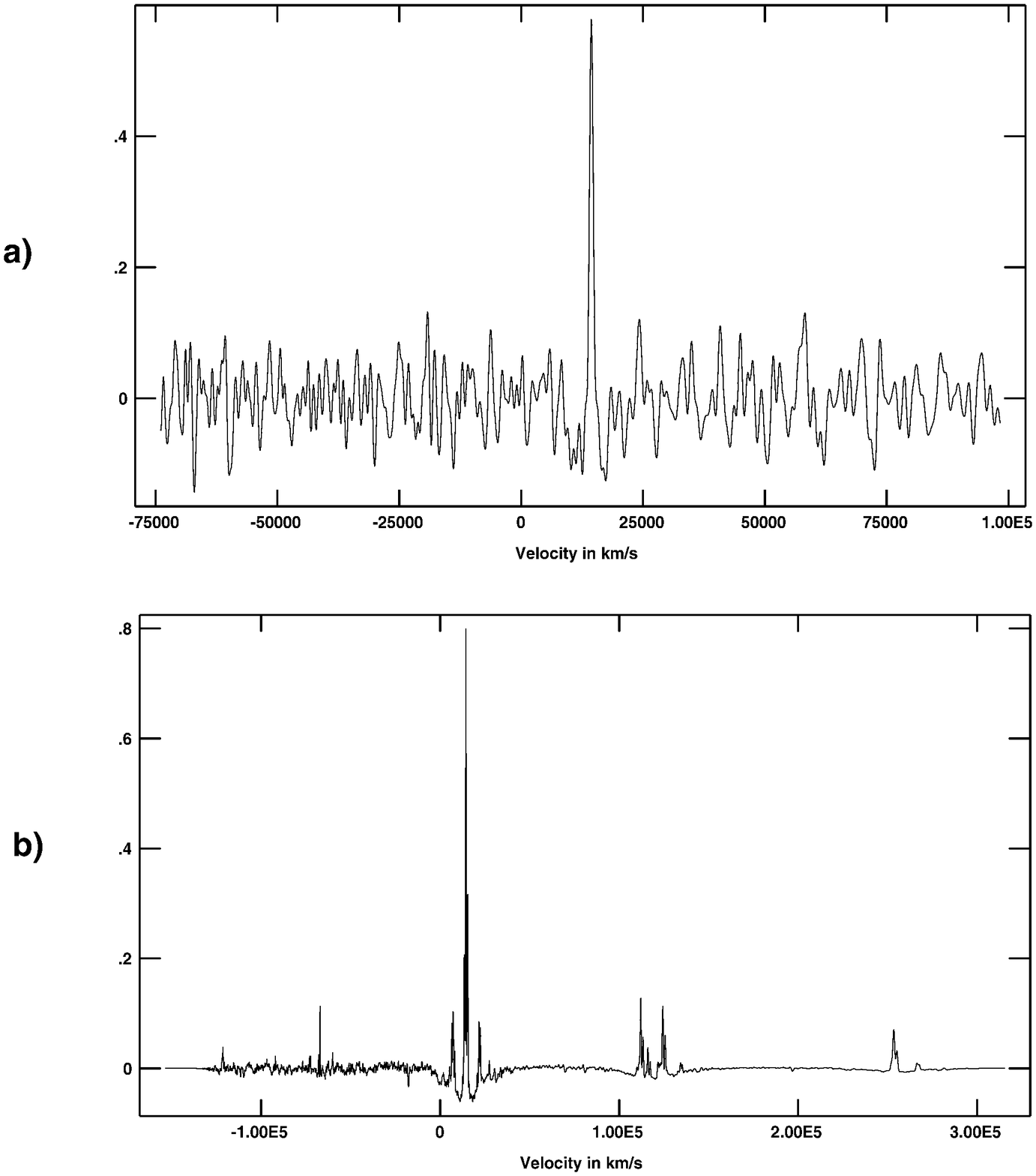}
     \caption{\label{xcsao10}
     Normalized, filtered cross correlations of the Fourier transforms of
     the object spectrum against the Fourier transforms of a) absorption
     and b) emission templates in {\bf xcsao}.}

     \end{figure}

The filtered transforms are then cross-correlated and normalized. If
{\it uxcor\_plot} is yes, the unfiltered correlation is displayed. If
{\it xcor\_plot} is yes, this result is displayed as in
Figure \ref{xcsao10}, and a specific peak may be selected using the
cursor if {\it curmode} is yes. In that case, the maximum value within
{\it pksrch} pixels of the cursor position is used. Otherwise, the
highest correlation peak between the velocities {\it minvel} and {\it
maxvel} is used. The redshift is calculated by fitting a parabola or
similar function specified by {\it pkmode} to the portion of the peak
above {\it pkfrac} of the maximum value of that peak. The R-value and
error are computed, and control returns to XCFIT to set up the
template for the next pass.

\begin{figure}
     \plotone{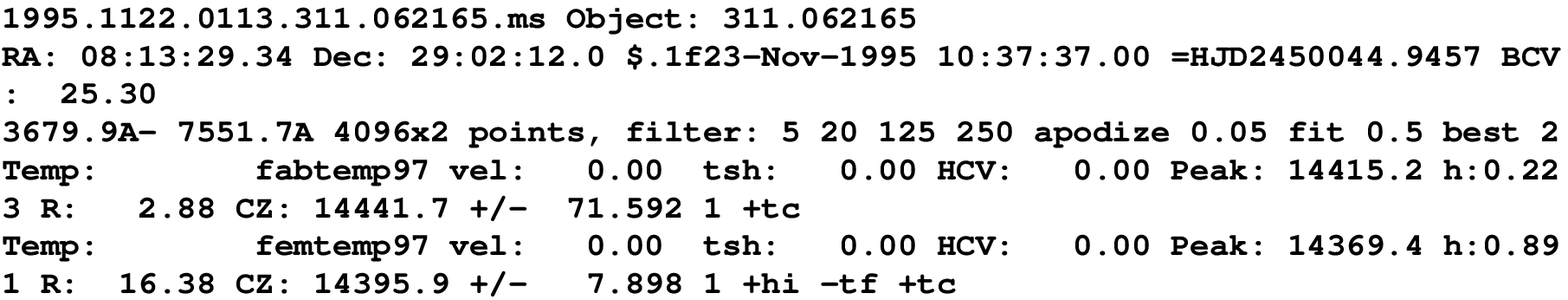}
     \caption{\label{xcsao11}
     {\bf xcsao} report of results of cross correlation of the object
     spectrum transforms against transforms of absorption and emission
     template spectra.}
     \end{figure}

After all of the template spectra have been correlated against an
object spectrum, the template with the highest R-value is
selected. The results are displayed as text to the devices specified
by {\it logfiles} in the format specified by {\it report\_mode}.
Figure \ref{xcsao11} shows the default report.

\begin{figure}
     \plotone{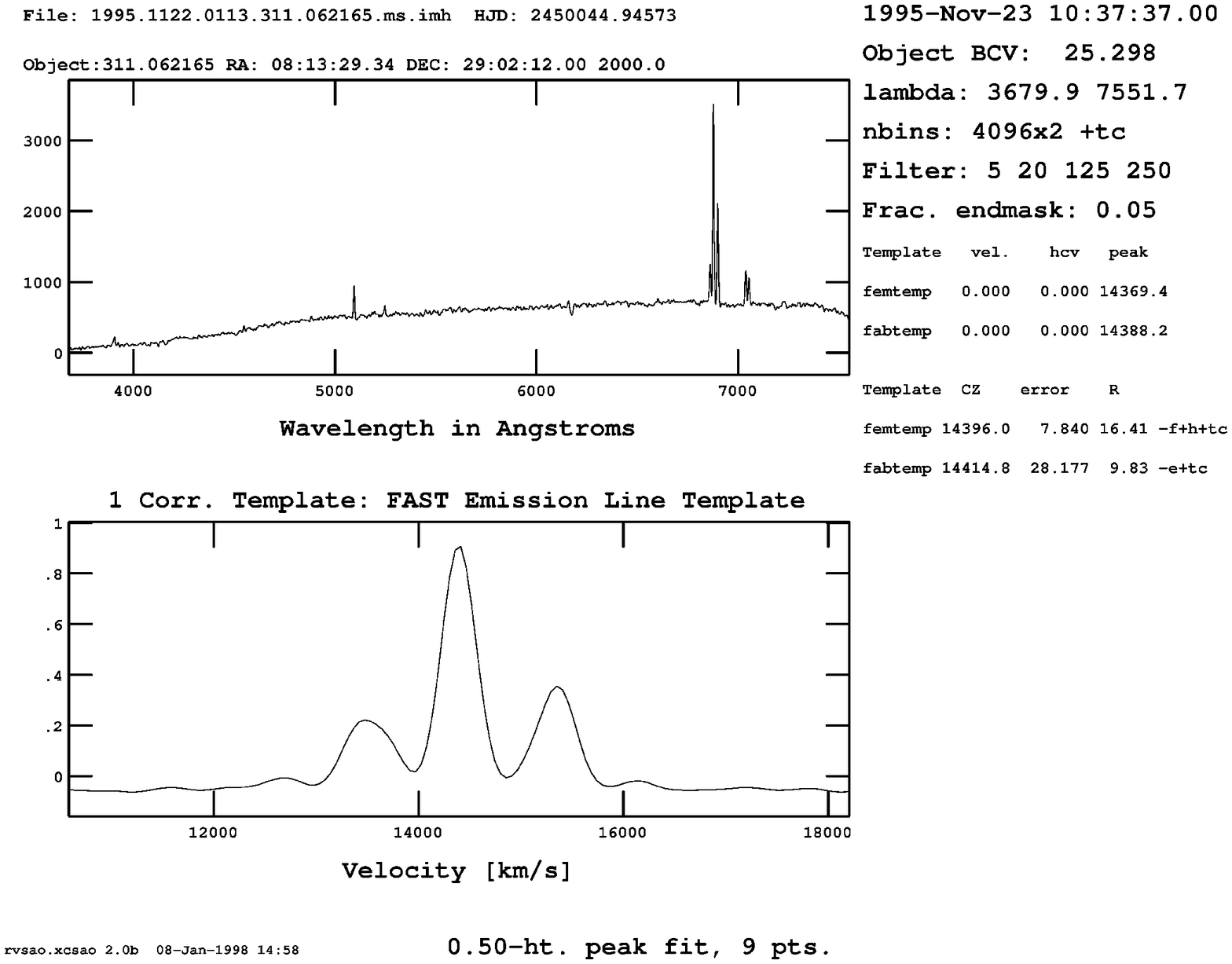}
     \caption{\label{xcsao12}
     {\bf xcsao} summary display for emission line cross-correlation.}
     \end{figure}

\begin{figure}
     \plotone{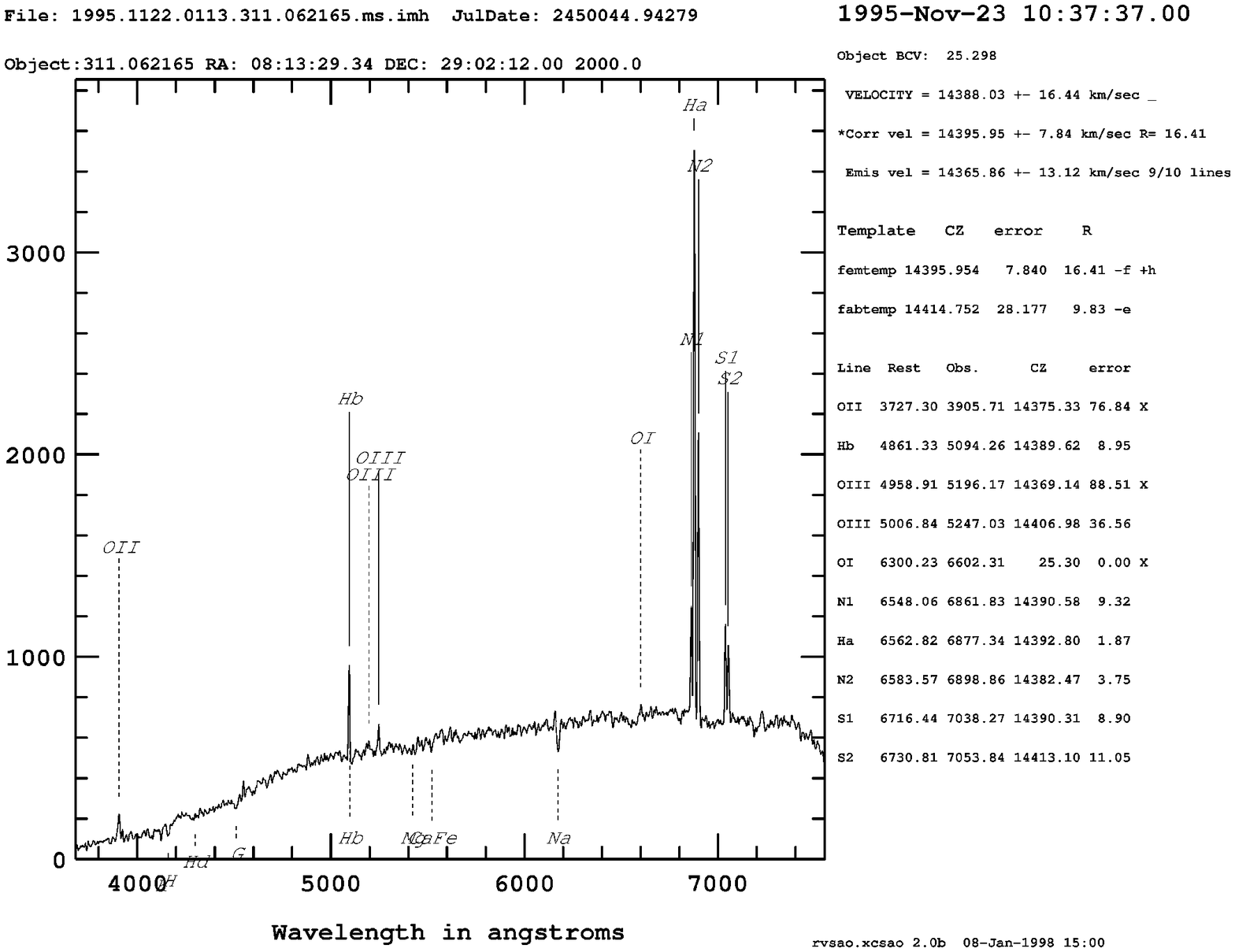}
     \caption{\label{xcsao13}
     {\bf xcsao} summary display with labeled lines for emission line
     cross-correlation.}
     \end{figure}

If {\it displot} is yes, the object spectrum and, optionally, the selected
correlation peak, are plotted to device in the format specified by
{\it dispmode}. Figure \ref{xcsao12} shows the summary graph if {\it
dispmode} is 1.  Figure \ref{xcsao13} shows the {\it dispmode} 2
summary graph, with emission and absorption lines labeled.  If {\it
hardcopy} is yes, the same graph is sent a printer.

If {\it nsmooth} $> 0$, the object spectrum is smoothed by a 1-2-1
sliding filter {\it nsmooth} times for display purposes only.  This
smoothing may be changed interactively using the g command in cursor
mode.  The filtered cross-correlation
with the best R-value is displayed centered on the redshift {\it cvel}
(in km/sec) with a width in km/sec of {\it dvel}. If {\it cvel} is
INDEF, the fit redshift is used; if {\it dvel} is INDEF, the width is
set to 20 times the peak width. If the correlation is not displayed,
absorption lines ({\it ablines}$=$yes) and/or emission lines ({\it
emlines}$=$yes) may be labeled from line lists in the directory {\it
linedir}, as shown in Figure \ref{xcsao13}.

\begin{figure}
     \plotone{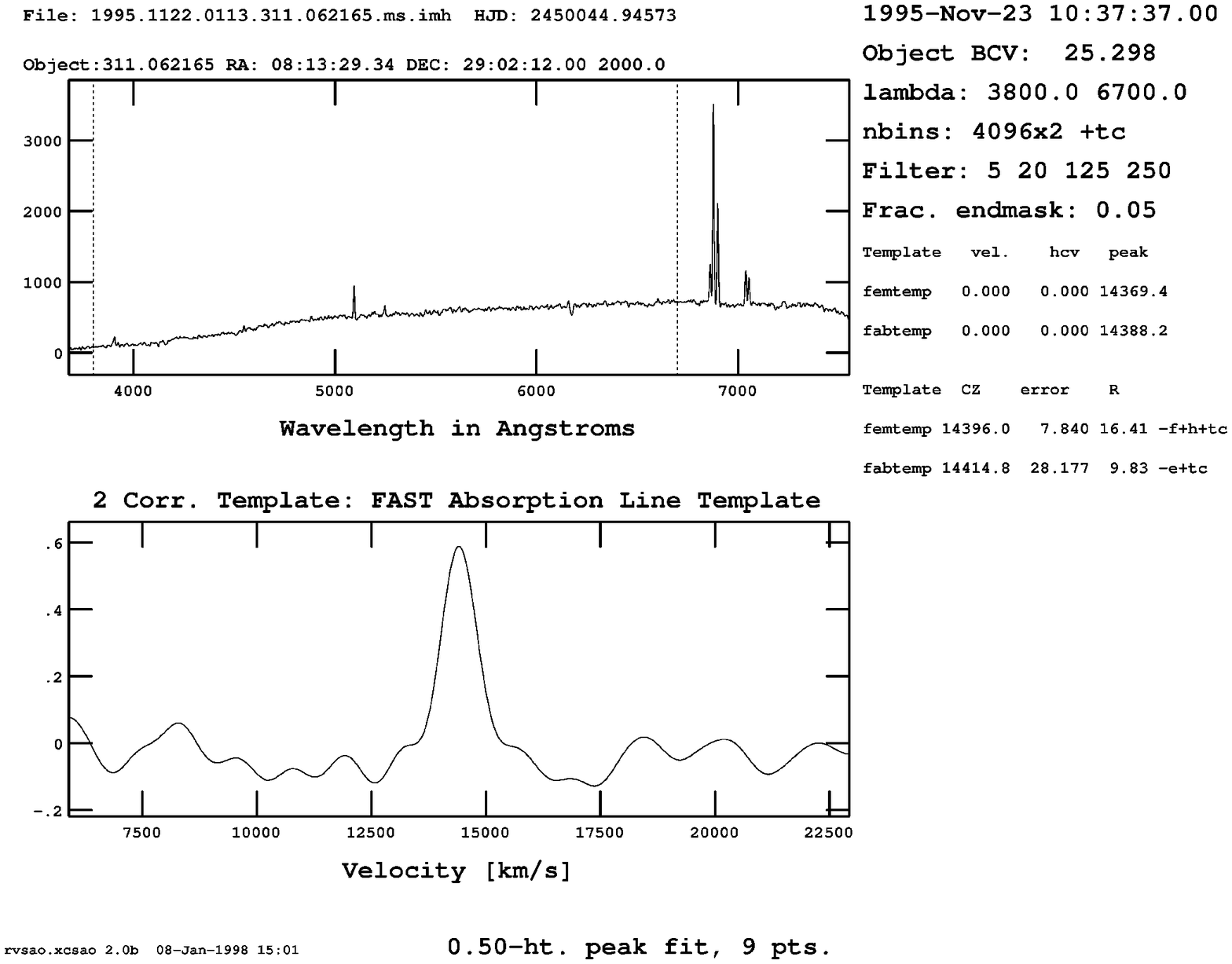}
     \caption{\label{xcsao14}
     {\bf xcsao} summary display for absorption cross-correlation.}
     \end{figure}

If {\it curmode} is yes, the user can interact with the display using the
terminal cursor to zoom in on portions of the spectrum, rerun the
cross-correlation, change the display format, edit the spectrum, or several
other functions.  For example, figure \ref{xcsao14} shows the correlation
result for the second best template, selected using the T cursor command.

\begin{figure}
     \plotone{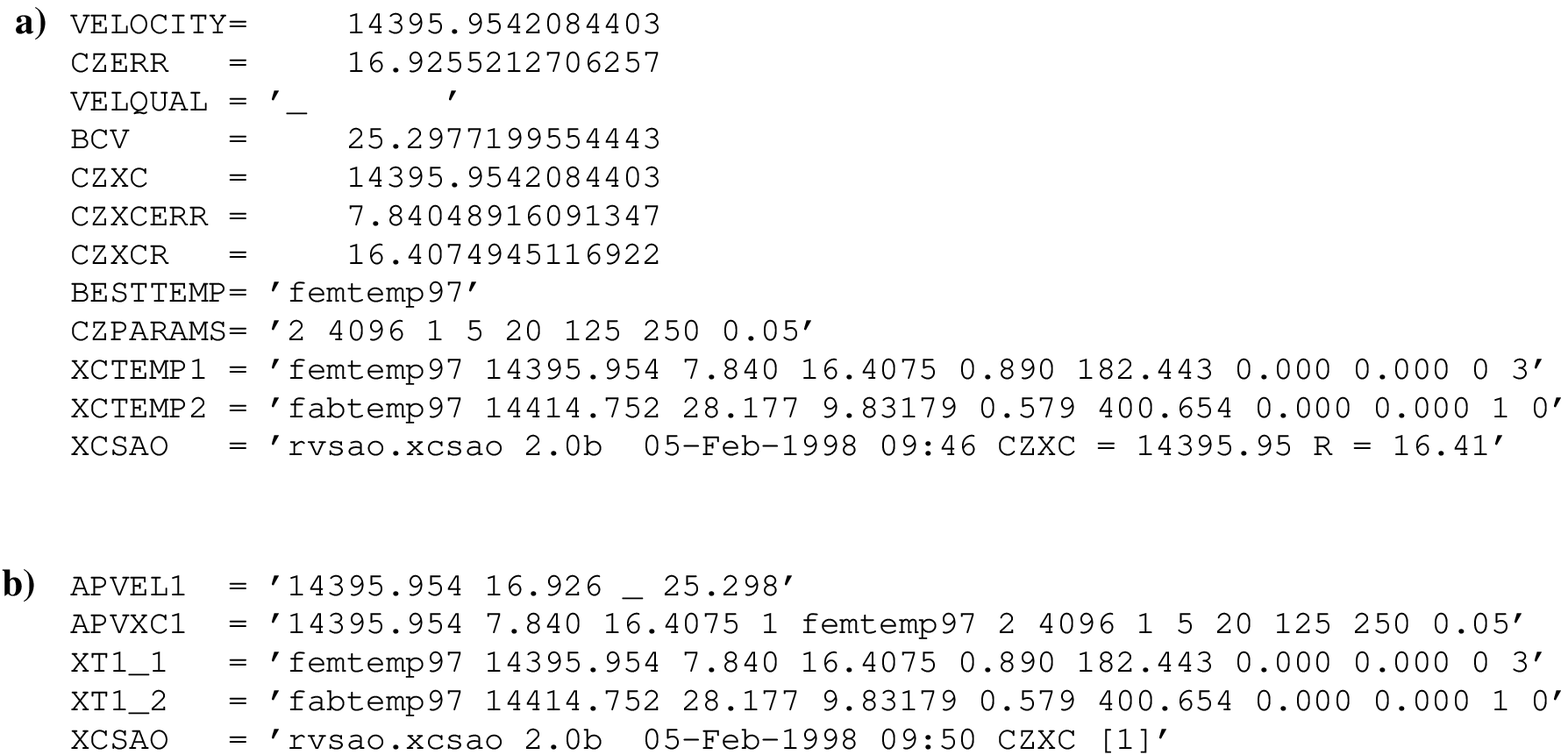}
     \caption{\label{xcsao15}
     Object spectrum header parameters added by {\bf xcsao}:
     For a single spectrum, as shown in a), separate keywords are used
     for the results of the correlation.  CZPARAMS gives the total number
     of templates, the number of points in the Fourier transforms, 2 if
     zero-padded, otherwise 1, and the four values for the cosine-bell filter
     used on the Fourier tranforms before they are correlated.  For multispec
     spectra, as shown in b) VELOCITY, CZERR, VELQUAL, and BCV are combined
     on one line, as APVELn, and the best correlation results (CZXC, CZXCERR,
     CZXCR), best template spectrum sequence number and name (BESTTEMP), and
     CZPARAMS are combined as APVXCn.  In either case there is one line of
     information per template, giving template name, redshift velocity,
     velocity error, R-value, peak height (0.0 to 1.0), peak width at
     half-max (km/sec), template velocity and barycentric correction,
     line-removal flag, and transform filter flag.}
     \end{figure}

If {\it save\_vel} is yes, cross-correlation redshift results are written
into the object spectrum image header in a form appropriate to the
spectrum format: two entries plus one per template if multispec; otherwise
one  value per keyword, as in Figure \ref{xcsao15}.

\subsection{\label{emsao}Fitting Redshifted Lines in a Spectrum in {\bf emsao}}

\begin{figure}
     \plotone{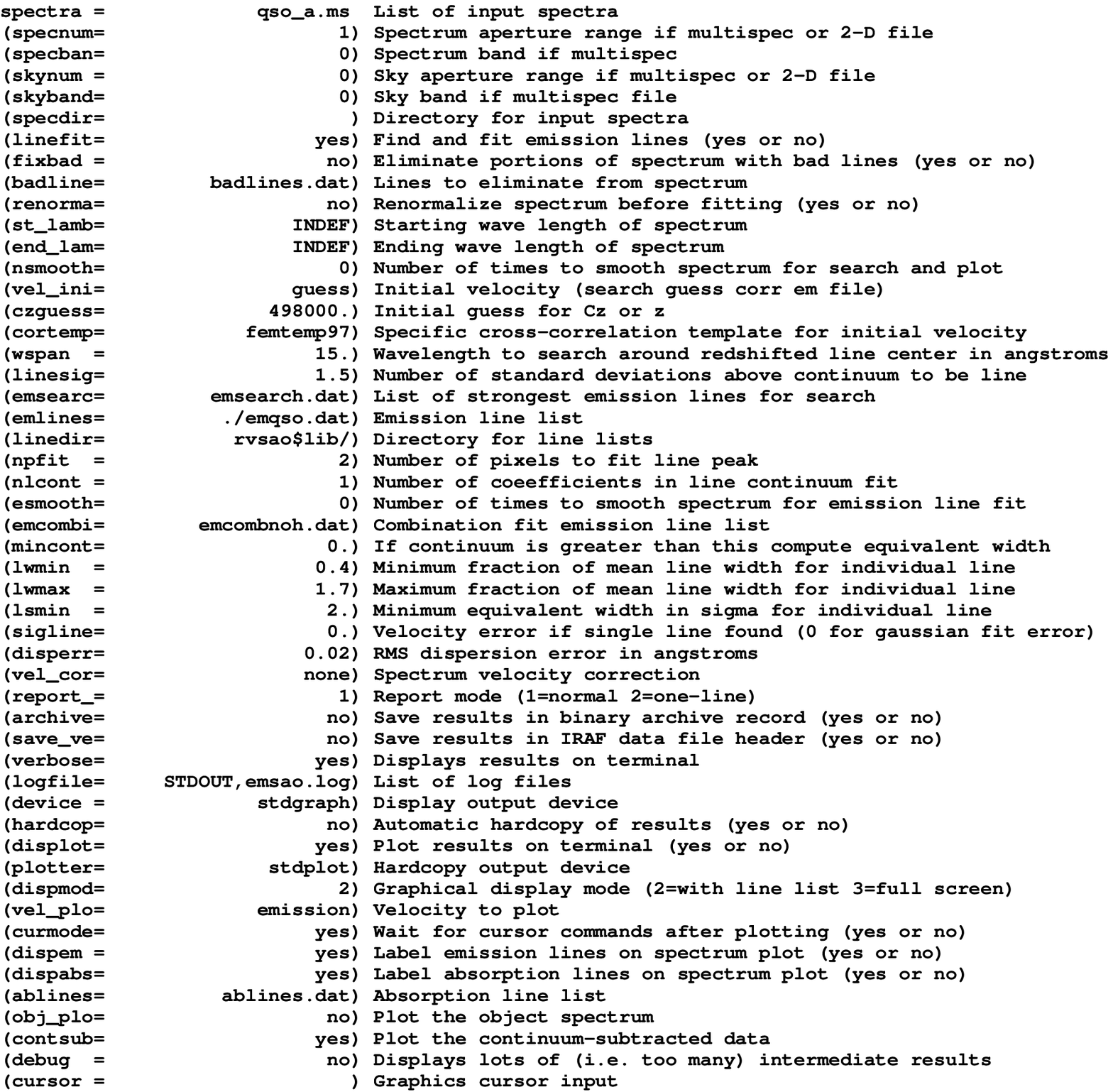}
     \caption{\label{emsao0} 
     Parameters for {\bf emsao}.}

     \end{figure}

While {\bf xcsao} can now use emission line templates, it is still useful
to measure redshifts of emission line spectra directly.  For large
surveys, the interactive determination of line centers and calculation
of redshifts by a program like IRAF's {\bf splot}, is simply too slow.  {\bf
emsao}, a companion to the cross-correlation task {\bf xcsao}, was
written to find emission lines automatically, compute redshifts for
each identified line, and combine them into a single radial velocity.
The results may be graphically displayed or printed, saved to a file,
and/or stored in the spectrum file header.  {\bf emsao} is designed to
run with minimal human intervention, but options may be set to allow manual
improvement of the line identifications and resulting redshift.  The
graphic cursor may be used to change fit and display parameters.
Figure \ref{emsao0} shows the full parameter list for {\bf emsao}.

For each object spectrum file from the input list of spectra, and/or each
aperture specified in the aperture list, the fitting subroutine, EMFIT,
is called for the specified spectrum image band.  Spectrum files are all read
from the specified directory unless a full pathname is given in the spectrum
list.  Relative pathnames may be used for spectra.  If a directory is not
set, the spectra are expected to reside in the current working directory.

\begin{figure}
     \plotone{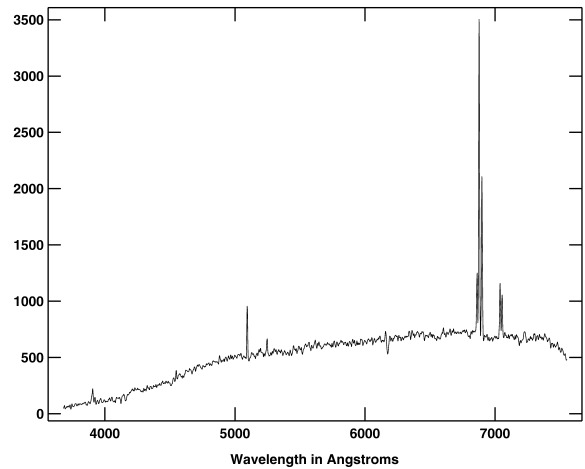}
     \caption{\label{emsao1} 
     {\bf emsao} spectrum display if obj\_plot is yes.}

     \end{figure}

After the spectrum is loaded, it is renormalized, if the {\it renormalize}
flag is yes. This should usually be done if the spectrum is in flux units.
If the {\it fixbad} flag is yes, regions specified in
the file named by {\it badlines} are replaced by straight lines.  The
spectrum is then smoothed {\it nsmooth} times.  If {\it obj\_plot} is
yes, the spectrum is plotted and the plot, as shown in figure
\ref{emsao1} is kept on the screen available for zooming and editing
until a "q" is typed.

To use a sky spectrum to compute noise statistics and to improve the
error computations, an aperture or band must be specified.  If {\it
skynum} is not zero, a sky spectrum is read from that multispec
aperture in the same file as {\it specnum}.  If {\it skyband} is not
zero, a sky spectrum is read from that multispec band in the same
file.  The sky spectrum, used to get the noise for error computations,
is plotted if {\it obj\_plot} is yes.  Previous results which have
been saved in the spectrum image header may be used by setting {\it
linefit} to no; in that case, all of the fitting below is skipped and
the results are displayed.

\begin{figure}
     \plotone{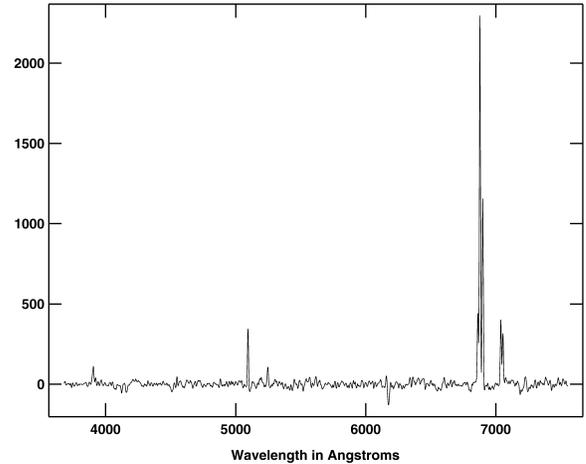}
     \caption{\label{emsao3} 
     The smoothed spectrum with the continuum subtracted is displayed 
     by {\bf emsao} if contsub\_plot is yes.}

     \end{figure}

The continuum, computed by the IRAF curve fitting subroutine which is
driven by the parameters set in the {\bf contpars} pset task, is
subtracted from the spectrum.  If {\it contsub\_plot} is yes, the
spectrum is plotted with the continuum removed as in figure
\ref{emsao3}.

\begin{figure}
     \plotone{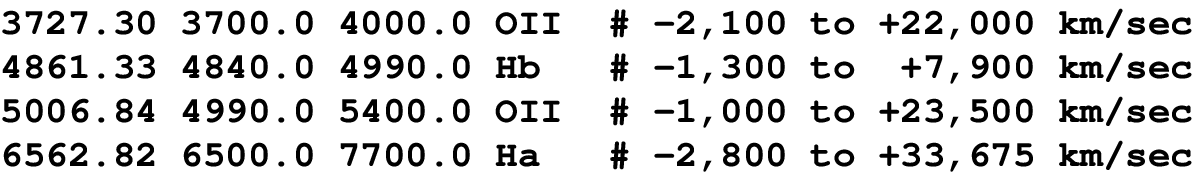}
     \caption{\label{emsao4}
     Emission lines for {\bf emsao}'s initial redshift guess from the
     file emsearch.dat.
     The first column is the line wavelength in Angstroms.  The second and
     third columns are the minimum and maximum observed wavelengths to which
     it could be shifted.  The fourth column is the name of the line.  The
     comment field gives the redshift range over which the emission line
     will be in the given wavelength range.}

     \end{figure}

The source of the initial redshift is specified by {\it vel\_init}.
If it is ``guess", the starting redshift is read from the parameter
{\it czguess}.  If it is ``search", one line in the spectrum is
identified by the program using the table specified by the {\it emsearch}
parameter, as shown in Figure \ref{emsao4}.  This table lists line centers in
Angstroms and the wavelength range over which each one should be the
strongest line.  It can be modified by the user to match the data.
The brightest line in each region is assumed to be the one in the
table, and its observed wavelength is saved.  The redshift of the
brightest of those lines is returned as an initial value to be refined
by looking at more lines.

If {\it vel\_init} is ``combination", the initial redshift velocity is
read from the spectrum header parameter VELOCITY; if ``correlation", it
is read from the spectrum header parameter CZXC; and if ``emission",
from CZEM.

\begin{figure}
     \plotone{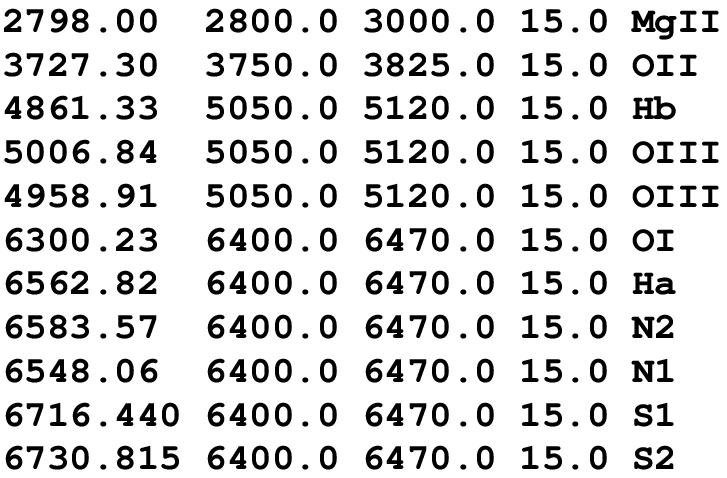}
     \caption{\label{emsao5}
     Emission lines for {\bf emsao} to look for in galaxy spectra from the
     file emlines.dat.  If Z=0, OI at 6300 angstroms is from the earth's
     atmosphere.  The first column is the line wavelength in Angstroms.
     The second and third columns are the minimum and maximum rest
     wavelengths from which to take continuum values while looking for
     the line.  The fourth column is the wavelength to fit on either side
     of the peak (for single line fits), and the last column is the name
     of the line.}

     \end{figure} 

Search regions are read for each line in the file specified by the
{\it emlines} parameter.  Figure \ref{emsao5} shows the contents of
such a file.  Each region is then shifted by the guessed redshift and
expanded in each direction by {\it wspan} Angstroms.  All emission
lines within each specified wavelength region are found.  A spectrum
pixel is assumed to be a line center if the pixel value is the max of
the {\it npfit} neighbors on either side, and greater than {\it
linesig} times the square root of the average counts in those pixels.
A second order fit is then made to the (2*{\it npfit})+1 points
centered on the peak to refine the center and peak height.  The
brightest line in each region is kept unless it has already been
identified.  Order matters--the brightest line in a region should be
listed first, so that if it is the only one present in overlapping
regions, it is correctly named.

\begin{figure}
     \plotone{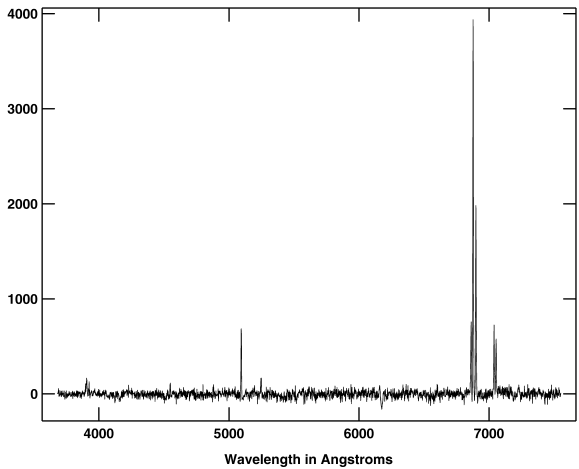}
     \caption{\label{emsao7}
     Unsmoothed spectrum with continuum removed, displayed by {\bf emsao}
     if {it contsub\_plot} is yes.}

     \end{figure}

Before line profiles are fit, a copy of the spectrum is smoothed {\it
esmooth} times using the same smoothing algorithm as is used before
the line search is conducted.  The parameter {\it esmooth} should be
left at zero unless the data are especially noisy.  It is best never to
go above 2.  Because local continuum values are important to the line fit,
the continuum is removed from this copy of the spectrum using the same
parameters as were used before conducting the line search.  If
{\it contsub\_plot} is yes, the spectrum with the continuum removed
is plotted as shown in figure \ref{emsao7}.

\begin{figure}
     \plotone{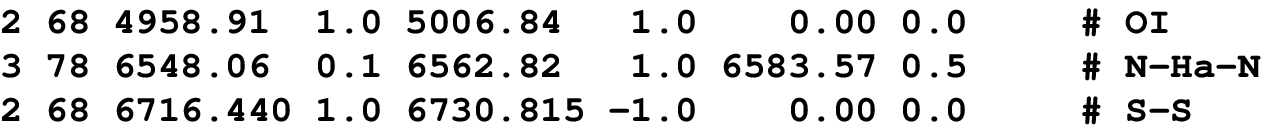}
     \caption{\label{emsao8}
     Emission line combinations for {\bf emsao} to use with galaxy spectra
     from the file emcomb.dat.  The first column is
     the number of lines to be fit together.  The second column is the
     additional wavelength range to fit blueward of the bluest peak and
     redward of the reddest peak.  Then there are center wavelength and
     relative height pairs for each of the two or three lines to be fit
     together.}

     \end{figure} 

Each identified line is checked to see if it is part of one of the
groups of close emission lines listed in the file specified by {\it
emcombine}, such as that shown in figure \ref{emsao8}.  If it is, all
lines in the combination will be simultaneously fit.  Those members of
a line combination which are not found are initialized at the redshift
of the most recently found line of the group. The missing line heights
are assumed proportional to that line according to the relative
heights in the {\it emcombine} file.  The lines are fit by one to
three Gaussians, along with an optional local continuum level using
0--3 additional coefficients as set by {\it nlcont}.
Redshift (computed from the wavelength of the center pixel
coordinate), width, height, and errors are returned for each line.

Each emission line is checked to see whether: 1) it has a fit center
greater than zero, 2) it is wider than lwmin times the mean width of
all of the identified lines which are found and 3) narrower than {\it
lwmax} times that mean width, 4) it has an equivalent width more than
{\it lsmin} times the error in the equivalent width, 5) it is too
close to the blue edge of the spectrum, 6) it is too close to the red
edge of the spectrum, or 7) the error in the center of the Gaussian is
zero.  If {\it dispmode} is 1 or {\it report\_mode} is 1, wavelengths
and redshifts are printed for each line, with an X followed by the
code for the test which failed at the end of entries which were
omitted from the fit.  If a line has been successfully fit, the
rejection can be overridden interactively (if {\it curmode} is yes)
using the + and - commands in cursor mode from the final display.  If
a line has been added or subtracted in cursor mode, a + or - at the
end of the entry indicates that fact.  A mean velocity is computed,
weighted by the square of the error in the line centers and returned.
If only a single line is found, the error is set to {\it sigline},
which should be set to the uncertainty in the dispersion function in
Angstroms.  This is usually significantly greater that the error in
the fit to the center of the Gaussian.

After all of the lines have been fit and a combined velocity has been
computed, a zero-point redshift is computed by adding the solar system
barycentric velocity correction, from a source specified by {\it
svel\_corr}.

\begin{figure}
     \plotone{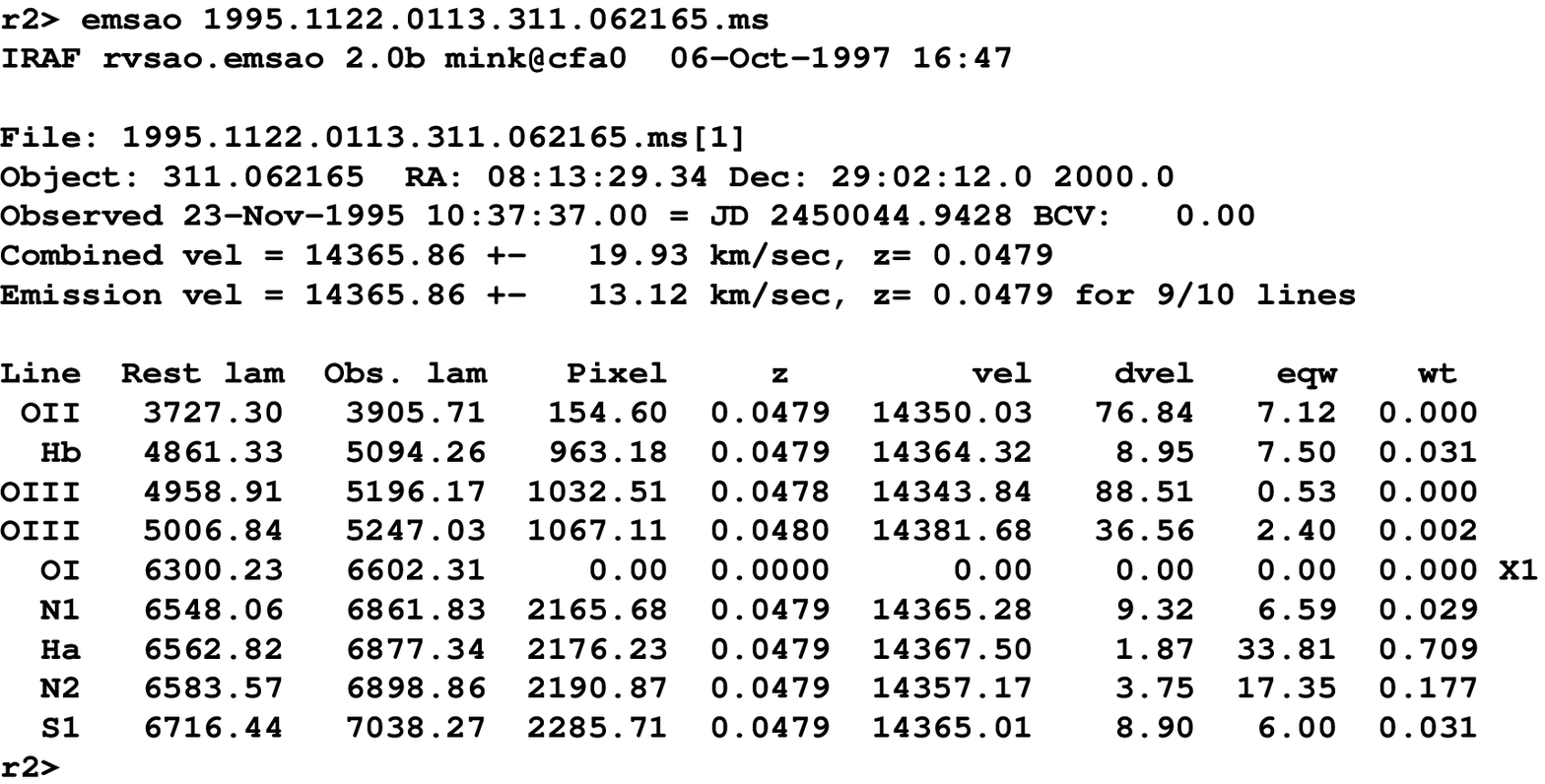}
     \caption{\label{emsao9}
     {\bf emsao} report for an emission line galaxy when {\it report\_mode}
     is 1.  For each emission line, the name, rest wavelength, observed
     wavelength, pixel number at the observed wavelength, redshift in
     fraction of the speed of light and km/sec and error in km/sec,
     equivalent width, and relative weighting factor are printed.  An "X",
     followed by a rejection code, after the last column indicates that
     the line was not used.}

     \end{figure}

The results are displayed as text to the devices specified by {\it
logfiles} in the format specified by {\it report\_mode}.  Options
include 1) one line per found emission line under a self-documenting
summary, 2) a single line report listing the names of the lines which
are found, and 3) a single line report listing a velocity for each
searched-for line.  Figure \ref{emsao9} shows the mode 1 report for the
spectrum we are following.

\begin{figure}
     \plotone{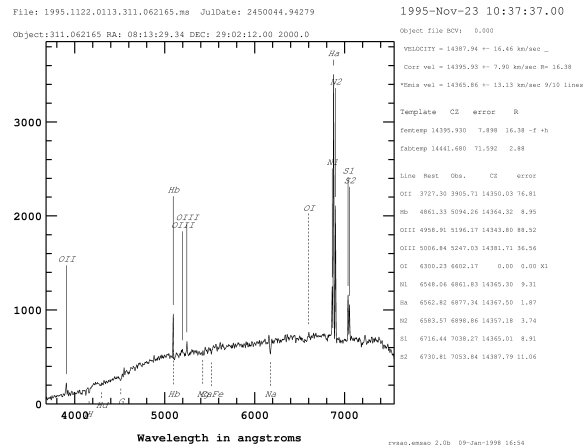}
     \caption{\label{emsao10}
     The {\bf emsao} summary display shows the spectrum and line fit results
     if {\it dispmode} is 1.}

     \end{figure}

\begin{figure}
     \plotone{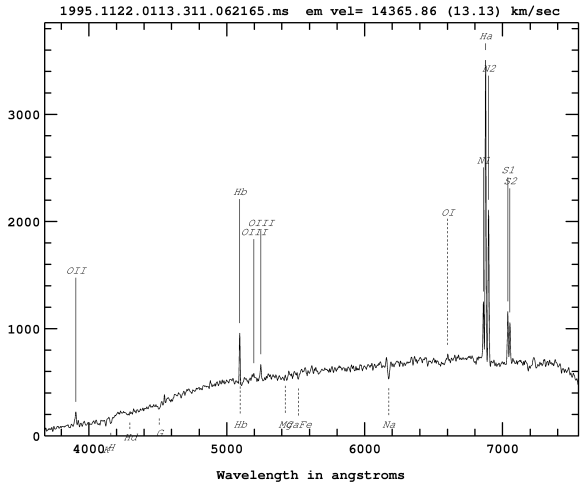}
     \caption{\label{emsao11}
     The {\bf emsao} summary display shows only the spectrum if {\it dispmode}
     is 2.}

     \end{figure}

If {\it displot} is yes, the spectrum is plotted to device in the
format specified by {\it dispmode}.  Figure \ref{emsao10} shows the
mode 1 report; figure \ref{emsao11} shows the mode 2 report.  If {\it
hardcopy} is yes, the same graph is automatically sent to a printer
as well.

If {\it nsmooth} is greater than zero, the displayed spectrum is
smoothed by a 1-2-1 sliding filter that many times.  Absorption lines
listed in the file {\it ablines} are labeled if {\it dispabs} is yes,
and emission lines listed in the file {\it emlines} are labeled if
{\it dispem} is yes.  Both files are found in the directory {\it linedir}.

If {\it curmode} is yes, the user can interact with the display using
the terminal cursor to zoom in on portions of the spectrum, identify
lines and refit the emission lines, change the display format, edit
the spectrum, or several other functions.

\begin{figure}
     \plotone{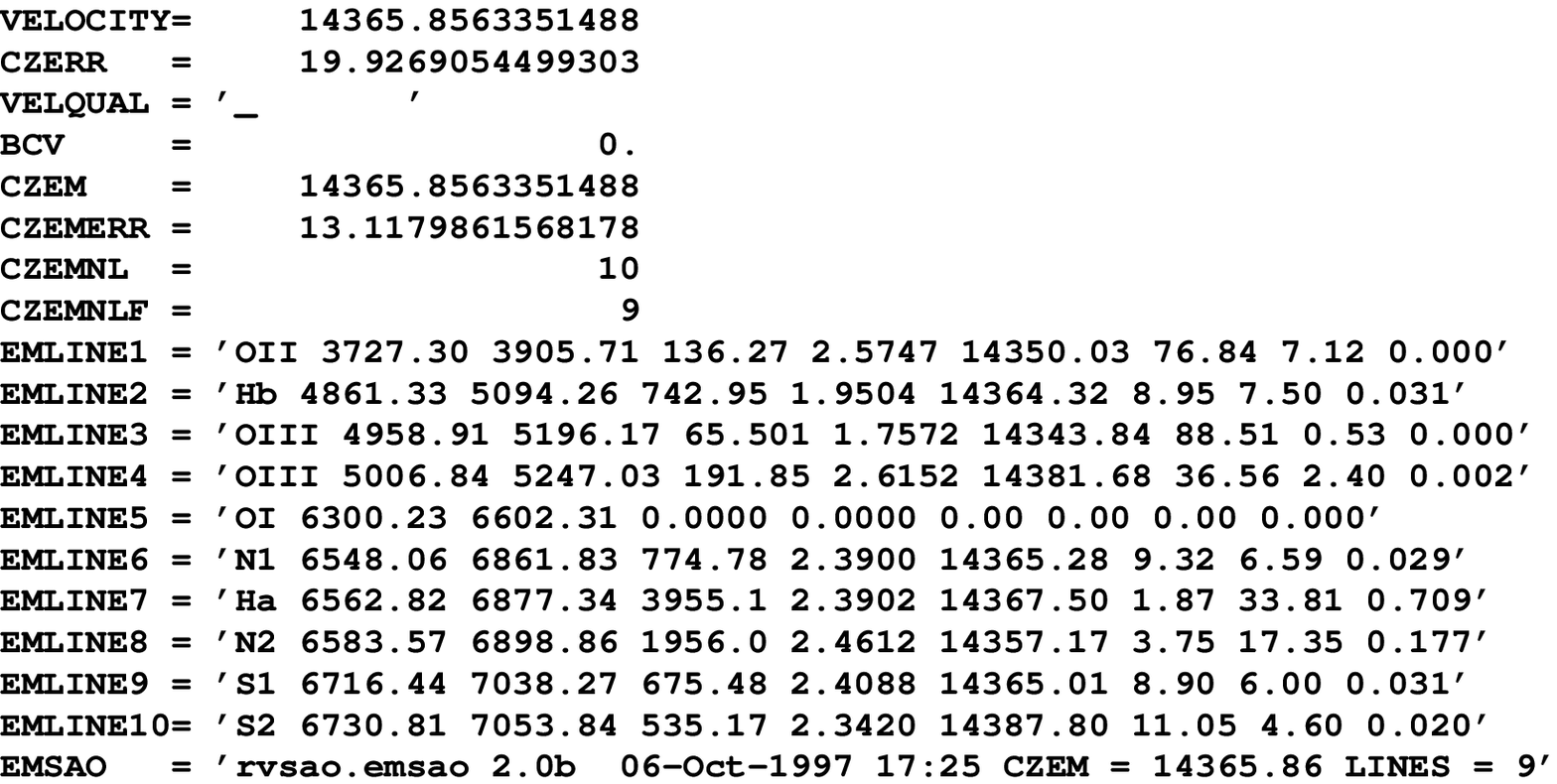}
     \caption{\label{emsao12}
     These keywords save the {\bf emsao} results in the spectrum's FITS header.
     For each emission line, the name, rest wavelength, observed wavelength,
     observed height and width, redshift and error in km/sec, equivalent
    width, and relative weighting factor are saved.}

     \end{figure}

If {\it save\_vel} is yes, emission line redshift results are written
into the spectrum image header in a form appropriate to the spectrum
format: two entries plus one per line if multispec; otherwise one,
value per keyword, as shown in Figure \ref{emsao12}.

\subsection{\label{linespec}Creating a Spectrum with {\bf linespec}}

\begin{figure}
     \plotone{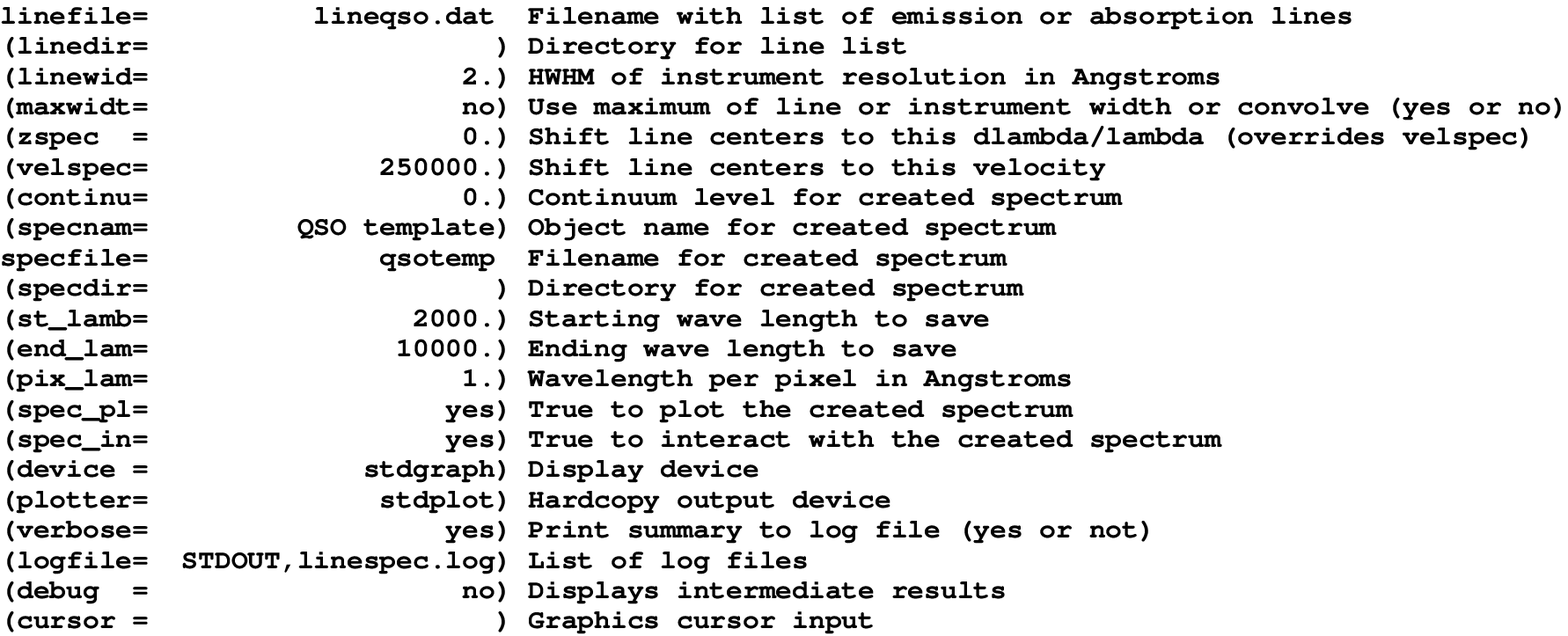}
     \caption{\label{linespec0} 
     Parameter list for {\bf linespec}}

     \end{figure}

\begin{figure}
     \plotone{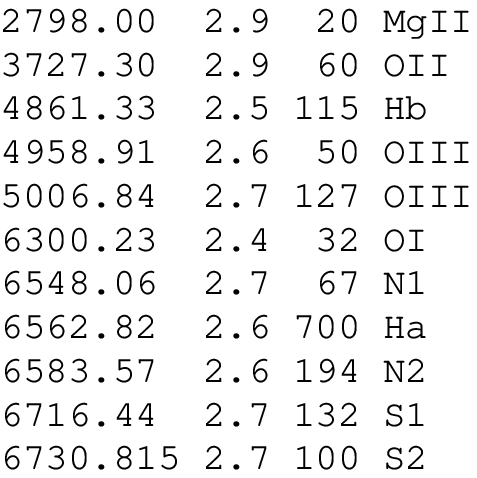}
     \caption{\label{linespec1}
     List of emission lines from {\bf linespec} input file. The first column
     is the center wavelength in Angstroms, the second is the half-width in
     Angstroms, the third is the height in arbitrary units to be used in the
     output spectrum. The last column identifies the line.}

     \end{figure}

{\bf linespec} is an IRAF task for making a spectrum from a list of
emission and/or absorption lines which is driven by the parameters shown
in Figure \ref{linespec0}.  It was written to create templates for use
by {\bf xcsao}.  An example of such a line list, with center
wavelengths, widths, and heights for each line which may appear in the
spectrum, is shown in figure \ref{linespec1}. The same line-defining
parameters are used as are output of {\bf emsao}, so low-noise templates
can be easily made. The line list is read from the file specified by
the {\it linefile} parameter in the directory designated by the
parameter, {\it linedir}.  If {\it linedir} is null, the file is
assumed to be in the current working directory.

A blank (all zero) spectrum, with the object name {\it specobj} and
file name {\it specfile} in the directory {\it specdir} is created. It
is linear in wavelength, with a resolution in Angstroms given by {\it
pix\_lambda}, and a range from {\it st\_lambda} to {\it
end\_lambda}. Spectral world coordinate system information is written
to the header, with the standard FITS parameters CRPIX1, CRVAL1, and
CDELT1. If {\it verbose} is yes, files specified by {\it logfiles} are
opened and a header is written.

\begin{figure}
     \plotone{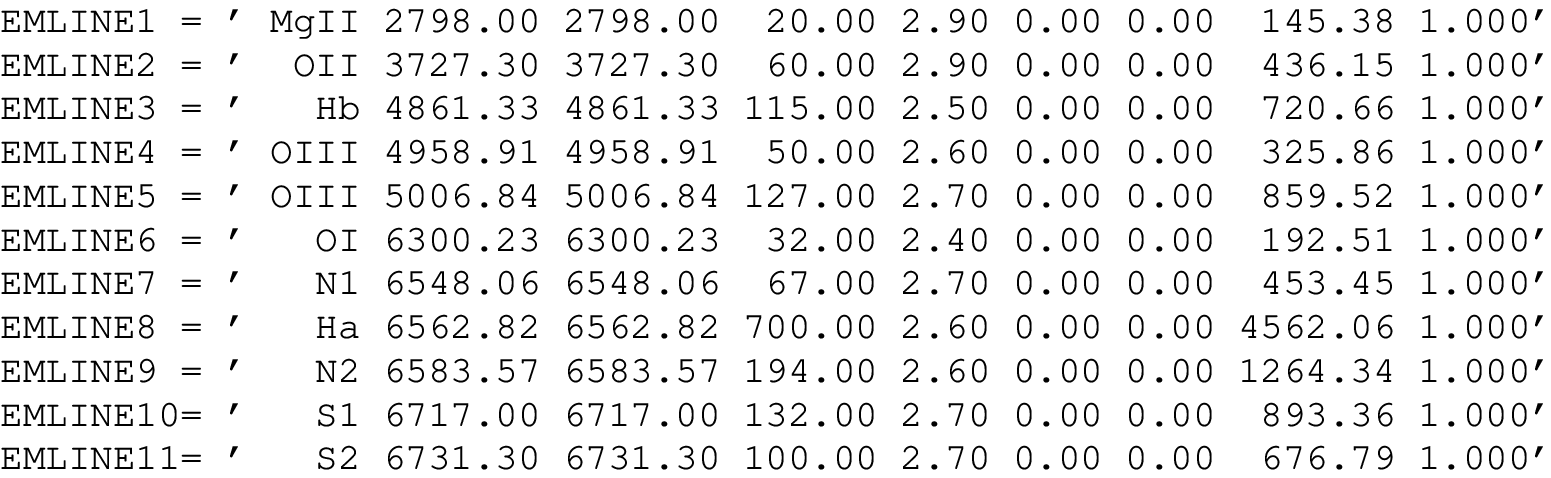}
     \caption{\label{linespec2}
     These multiple-valued keywords are added to the header of the synthesized
     spectrum to show what lines were added. For each emission line, the name,
     center, redshifted center, height, half-width, velocity, and equivalent
     width.}

     \end{figure}

The center of each line in the table is redshifted according to the
parameters {\it zspec} and {\it velspec}.
{\it velspec} (cz$=$apparent Doppler shifting velocity) is used unless
{\it zspec} $(\Delta \lambda / \lambda)$ is not zero, in which case
{\it zspec} is used for the redshift.
The {\it linewidth}, if it is tabulated in kilometers per second, is
converted to Angstroms at the shifted line center. The line width is
also broadened appropriately if the line is redshifted. The redshift
velocity is put in the spectrum header using the VELOCITY keyword. If
{\it maxwidth} is yes, the {\it linewidth} parameter is used as the
width of the line if it is greater than the tabulated width.  If {\it
maxwidth} is no, the width from the table is used, and the {\it
linewidth} smoothing is done later. For each line, a Gaussian at the
shifted center wavelength, half-width, and tabulated height is added
to the spectrum.  As each line is added to the spectrum, the line
definition is written to the spectrum's header, producing a table such
as that shown in figure \ref{linespec2}.  After all of the lines are
computed, a continuum level specified by the parameter {\it continuum}
is added to the spectrum.

\begin{figure}
     \plotone{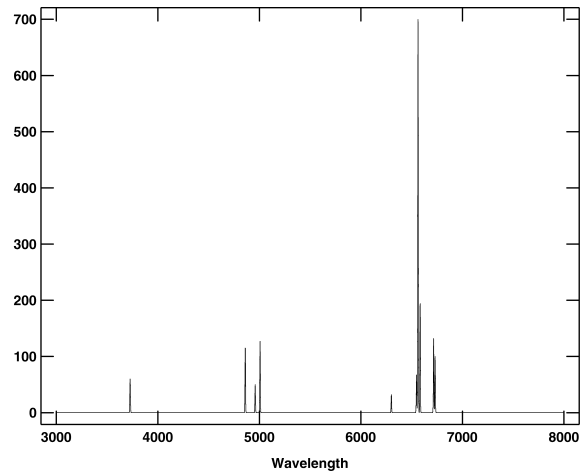}
     \caption{\label{linespec3}
     Graph of an emission line spectrum produced by {\bf linespec}.}

     \end{figure}

\begin{figure}
     \plotone{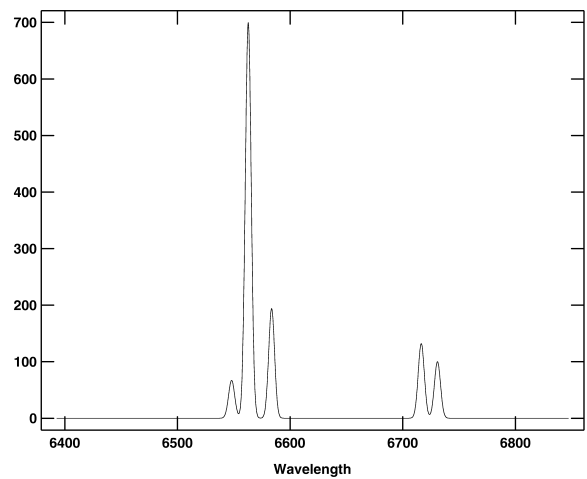}
     \caption{\label{linespec4}
     Graph of hydrogen alpha region of the same spectrum displayed using
     the "z" cursor command in {\bf linespec}.}

     \end{figure}

The spectrum is plotted to a graphics device, as shown in figure
\ref{linespec3}, if {\it spec\_plot} is yes. After the graph is
displayed to the terminal, if {\it spec\_int} is yes, the user must
interact with it using single character cursor commands, such as the
``z" for zoom command, the result of which is shown in figure
\ref{linespec4}.  If the ``@" command is given in this mode, the graph
is immediately sent to the device specified by {\it plotter}.  The
``q" command must be typed to leave the graph and proceed.

If {\it maxwidth} is no, the resolution of the spectrograph is
simulated by convolving the entire spectrum with a Gaussian of height
1.0 and sigma (half width at half of maximum) of {\it linewidth}. If
{\it spec\_plot} is yes, the spectrum is displayed again.

The {\bf linespec} version and the date the program is being run are written
to the spectrum image header, the spectrum is written, and the file is closed.

\subsection{\label{sumspec}Creating a Composite Spectrum with {\bf sumspec}}

\begin{figure}
     \plotone{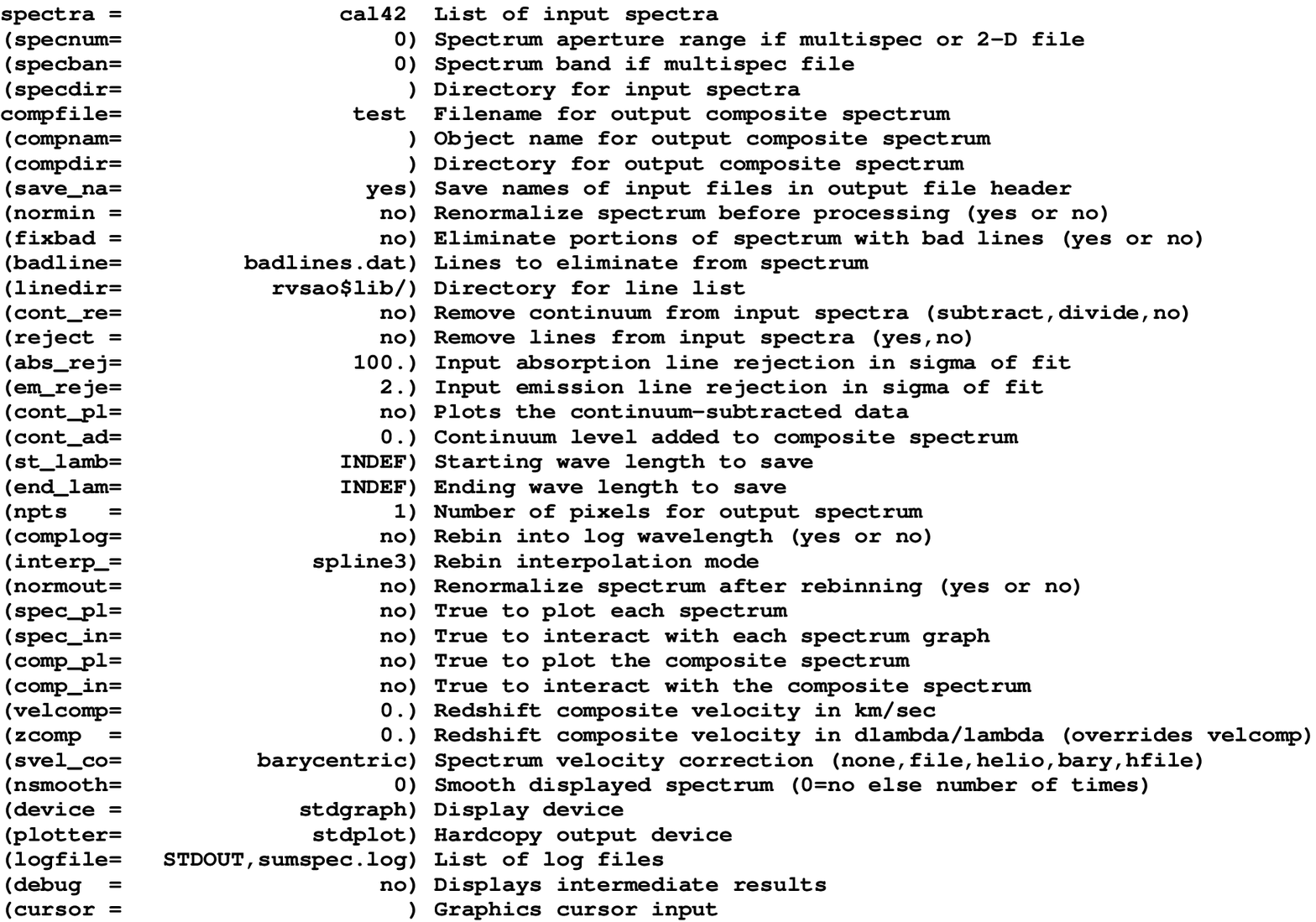}
     \caption{\label{sumspec0} 
     Parameter list for {\bf sumspec}.}

     \end{figure}

{\bf sumspec} is an IRAF task for making a composite spectrum which is adjusted
for velocity. It rebins input spectra to a common wavelength or log-wavelength
range and resolution, with optional continuum removal and normalization.
While it was originally intended for making templates to be used by the
{\bf xcsao} cross-correlation task, it has turned out to be very useful for
rebinning individual spectra as well as combining multiple spectra. 
The parameter list for {\bf sumspec} is shown in Figure \ref{sumspec0}.

The spectra to be added are specified by the list of files in the
parameter {\it spectra} and/or the list of spectrum numbers in {\it
specnum}. Unless a pathname is specified as part of the filename in
{\it spectra}, each spectrum is expected to be found in the directory
{\it specdir}. All spectra are taken from the band, {\it specband}, in
multispec spectra. Files specified by {\it logfiles} are opened to
receive logging information. If {\it debug} is yes, additional
information about the input files and the progress of the program is
written to STDERR.

If either {\it st\_lambda} or {\it end\_lambda} is set to INDEF, all
of the spectrum headers are read to find the limits of the overlap of
their wavelength coverage, and the missing limit(s) of the output
spectrum are set accordingly.

Before the first spectrum in the list is added, a blank (all zero)
spectrum, with the object name {\it compobj} and file name {\it
compfile} in the directory {\it compdir} is created. If {\it complog} is
yes, it is linear in log-wavelength; otherwise, it is linear in wavelength.
Either way, the output spectrum contains {\it npts} data points between
{\it st\_lambda}, or its log, and {\it end\_lambda}, or its log. World
coordinate system information is then written to the header.

\begin{figure}
     \plotone{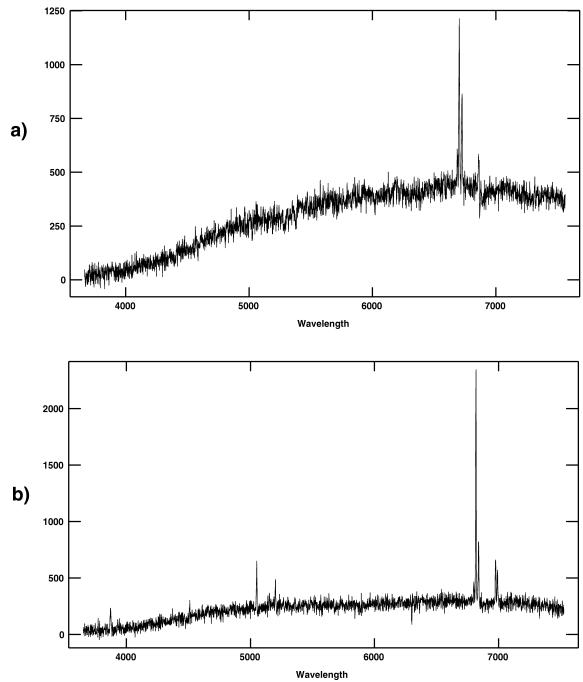}
     \caption{\label{sumspec1}
     Input spectra for {\bf sumspec}.}

     \end{figure}

Spectra are read from the input list one at a time and renormalized if
{\it renormalize} is set to yes. This should be done if the spectra
are fluxed or whenever pixel values are much bigger or smaller than
their variation.  If {\it spec\_plot} is yes, each spectrum is plotted
to the device specified by the parameter {\it device}, as in figure
\ref{sumspec1}. If {\it spec\_int} is yes, the user can interact with
that graph using single character cursor commands and must type a ``q" to
proceed with the task. If the ``@" command is given in this mode, the
graph is sent to the device specified by {\it plotter}.

Each spectrum is rebinned using an interpolation mode specified by
{\it interp\_mode} into a {\it npts}-pixel wavelength-linear spectrum
covering the range computed above. If {\it complog} is yes, the
rebinned spectrum is linear in log-wavelength; if {\it complog} is no,
the rebinned spectrum is linear in wavelength. If specplot is yes, the
rebinned spectrum also is plotted. 

If {\it s\_contin} is set to subtract or divide instead of no, the
IRAF curve fitting subroutines, are used to fit a continuum to each
input spectrum.  The CONTSUM task, described in the following section,
sets all of the appropriate parameters.

As it is rebinned, each input spectrum is redshifted according to the
parameters {\it ztemp} ($z = \Delta \lambda / \lambda)$, if it is not zero,
or {\it veltemp} (cz = apparent Doppler shifting velocity). A velocity
correction to the solar system barycenter is removed according to the
{\it svel\_corr} parameter.  Set it to none if the input spectra have not
been shifted. If ``file", BCV is used if present in the file header, or
else HCV. If ``hfile", the header parameter HCV is always used. If
neither is found, no correction is made.  If ``heliocentric" or
``barycentric" corrections are chosen, position and time parameters are
read from the spectrum data file header, and the correction is computed
as described in section \ref {bcv} below.

Emission and/or absorption lines may be removed from each input
spectrum if {\it reject} is yes. If the spectrum header parameter
SUBCONT is present, its value overrides that of {\it reject}. The
parameters {\it absrej} and {\it emrej} set the lower and upper
acceptable limits for input spectrum pixels in standard deviations of
the continuum fit to the spectrum.

\begin{figure}
     \plotone{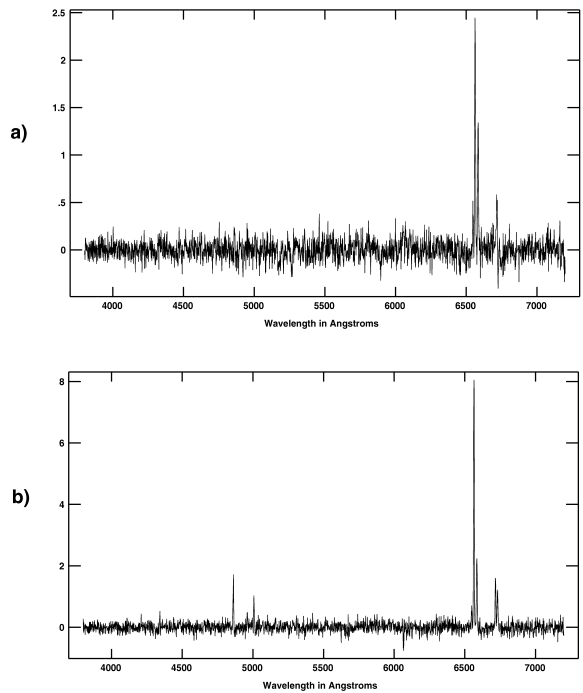}
     \caption{\label{sumspec4}
     Rebinned {\bf sumspec} input spectra with the continuum removed.}

     \end{figure}

Graphs of the continuum-removed, apodized data are displayed, as in figure
\ref{sumspec4} if the {\it cont\_plot} parameter is yes. 

\begin{figure}
     \plotone{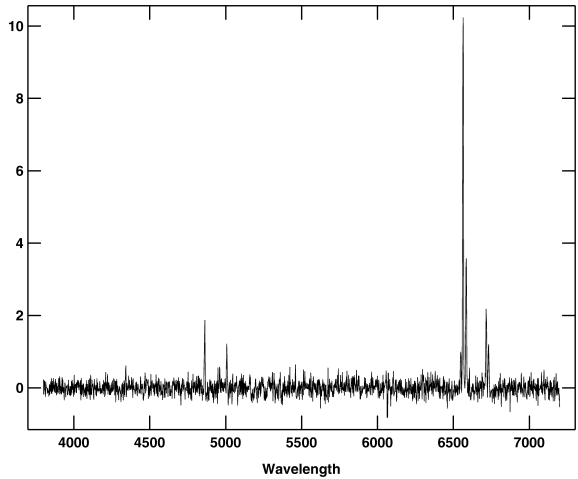}
     \caption{\label{sumspec5}
     Output composite of two input spectra plotted by {\bf sumspec}.}

     \end{figure}

The composite spectrum is plotted to the device specified by the
parameter {\it device}, as in figure \ref{sumspec5}, if {\it
comp\_plot} is yes. If {\it comp\_int} is yes, the display is held to
allow the user to interact with it using single character cursor
commands. If the ``@" command is given in this mode, the graph is sent to
the device specified by plotter. A ``q" command allows the task to proceed.

\begin{figure}
     \plotone{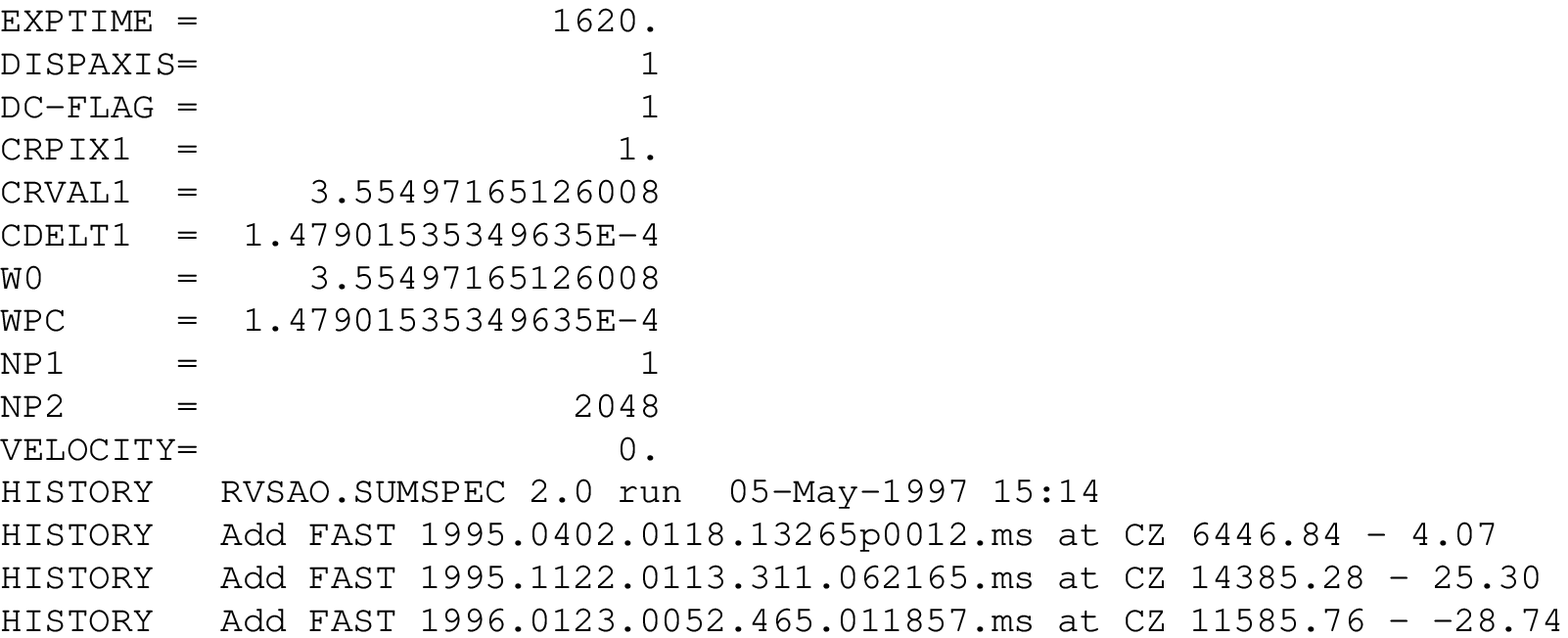}
     \caption{\label{sumspec6}
     These parameters are set in the output spectrum produced by
     {\bf sumspec}.}

     \end{figure}

After each spectrum is added, the {\bf sumspec} version and the current date
are written to the output spectrum image header and that spectrum
image file is written. After the last spectrum is added the file is
closed. Figure \ref{sumspec6} shows the parameters added to the
spectrum header. EXPTIME is the total exposure time from all of the
input images. A one-dimensional image is produced, so DISPAXIS is
always 1. DC-FLAG is 1 if the output spectrum is log-wavelength, in
which case the log-wavelength of the first pixel is given in both
CRVAL1 and W0, and the log-wavelength per pixel is given in CDELT1 and
WPC. The VELOCITY is set by the {\it velcomp} or {\it zcomp}
parameter. If no velocity is specified, it is left at 0. If {\it
savenames} is yes, the instrument, filename, input redshift velocity
and barycentric velocity correction are written to the header.

\subsection{\label{contpars}Removing a Spectrum's Continuum using {\bf contpars} 
or CONTSUM}

\begin{figure}
     \plotone{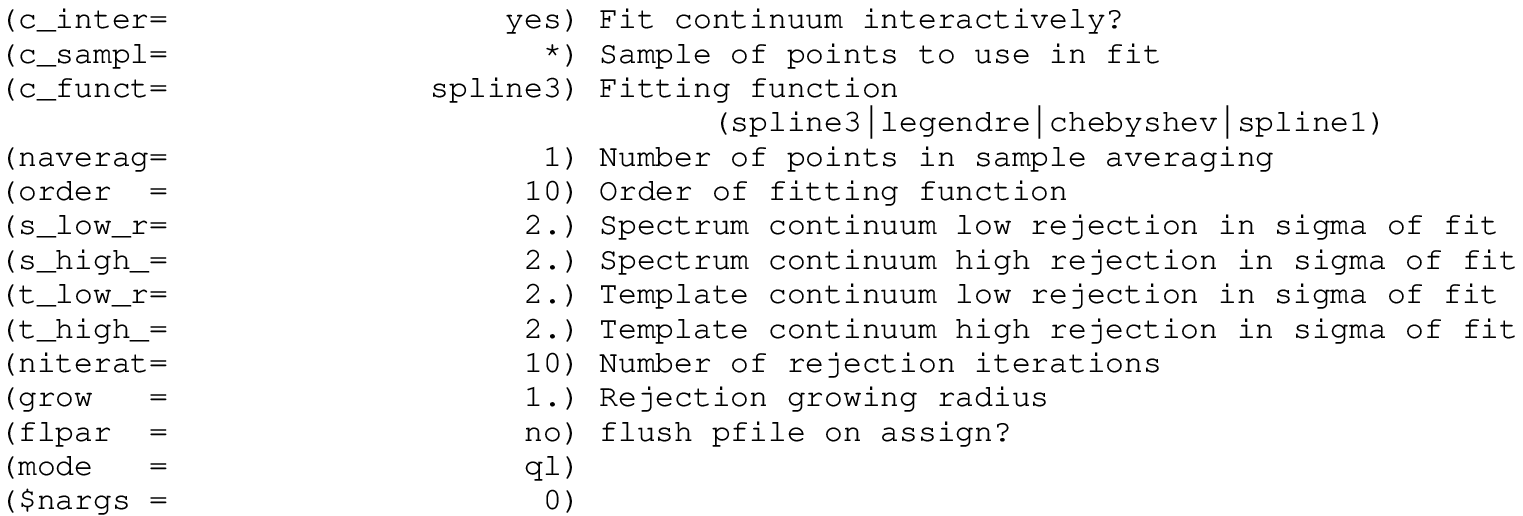}
     \caption{\label{contpars1}
     {\bf contpars} parameters as shown by IRAF eparam task.}

     \end{figure}

{\bf contpars} and CONTSUM are IRAF parameter files which are used by a
single subroutine to fit and remove a continuum from object or
template spectra.  The subroutine uses the IRAF interactive curve
fitting subroutine package and is based on the technique used in the
IRAF RV package (Fitzpatrick 1993).  The parameters which are set are
shown in figure \ref{contpars1}.  The same subroutine is called prior
to cross-correlation by {\bf xcsao}, an emission line search by {\bf emsao}, and
before or after summation of spectra by {\bf sumspec}.  If the SUBCONT
header keyword in a template spectrum is set to F, no continuum is fit or
removed from that template spectrum.  If the keyword is T or not present, the
continuum is fit and removed.

\begin{figure}
     \plotone{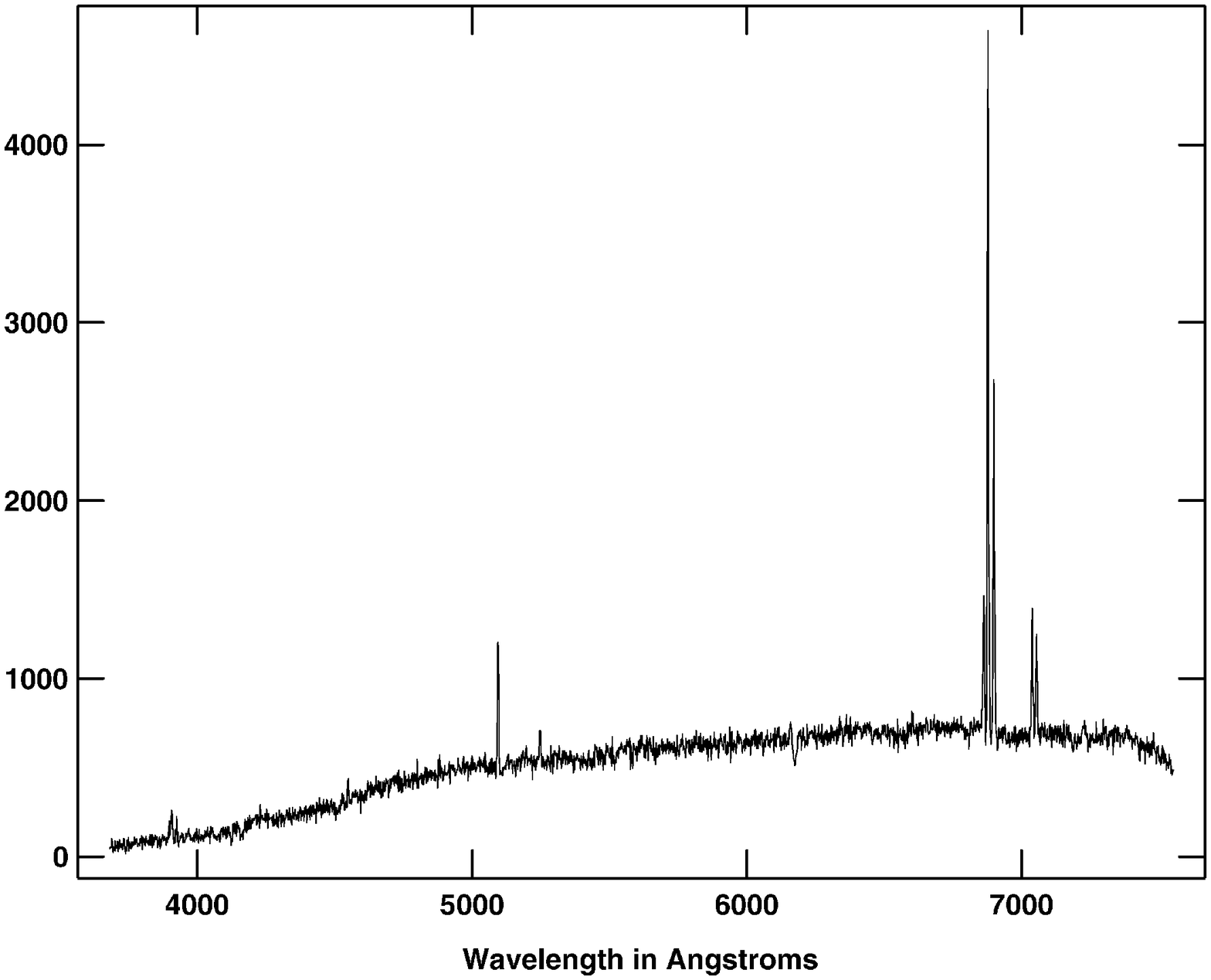}
     \caption{\label{contpars2}
     Input spectrum before the continuum is removed according to
     {\bf contpars}.}

     \end{figure} 

\begin{figure}
     \plotone{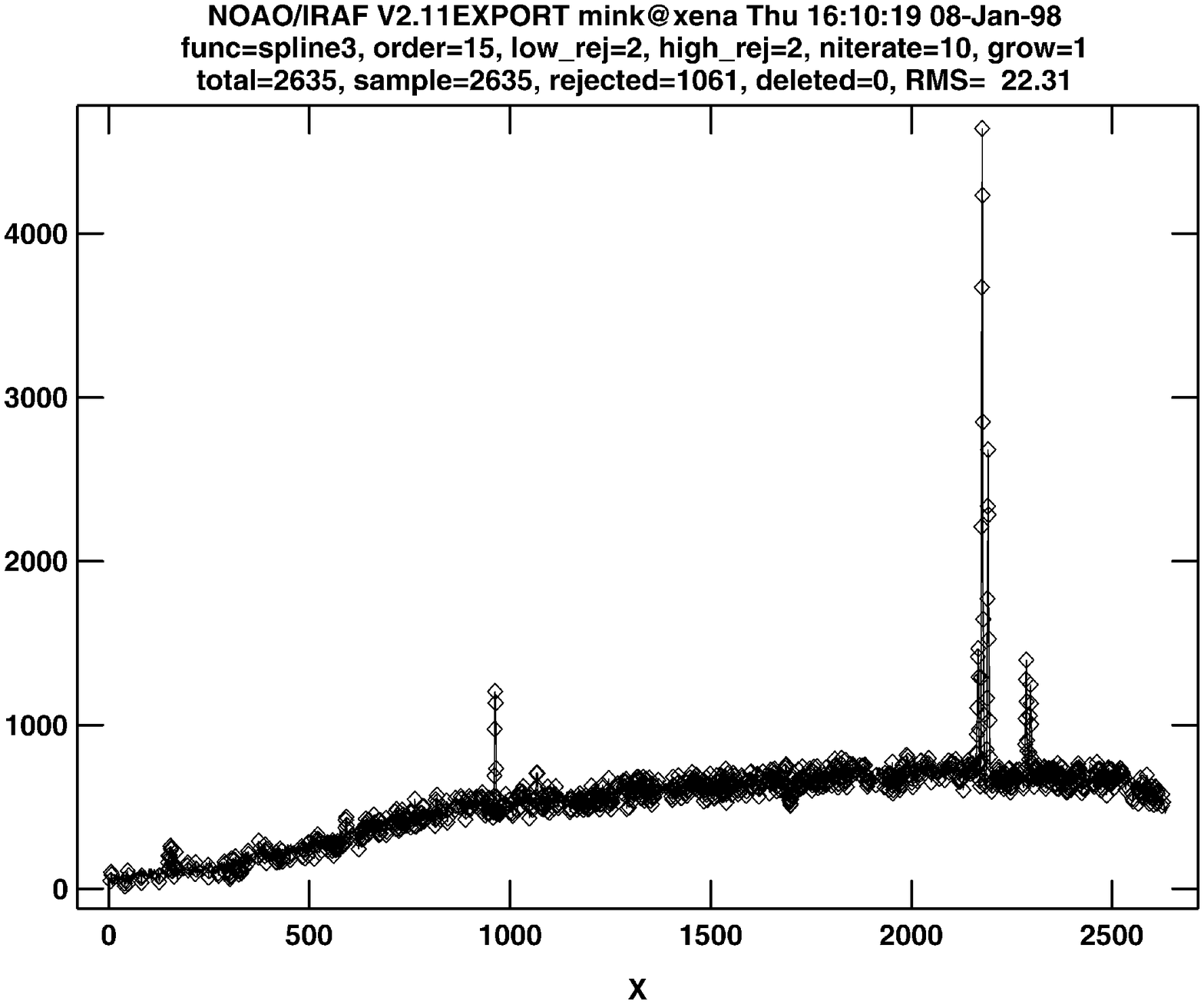}
     \caption{\label{contpars3}
     On this display shown during interactive continuum fitting, spectrum
     pixels rejected from the fit are marked with diamonds.}

     \end{figure}

\begin{figure}
     \plotone{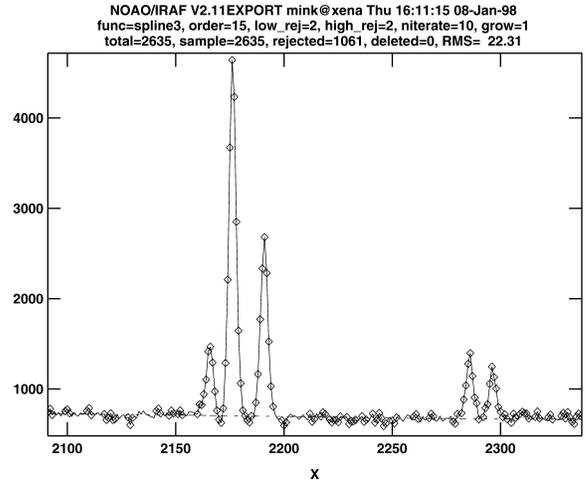}
     \caption{\label{contpars4}
     Here is a closeup of the region around H$\alpha$, showing the continuum
     which has been fit and giving a better view of the rejected pixels.}

     \end{figure}

For a spectrum, such as that in figure \ref{contpars2}, the continuum
is fit using the IRAF curve fitting subroutine. It may be used
interactively by setting the {\it c\_interactive} parameter to yes. In
that case, a graph of the fit and the spectrum is displayed as shown
in figures \ref{contpars3} and \ref{contpars4}.  The result may be
examined or refit after individual pixels are removed. A "q" command
is needed to exit from the interactive graph.

The {\it c\_function} parameter specifies the function to be used to
fit the continuum. If the continuum is irregular, it is best to use
the spline3, but spline1, Legendre, or Chebyshev polynomials may be
used as well. The parameter {\it order} sets the order of the fit.
The portion of the the spectrum to be fit is specified, in image
section format, by the parameter {\it sample}, Before the fit, the
spectrum is averaged in groups of {\it naverage} pixels.

Pixels with values outside of the limits {\it s\_low\_reject} and {\it
s\_high\_reject}, for spectra, and {\it t\_low\_reject} and {\it
t\_high\_reject}, for templates are rejected.  Separate limits are
used because composite template spectra will have much higher
signal-to-noise levels than object spectra and limits may need to be
set differently to fit a good continuum.  These limits are specified
in standard deviations of the fit.  To make sure that the wings of
lines are thoroughly eliminated, this rejection process is repeated
{\it niterate} times. Each time, a number of pixels, specified by the
{\it grow} parameter, are rejected on each side of the rejected pixel.

\begin{figure}
     \plotone{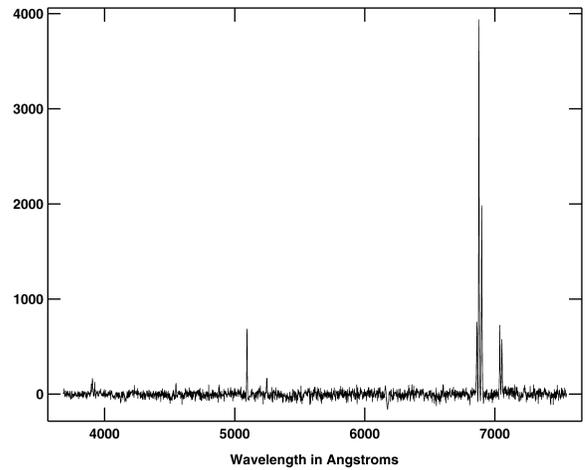}
     \caption{\label{contpars5}
     Figure \ref{contpars2} spectrum with its continuum removed.}

     \end{figure}

Normally, the continuum which is fit is subtracted from the spectrum,
preserving the signal-to-noise of a spectrum when it is dominated by
photon noise.  To remove the continuum of both template and object
spectra by division, the template spectrum header keyword DIVCONT is set
to T.  In that case, noise in portions of the spectrum with low
signal-to-noise levels will be amplified.  If {\it
emsao.contsub\_plot}, {\it xcsao.contsub\_plot}, or {\it
sumspec.cont\_plot} is yes, the spectrum is plotted with the continuum
removed, as shown in figure \ref{contpars5}.

\subsection{\label{bcv}Correcting Radial Velocities to the Solar System Barycenter}

\begin{figure}[t]
     \plotone{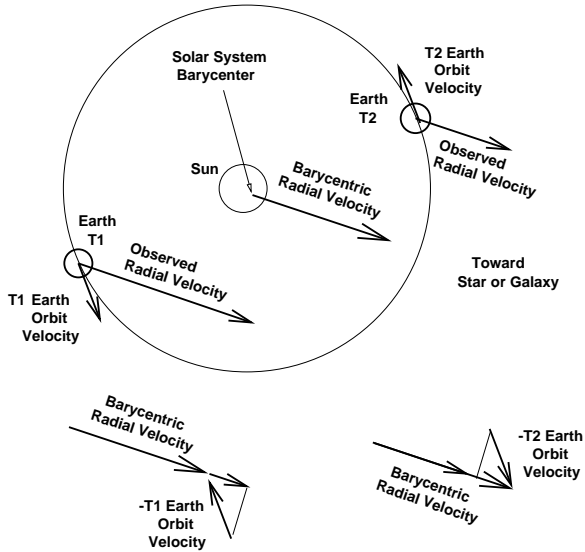}
     \caption{\label{bcvcorr0}
     The correction of a radial velocity observed from the earth to
     that which would be observed from the center of mass of the solar
     system is shown at two different times in the earth's orbit.}

     \end{figure}

To compare redshift velocities observed when the earth is at different
positions in its orbit, the velocity of the sun relative to the earth
is added to the redshift velocity.  This heliocentric velocity
correction is more accurately done to the true stationary point of the
solar system barycenter, its center of mass.  This barycentric
velocity correction is shown in figure \ref{bcvcorr0}.  A second,
smaller correction is added for the motion of the observer relative to
the center of the earth as it rotates. This vector velocity is
projected in the direction of the observed object, and that component
is saved.

The {\bf xcsao}, {\bf emsao}, and {\bf sumspec} tasks of the RVSAO
package compute this correction using subroutines which read the time
of observation, object position, and observatory position from the
spectrum image header.  Several common alternative keywords are built
into those subroutines.  RA, DEC, and EPOCH give the right ascension,
declination, and equinox of the observed object. SITELONG, SITELAT,
and SITEELEV for longitude, latitude, and altitude, respectively give
the observer's location on the earth. OBS-DATE yields the observation
date and UTMID, the midtime of the observation.  If UTMID is not found
in the header, UT, assumed to be the end time of the observation, or
UTOPEN, the start time of the observation are used in combination with
EXPOSURE or EXPTIME, which give the duration of the observation in
seconds to compute a midtime.  The requested heliocentric or
barycentric correction is then calculated at that midtime.

Since there is really no standard for the meaning of these keywords, a
separate task, {\bf bcvcorr}, has been added to RVSAO to allow several
alternate ways of specifying these three major pieces of information.
{\bf bcvcorr} can write its result to the header of the image which it is
processing; the other RVSAO tasks will use this value when their {\it
svel\_corr} parameter is set to ``file". First, the sky direction of
the object is set. The right ascension, declination, and equinox of
the object's sky position are read from the keywords specified by the
{\it keyra}, {\it keydec}, and {\it keyeqnx} parameters. If those are
not all found, the position is read directly from the parameters {\it
ra}, {\it dec}, and {\it equinox}.

The time for the velocity correction is read from a Julian date
keyword specified by {\it keyjd}. If none is found, the date is taken
from the header keyword specified by {\it keydate}. The UT midtime is
taken from the keyword in {\it keymid}. If that is not found, the
observation midtime is computed by finding the exposure duration from
the {\it keyexp} keyword value and adding half of it to the start time
from the {\it keystart} keyword value or subtracting half of it from
the end time in the {\it keyend} keyword value.  If no time can be
found in the header, the Julian date is read directly from the {\it
gjd} parameter, if it is greater than zero, or the midtime is taken
from the {\it year}, {\it month}, {\it day}, and {\it ut} parameters.

The heliocentric Julian Date (HJD) is the time at which the light from
the object for this observation reachs the sun. It is needed if
multiple radial velocity observations of an object are to be compared
accurately.  The HJD is not used in computation of the velocity correction,
which is dependent on the time of the observation at the earth. It may
be read from the header keyword {\it keyhjd} or set by the parameter
{\it hjd} (only if $>$ 0).  If none is set, the HJD is computed from the
observation time.

If the {\it obsname} parameter is ``file", the observatory name and
position is read from the image header using the keywords in {\it
keyobs}, for the name, {\it keylat}, for the latitude, {\it keylong},
for the longitude, and {\it keyalt}, for the altitude. Otherwise, the
value of {\it obsname} is used to get a position from IRAF's
observatory database. If the string is not found there, the longitude,
latitude, and altitude are given by the parameters {\it obslong}, {\it
obslat}, and {\it obsalt}.

\begin{figure}[t]
     \plotone{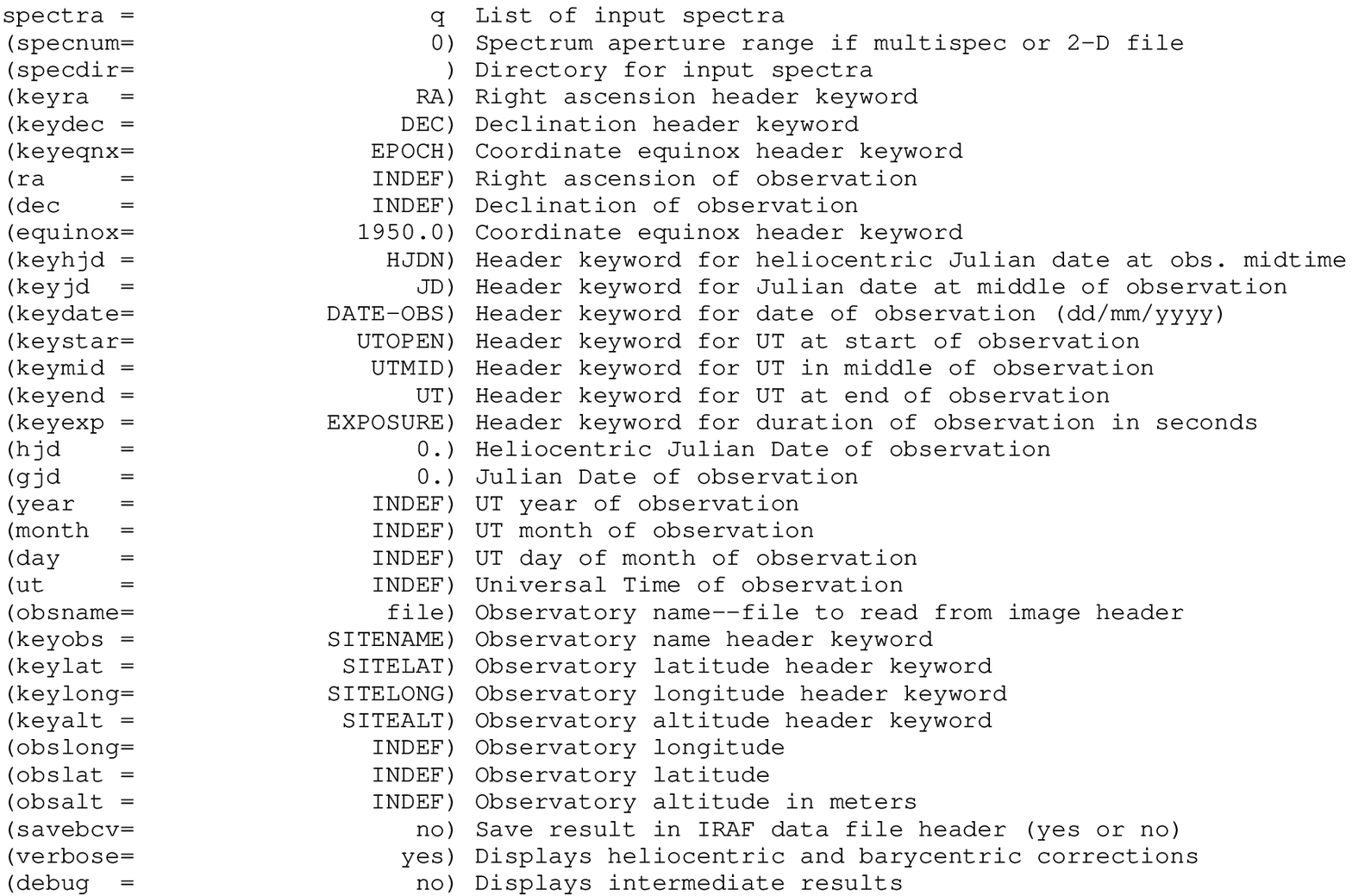}
     \caption{\label{bcvcorr1}
     {\bf bcvcorr} parameters as shown by IRAF eparam task.}

     \end{figure}

If {\it verbose} is yes, the information in Figure \ref{bcvcorr1} is
printed to standard output. If {\it savebcv} is yes, several spectrum
image header keywords are set.  BCV is the barycentric velocity
correction in km/sec.  The midtime of the observation is stored three
ways: UTMID, the Universal Time, GJDN, the geocentric Julian Date, and
HJDN, the heliocentric Julian Date.  These times are displayed by
{\bf xcsao} and {\bf emsao} as the time of observation.

\begin{figure}[t]
     \plotone{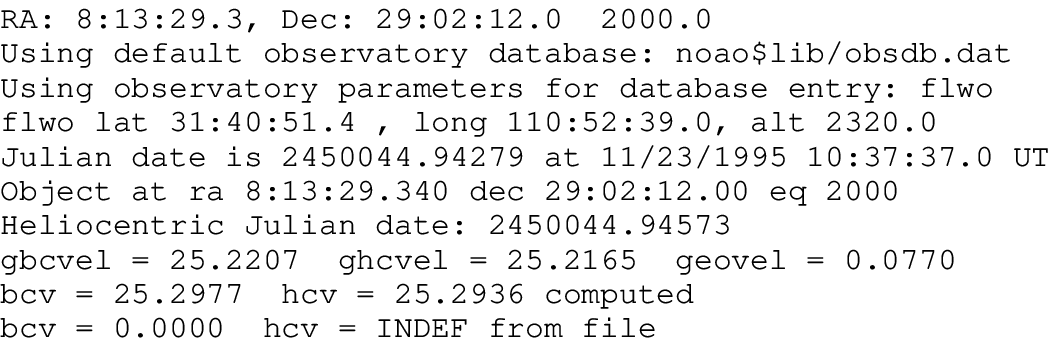}
     \caption{\label{bcvcorr2}
     Output from {\bf bcvcorr}}

     \end{figure}


\begin{thebibliography}{}

\bibitem [Babcock 1955]{bab55} Babcock, H.W., 1955, Ann. Report of the
Director of Mount Wilson and Palomar Observatories, 1954/55, p. 27.

\bibitem [Baranne, et al.,\ 1979]{bar79} Baranne, A., Mayor, M., and
Poncet, J.L. 1979. Vistas in Astronomy, 23, 279.

\bibitem [Barton, et al.,\ 1998]{bar98} Barton, et. al.,\ 1998, in
preparation.

\bibitem [Blackman and Tukey 1958]{bla58} Blackman, R.B. and Tukey,
J.W., 1958, The Measurement of Power Spectra, New York: Dover.

\bibitem [Bottinelli, et al.,\ 1990]{bot90} Bottinelli, L.,
Gouguenheim, L., Fouque, P., and Paturel, G., 1990, \aaps , 82, 391.

\bibitem [Brault and White 1971]{bra71} Brault, J.M. and White,
O.R. 1971, \aap , 13, 169.

\bibitem [Bromley, et al.,\ 1997]{bro97} Bromley, B.C., Press, W.H.,
Lin, H., and Kirshner, R.P., 1997, submitted to \apj .

\bibitem [Cooley and Tukey 1965]{coo65} Cooley, J.W. and Tukey, J.W.,
1965, Mathematics of Computation, 19, 297.

\bibitem [Da Costa et al.,\ 1994]{dac94} da Costa, L.N., Geller, M.J.,
Pellegrini, P.S., Latham, D.W., Fairall, A.P., Marzke, R.O., Willmer,
C.N.A., Huchra, J.P., Calderon, J.H., Ramella, M., and Kurtz, M.J.,
1994, \apjl , 424, 1.

\bibitem [Doppler 1841]{dop41} Doppler, C., 1841, K\"onigliche
b\"ohmische Gesellschaft der Wissenschaft Abhandlungen, ii, p. 465.

\bibitem [Evershed 1913]{eve13} Evershed, J., 1913, Kodaikanal
Bulletin, 3, 17, (No. 32).

\bibitem [Fabricant et al.,\ 1994]{fab94} Fabricant, D. G., Hertz,
E. H., and Szentgyorgyi, A. H., 1994 in Proc. SPIE Vol. 2198,
Instrumentation in Astronomy VIII, D. L. Crawford;
E. R. Craine; Eds., p. 251.

\bibitem [Fabricant et al.,\  1998]{fab98} Fabricant, D.G., Cheimets, P.,
Caldwell, N. and Geary, J. 1998, 
\pasp , 110, 79.

\bibitem [Fellgett 1955]{fel55} Fellgett, P., 1955, Optica Acta, 2, 9.

\bibitem [Fitzpatrick 1993]{fit93} Fitzpatrick, M. J., 1993 
in Astronomical Data Analysis Software and Systems II, ASP Conference
Series vol. 52, R.J. Hanisch, R.J.V. Brissenden, and J. Barnes, eds.,
p. 472.

\bibitem [Fizeau 1848]{fiz48} Fizeau, M.H., 1848, lecture before the
Soci\'et\'e Philomathique, 23. December 1848.

\bibitem [Fizeau 1870]{fiz70} Fizeau, M.H., 1870, Annales de Chimie et
de Physique, 19, 217.

\bibitem [Geller et al.,\ 1997]{gel97} Geller, M.J., Kurtz, M.J.,
Wegner, G., Thorstensen, J.R., Fabricant, D.G., Marzke, R.O., Huchra,
J.P., Schild, R.E., and Falco E.E., 1997, \aj , 114, 2205.

\bibitem [Geller, et al.,\ 1998]{gel98} Geller, M.J., et al., 1998,
15R Survey, in preparation.

\bibitem [Glazebrook, et al.,\ 1997]{gla97} Glazebrook, K., Offer,
A.R., and Deelet, K., 1997,  \apj , 492, 98.

\bibitem [Griffin 1967]{gri67} Griffin, R.F., 1967, \apj , 148, 465.

\bibitem [Hassab and Boucher 1979]{has79} Hassab, J.C. and Boucher,
R.E., 1979, IEEE Trans. Acoustics, Speech and Signal Processing,
ASSP-27, 922.

\bibitem [Huchra et al.,\ 1995]{huc95} Huchra, J.P. Geller, M.J.
and  Corwin, H.J. 1995, \apjs , 99, 391. 

\bibitem [Huggins 1868]{hug68} Huggins, W., 1868, Philosophical
Transactions of the Royal Society of London, 158, 529.

\bibitem [Kurtz 1982]{kur82} Kurtz, M.J., 1982, Automated Spectral
Classification, PhD Thesis, Dartmouth College, Hanover.

\bibitem [Kurtz, et al.,\ 1992]{kur92} Kurtz, M.J., Mink, D.J., Wyatt,
W,F., Fabricant, D.G., Torres, G., Kriss, G.A., and Tonry, J.L., 1992,
in Astronomical Data Analysis Software and Systems I, ASP Conference
Series vol. 25, D.M. Worrall, C. Biemesderfer, and J. Barnes, eds.,
p. 432.

\bibitem [Kurtz and La Sala 1991]{kur91} Kurtz, M.J. and La Sala, J.,
1991, in Objective-Prism and Other Surveys, A.G.D. Phillip and
A.R. Upgren, eds., Schenectady: L. Davis Press, p. 133.

\bibitem [Kurucz 1992]{kuru92} Kurucz, R.L., 1992, in Proc. IAU
Symp. 149 The Stellar Population of Galaxies, B. Barbuy and
A. Renzini, eds., Dordrecht: Reidel, p. 225.

\bibitem [Lacy 1977]{lac77} Lacy, C.H. 1977, \apj , 218, 444.

\bibitem [La Sala and Kurtz 1985]{las85} La Sala, J. and Kurtz, M.J.,
1985, \pasp , 97, 605.

\bibitem [Latham 1982]{lat82} Latham, D.W., 1982, in Proc. IAU
Coll. 67 Instrumentation for Astronomy with Large Optical Telescopes,
C.M. Humphries, ed., Dordrecht: Reidel, p. 259.

\bibitem [Latham 1985]{lat85} Latham, D.W., 1985, in Proc. IAU
Coll. 48 Stellar Radial Velocities, A.G.D. Phillip and D.W. Latham,
eds., Schenectady: L. Davis Press, p. 21.

\bibitem [Latham 1992]{lat92} Latham, D.W., 1992, in Proc. IAU
Coll. 135 Complementary Approaches to Binary and Multiple Star
Research, ASP Conference Series, Vol. 32, H. McAlister, and
W. Hartkopf (eds.), p. 110.

\bibitem [Latham, et al.,\ 1996]{lat96} Latham, D.W., Nordstr\"om, B.,
Andersen, J., Torres, G., Stefanik, R.P., Thaller, M, and Bester,
M.J., 1996, \aap , 314, 864.

\bibitem [Maker et al.,\ 1982]{mak82} Maker, S.,
Kurtz, M.J. and La Sala, J. 1982, {\it The
REDUCE/INTERACT Data Reduction System} (Hanover: Dartmouth College
Department of Physics and Astronomy).

\bibitem [Mink and Wyatt 1992]{min92} Mink, D.J. and Wyatt, W.F.  in
Astronomical Data Analysis Software and Systems I, ASP Conference
Series vol. 25, D.M. Worrall, C. Biemesderfer, and J. Barnes, eds.,
p. 439.

\bibitem [Mink and Wyatt 1995]{min95} Mink, D.J. and Wyatt, W.F. 1995,
in Astronomical Data Analysis Software and Systems IV, ASP Conference
Series, Vol. 77, R.A. Shaw, H.E. Payne, and J.J.E. Hayes, eds., p.
496.

\bibitem [Morse et al., 1991]{mor91} Morse, J.A., Mathieu, R.D., and
Levine, S.E., 1991, \aj , 101, 1495.

\bibitem [Nordstr\"om, et al.,\ 1994]{nor94} Norstr\"om, B., Latham,
D.W., Morse, J.A., Millone, A.A.E., Kurucz, R.L., Andersen, J., and
Stefanik, R.P., 1994, \aap , 287, 338.

\bibitem [Oppenheim and Schafer 1975]{opp75} Oppenheim, A.V. and
Schafer, R.W., 1975, digital Signal Processing, Englewood Cliffs:
Prentice-Hall. 

\bibitem [Press 1995]{pre95} Press, W.H., 1995, Unpublished colloquium
talk, Harvard-Smithsonian Center for Astrophysics.

\bibitem [Press 1997]{pre97} Press, W.H., 1997, Personal
communication.

\bibitem [Quintana, et al.\ 1996]{quin96} Quintana, H., Ramirez, A.,
and Way, M.J., 1996, \aj , 112, 36.

\bibitem [Rybicki and Press 1995]{ryb95} Rybicki, G.B. and Press, W.H., 
1995, \prl , 74, 1060.

\bibitem [Sargent, et al.,\ 1977]{sar77}  Sargent, W.L.W., Schechter,
P.L., Boksenberg, A., and Shortridge, K. 1977, \apj , 212, 326.

\bibitem [Simkin 1974]{sim74} Simkin, S.J., 1974, \aap , 31, 129.

\bibitem [Shectman et al.,\ 1996]{she96} Shectman, S.A.,
Landy, S.D., Oemler, A., Tucker, D.L., Lin, H.,
Kirshner, R.P., Schechter, P.L. 1996, \apj ,  470, 172

\bibitem [Tody 1986]{tod86} Tody, D. 1986, in Proc. SPIE
Instrumentation in Astronomy VI, ed. D.L. Crawford, 627, 733

\bibitem [Tody 1993]{tod93} Tody, D. 1993, in Astronomical Data
Analysis Software and Systems II, A.S.P. Conference Ser., Vol 52,
eds. R.J. Hanisch, R.J.V. Brissenden, and J. Barnes, 173.

\bibitem [Tokarz and Roll 1997]{tok97} Tokarz, S.P. and Roll, J., 1997,
Astronomical Data Analysis Software and Systems VI, A.S.P. Conference
Series, Vol. 125,  G. Hunt and H. E. Payne, eds., p. 140.

\bibitem[Tonry and Davis 1979]{ton79} Tonry, J.L. and Davis, M. 1979, \aj ,
43, 393, TD79.

\bibitem [Tonry and Wyatt 1988]{ton88} Tonry, J.L. and Wyatt, W.F.,
1988, CFA Z-Machine Data Analysis Software, Cambridge: Smithsonian
astrophysical Observatory.

\bibitem [Vettolani et al.,\  1997]{vet97} Vettolani, G., Zucca, E.,
Zamorani, G., Cappi, A., Merighi, R., Mignoli, M., Stirpe, G.M., 
MacGillivray, H., Collins, C., Balkowski, C., Cayette, V., Maurogordato, S.,
Proust, D., Chincarini, G., Guzzo, L., Maccagni, D., Scaramella, R.,
Blanchard, A., and Ramella, M.,  
1997, \aap, 325, 954.

\bibitem [Zucker and Mazeh 1994]{zuk94} Zucker, S. and Mazeh, T.,
1994, \apj , 420, 806

\end{thebibliography}
\end{document}